\documentclass{ws-ijmpa}
\usepackage[super,compress]{cite}
\usepackage{graphicx}
\usepackage{amsmath,amssymb,graphicx,array,calc,epsfig}
\usepackage{multirow,bigdelim}
\usepackage{arydshln} 
\usepackage{slashbox,pict2e}
\usepackage{longtable}
\usepackage{color}
\usepackage[
      colorlinks=true,
      linkcolor=red,  
      urlcolor=blue,    
      filecolor=red,     
      linkcolor=blue,
      citecolor = red,
      pdfstartview=FitV,
      pdftitle={},
        pdfauthor={Paolo Benincasa},
        pdfsubject={},
        pdfkeywords={},
        pdfpagemode=None,
        bookmarksopen=true    
      ]{hyperref}
\usepackage{epstopdf}
\usepackage{ccaption}
\usepackage{makecell}
\begin{document}
\markboth{Paolo Benincasa}
{New structures in scattering amplitudes: A review}

%
\catchline{}{}{}{}{}
%

\title{NEW STRUCTURES IN SCATTERING AMPLITUDES: A REVIEW
}

\author{PAOLO BENINCASA}

\address{Departamento de F{\'i}sica de Part{\'i}culas, \\
Universidade de Santiago de Compostela\\
E-15782 Santiago de Compostela, Spain\\
\&\\
Instituto de F{\'i}sica Te{\'o}rica, \\
Univerisdad Aut{\'o}noma de Madrid / CSIC \\
Calle Nicolas Cabrera 13, Cantoblanco 28049, Madrid, Spain
\\
paolo.benincasa@usc.es}

%
%
\maketitle


\begin{abstract}
We review some recent developments in the understanding of field theories in the perturbative regime. In particular, we discuss the notions of analyticity,
unitarity and locality, and therefore the singularity structure of scattering amplitudes in general interacting theories. We describe their tree-level structure
and their on-shell representations, as well as the links between the tree-level structure itself and the structure of the loop amplitudes. Finally, we describe
the on-shell diagrammatics recently proposed both on general grounds and in the remarkable example of planar supersymmetric theories. This view is partially
based on lectures given at: Dipartimento di Fisica and INFN, Universit{\`a} di Bologna; Departamento de F{\'i}sica de Part{\'i}culas, Universidade de Santiago de 
Compostela; and as part of the program Strings@ar Lectures on Advanced Topics of High Energy Physics held at the IAFE.

\end{abstract}


\tableofcontents

\section{Introduction}\label{sec:Intro}	

Standard approach in quantum field theory is strongly based on two crucial assumptions: locality and unitarity. These features have been codified in the Feynman 
diagrammatics, which has been the main tool to get an handle on perturbation theory and, thus, for the computation of scattering processes. If on one side it is
undeniable that this approach allowed us to make progress in our understanding of perturbative physics, on the other side it is a fair question to ask
to which extent we really understand it.

Demanding to have manifest locality and unitarity, as it happens in the Lagrangian formulation which Feynman's rules come from, leads to the introduction of
gauge redundancies with the side effect of hiding many properties of the theory. This is evident from the analysis of scattering amplitudes. The typical example
of the complications induced by this approach is the Parke-Taylor formula for $n$-gluon MHV amplitudes at tree level \cite{Parke:1986gb}, which is a single 
function of holomorphic Lorentz invariants. This structure is not really intuitive if we look at the Feynman representation for this amplitudes: especially
for an arbitrary number $n$ of external gluons, one would need to sum over a high number of diagrams to obtain this really simple result -- given that individual 
Feynman diagrams break gauge invariance, at most it is possible to make a smart gauge choice to try to simplify, at least partially, the computation. Thus,
the computation turns out to be cumbersome and, even once the final result is obtained, it is not clear if there is anything underlying which leads to such
a simple formula. 

A further suggestion that Feynman's representation ends up hiding interesting structures of the scattering amplitudes was already given by Berends-Giele-like
recursion relations \cite{Berends:1987me, Mangano:1987xk, Berends:1988zp, Berends:1989hf, Kosower:1989xy, Mangano:1990by}, and then it started to become
more and more evident once the {\it unitarity based methods} \cite{Bern:1994zx, Bern:1994cg, Bern:1995db, Bern:1996je, Bern:1996fj, Bern:1996ja} as well as new 
tree-level representation were introduced for pure Yang-Mills theory \cite{Cachazo:2004kj, Britto:2004aa, Britto:2005aa}.

In particular, the so-called BCFW-construction \cite{Britto:2005aa} provided, and still provides, a very powerful and general method to learn about the structure
of scattering amplitudes. The main idea is to introduce a one-parameter deformation of the complexified momentum space, to study the analytic structure of the
function that the deformation generates, and finally to reconstruct the physical amplitudes through the analysis of the singularities in this parameter. This procedure 
allowed to reveal a recursive structure not only for Yang-Mills tree-level amplitudes \cite{Britto:2004aa, Britto:2005aa}, but also for tree-level amplitudes in 
general relativity \cite{Bedford:2005yy, Benincasa:2007qj}, in theories with different species of particles whose highest spin is $1$ or $2$ \cite{Cheung:2008dn} and 
in maximally supersymmetric theories \cite{ArkaniHamed:2008gz}.

Actually, as far as the last class of theories is concerned, $\mathcal{N}\,=\,4$ supersymmetric Yang-Mills theory (in the planar sector) turned out to have a
very rich structure. In particular, it admits a {\it dual} description as a supersymmetric Wilson loop \cite{Alday:2010zy, Mason:2010yk, CaronHuot:2010ek} and, as a 
consequence, it is endowed with further symmetries: the dual (super)conformal \cite{Drummond:2006rz} and the Yangian symmetries \cite{Drummond:2009fd}. These are 
feature which gets completely obscured in the standard Feynman diagram language.

Furthermore, the very same theory (again in the planar sector) is characterized by amplitudes which have an all-loop BCFW-like recursive structure 
\cite{ArkaniHamed:2010kv}. This structure is intimately connected to the fact that these amplitudes have a natural definition on the (positive) Grassmannian
\cite{ArkaniHamed:2009dn, ArkaniHamed:2009sx, ArkaniHamed:2010kv}. Not only. If one begins with the Grassmannian formulation, both the dual superconformal and
Yangian symmetries appear are manifest \cite{ArkaniHamed:2009go}, with no mention of neither locality nor unitarity.

The BCFW-construction provides a fully on-shell representation of the theory, so that gauge-invariance (if any) is preserved by each individual diagram. However,
there is a price one has to pay: each term in the recursion relations show spurious poles, that indeed cancel upon the sum of all the diagrams. As a consequence,
each single BCFW-diagram breaks locality (that is anyway restored once the sum is performed).

Furthermore, the existence of such a recursion relations means that, if one iterates the procedure, one can end up having the amplitudes expressed in terms of products
of just three-particle amplitudes. More precisely, one can find the amplitude expressed in terms of products of some minimal amplitude. Despite of this, one
can introduce some massive particle and, therefore, some effective three-particle couplings which splits the higher particle minimal amplitudes (which anyhow
needs to be recovered in the suitable large mass limit). The most important implication of all this is that any theory which admits an on-shell representation
needs only the data encoded in the three-particle amplitudes to be totally determined. What is special about the three-particle amplitudes is that they not only
do not vanish in the complexified momentum space, but also are fixed by Poincar{\'e} invariance.

One feature which a theory typically needs to be endowed with is that all its scattering amplitudes need to have a good complex-UV behavior, {\it i.e.} they need
to vanish as the momenta are taken to infinity along some complex direction. What about those amplitudes which do not fulfill this condition? This is
a long-standing issue and it has not been completely explored yet, with the exception of some very specific case \cite{Feng:2009ei, Feng:2010ku} and a more
general approach based on the knowledge of a subset of the zeros of the amplitudes  \cite{Benincasa:2011kn}. Even if the latter still suffers of some issue,
it reveals a very interesting property: at tree level, even theories which do not vanish in the complex UV are still characterized with the very same structure
and satisfy a dressed version of the standard BCFW-representation. In other words, even in those cases, the three-particle amplitudes can determine the full
tree-level. Actually, the three-particle amplitudes themselves contain information about the possibility of admitting a standard BCFW-representation
\cite{Schuster:2008nh, Benincasa:2011pg} and, thus, about the complex-UV behavior. Another amazing result of this point of view is that features such as the gauge 
algebra \footnote{In relation to gauge theories, a total on-shell perspective on $U(1)$-decoupling relation, KK \cite{Kleiss:1988ne} and BCJ \cite{Bern:2008qj} 
relations was provided \cite{Feng:2010my}}, supersymmetry or also no-go theorems about the couplings emerge naturally from simple consistency conditions 
\cite{Benincasa:2007xk, Benincasa:2011pg, McGady:2013sga}.

Thus, a very natural idea to start from the three-particle amplitudes, which are determined from first principles ({\it i.e.} the invariance under
the isometry group of the space-time), and try to build higher point objects by gluing them together in a consistent fashion under a minimal amount of assumptions.
This direction has been developed in the seminal work of Arkani-Hamed et al \cite{ArkaniHamed:2012nw} where a first principle direct connection between 
scattering amplitudes and the Grassmannian was established. The on-shell diagrams constructed by suitably gluing together the three-particle amplitudes 
actually represent physical processes and scattering amplitudes in planar $\mathcal{N}=4$ SYM can be computed through them, with no mention about virtual particles.
Strikingly, on one side the on-shell diagrams can be characterized by permutations, and on the other side each on-shell diagram can be associated to a particular 
{\it configuration} among the boundaries of the positive Grassmannian: in this formulation, the conformal and dual conformal symmetries are mapped into each other
by a mapping of permutations.

Even if neither locality nor unitarity are taken as guiding principles, in a sense they are built in the way that the three-particle building blocks are 
glued together. As a consequence, this formulation, as it stands, does not really show locality and unitarity as emergent properties. More progress in this
directions have been recently made with the definition of a new object, the {\it amplituhedron} \cite{Arkani-Hamed:2013jha}, which is a generalization
of the positive Grassmanian. In this context, the amplitudes are identified as the ``volume'' of such an object, and both locality and unitarity emerges as
consequence of positivity.

In this review, we intend to describe, with some level of pedagogy, some of these developments, focusing on those -- and the related issues -- which are general
to any field theory or, anyway, have a potential generalization. Its spirit is to try to summarize, in a self-contained way, some of the salient features pointing to a
general redefinition of (perturbative) interacting theories in terms of a minimal amount of assumptions and, thus, with a deeper understanding of their features. The 
hope is that it can provide a certain degree of complementarity with the already existing reviews on the subject \cite{Brandhuber:2011ke, Dixon:2011xs, Feng:2011np, 
Elvang:2013cua}.

The structure of this review goes as follows: 
In Section \ref{sec:Smatr}, we provides some generalities about scattering amplitudes, with an extensive discussion of the Lorentz little group as well as three
pillars of our current understanding: analyticity, unitarity and locality. Section \ref{sec:3ptAmpl} is entirely devoted to the three-particle amplitudes.
Section \ref{sec:HPampl} describes the pillars of the construction of high point on-shell objects, starting with a general discussion of the momentum space
deformation, and providing a notion of constructibility and tree-level consistency \footnote{As far as the tree level is concerned, new progress has been 
recently obtain for amplitudes in gauge theory and gravity in arbitrary space-time dimensions \cite{Hodges:2012ym, Cachazo:2012da, Cachazo:2013zc, Cachazo:2013iaa,
Cachazo:2013gna, Cachazo:2013hca, Cachazo:2013iea, Monteiro:2013rya, Dolan:2013isa}.}. In particular, we underline how already at this level many features of a 
theory can be considered as emergent. Section \ref{sec:TreeLoop} deals with an analysis of the general loop structure and the connection between tree- and
loop-level properties. Finally, in Section \ref{sec:OnShDiag} we present the recently proposed on-shell diagrammatics, taking the point of view of a general 
interacting theory. We also discuss, as a specific case, the planar supersymmetric theories from this point of view. We decided not to include the
Grassmannian formulation of planar $\mathcal{N}=4$ SYM and the connection to the on-shell diagrammatics: on one side, as we said from the very beginning, we want to 
take the point of view of a generic interacting theory and, for the moment, this formulation is still too theory-specific, while on the other side, it deserves
a more extensive treatment of what could have been reserved here, so we refer to the original papers as well as to the review \cite{Elvang:2013cua}.


\section{The S-matrix: Generalities}\label{sec:Smatr}

Let us consider scattering processes in asymptotically Minkowski space-times. Supposing the existence of asymptotic states, they are defined as the irreducible 
representations of the space-time isometry group, in this case the Poincar{\'e} group $\Pi^{\mbox{\tiny $(D)$}}\,=\,\mathcal{T}^{\mbox{\tiny $D$}}\,\ltimes\,SO(D-1,1)$,
with $\mathcal{T}^{\mbox{\tiny $D$}}$ and $SO(D-1,1)$ being respectively the $D$-dimensional translations and the $D$-dimensional Lorentz group.
Given that the generators $\hat{P}_{\mu}$ of $\mathcal{T}^{\mbox{\tiny $D$}}$ commute, we can take the asymptotic states to be the direct product of eigenstates of 
$\hat{P}_{\mu}$ with eigenvalues $p_{\mu}$. Then, in a scattering process of $n$ particles, the S-matrix elements provide the transition amplitudes for 
$n_{\mbox{\tiny in}}$ initial states to produce $n_{\mbox{\tiny out}}\,=\,n-n_{\mbox{\tiny in}}$ final states
\begin{equation}\label{Smatrix}
 {}_{\mbox{\tiny out}}
  \langle 
   p^{\mbox{\tiny $(1)$}}\ldots p^{\mbox{\tiny $(n-n_{\mbox{\tiny in}})$}}|p^{\mbox{\tiny $(\sigma_1)$}}\ldots p^{\mbox{\tiny $(\sigma_{n_{\mbox{\tiny in}}})$}}
  \rangle_{\mbox{\tiny in}}\:=\:
  \langle
  p^{\mbox{\tiny $(1)$}}\ldots p^{\mbox{\tiny $(n-n_{\mbox{\tiny in}})$}}|\hat{S}|p^{\mbox{\tiny $(\sigma_1)$}}\ldots p^{\mbox{\tiny $(\sigma_{n_{\mbox{\tiny in}}})$}}
  \rangle,
\end{equation}
where the S-matrix operator $\hat{S}$ is unitary, ${}_{\mbox{\tiny out}}\langle p^{\mbox{\tiny $(1)$}}\ldots p^{\mbox{\tiny $(n-n_{\mbox{\tiny in}})$}}|$ and
$|p^{\mbox{\tiny $(\sigma_1)$}}\ldots p^{\mbox{\tiny $(\sigma_{n_{\mbox{\tiny in}}})$}}\rangle_{\mbox{\tiny in}}$ are respectively the state at infinite future
and past, while $\langle p^{\mbox{\tiny $(1)$}}\ldots p^{\mbox{\tiny $(n-n_{\mbox{\tiny in}})$}}|$ and 
$|p^{\mbox{\tiny $(\sigma_1)$}}\ldots p^{\mbox{\tiny $(\sigma_{n_{\mbox{\tiny in}}})$}}\rangle$ are the eigenstates of the momentum operator $\hat{P}_{\mu}$ spanning
respectively the whole future and past Hilbert spaces $\mathcal{H}_{\mbox{\tiny out}}$ and $\mathcal{H}_{\mbox{\tiny in}}$.

The S-matrix operator can be conveniently written as $\hat{S}\,=\,\hat{\mathbb{I}}+i\hat{T}$, with the operator $\hat{T}$ defining the scattering amplitude
\begin{equation}\label{ScatAmpl}
 \langle
 p^{\mbox{\tiny $(1)$}}\ldots p^{\mbox{\tiny $(n-n_{\mbox{\tiny in}})$}}|i\hat{T}|p^{\mbox{\tiny $(\sigma_1)$}}\ldots p^{\mbox{\tiny $(\sigma_{n_{\mbox{\tiny in}}})$}}
 \rangle\,=\,
 M_{n}\left(\left\{p^{\mbox{\tiny $(\sigma_1)$}}\ldots p^{\mbox{\tiny $(\sigma_{n_{\mbox{\tiny in}}})$}}\right\}\,\rightarrow\,
    \left\{p^{\mbox{\tiny $(1)$}}\ldots p^{\mbox{\tiny $(n-n_{\mbox{\tiny in}})$}}\right\}\right).
\end{equation}
The unitarity condition $\hat{S}\hat{S}^{\dagger}\,=\,\hat{\mathbb{I}}\:=\:\hat{S}^{\dagger}\hat{S}$ in terms of the operator $\hat{T}$ reads
\begin{equation}\label{Tunit}
 -i\left(\hat{T}-\hat{T}^{\dagger}\right)\:=\:\hat{T}^{\dagger}\hat{T}.
\end{equation}
Taking as convention that all the states are incoming, the object of interest is the scattering amplitude 
$M_{n}\,=\,M_{n}(p^{\mbox{\tiny $(1)$}},\,\ldots\,p^{\mbox{\tiny $(n)$}})$, and all the different processes can be computed from it by analytic continuation.
This object must be invariant under the Poincar{\'e} group. With the assumption that one-particle states exist, {\it i.e.} it is possible to define operators
which act on the whole scattering amplitude as they do on the one-particle states, the action of the space-time translations on $M_n$ is given by
\begin{equation}\label{Mtransl}
 M_n(p^{\mbox{\tiny $(1)$}},\,\ldots,\,p^{\mbox{\tiny $(n)$}})\:=\:
  e^{ix\cdot\sum_{i=1}^{n}p^{\mbox{\tiny $(i)$}}}M_n(p^{\mbox{\tiny $(1)$}},\,\ldots,\,p^{\mbox{\tiny $(n)$}})
\end{equation}
From such a relation, invariance under translations implies that the scattering amplitudes have a support on a $\delta$-function enforcing momentum conservation.
Lorentz invariance, instead, implies that the scattering amplitude is a function of Lorentz invariant combination of the momenta.

\subsection{The Lorentz little group}\label{subsec:LG}

Since now on, we specialize to four-dimensions, unless otherwise specified \footnote{For a discussion about dimensions different than four, see 
\cite{Cheung:2009dc, Boels:2009bv, Gang:2010gy}.}. There is a subgroup of the Lorentz group which deserves a particular discussion. It is
the subgroup of the Lorentz transformations which leaves the momentum of a given particle unchanged. These transformations can be classified by the two Casimir
operators of the Poincar{\'e} group, $\hat{P}^2$ and $\hat{W}^2$, with $\hat{W}^{\mu}$ being the Pauli-Lubanski pseudo-vector, which satisfies the commutation
relations
\begin{equation}\label{Wcomm}
 [\hat{W}^{\mu},\,\hat{P}^{\nu}]\,=\,0,\; [\hat{L}_{\mu\nu},\hat{W}_{\rho}]\:=\:i\left(\eta_{\nu\rho}\hat{W}_{\mu}-\eta_{\mu\rho}\hat{W}_{\nu}\right),
 \;
 [\hat{W}^{\mu},\,\hat{W}^{\nu}]\:=\:i\epsilon^{\mu\nu\rho\sigma}\hat{W}_{\rho}\hat{P}_{\sigma}.
\end{equation}
In order for an operator $e^{\frac{i}{2}\omega^{\mu\nu}\hat{L}_{\mu\nu}}$, with $\hat{L}_{\mu\nu}$ and $\omega^{\mu\nu}$ being respectively the
Lorentz generators and an arbitrary antisymmetric tensor, to keep a momentum invariant,
it needs to have the form $e^{-iv_{1}^{\mu}p_{\mu}}e^{-i\frac{1}{2}v_{2}^{\mu}\hat{W}_{\mu}}$, {\it i.e.} the little group transformations are generated by
$\hat{W}^{\mu}$. As it is manifest from the third commutator in (\ref{Wcomm}), the generators of the Little group satisfy an $so(3)$ algebra in the massive case
and an $iso(2)$ algebra in the massless one.

In this last case, it is always possible to go to a frame such that $p^{\mu}\,=\,(E,\,0,\,0,\,E)\,\equiv\,\mathfrak{p}^{\mu}$, so that the Pauli-Lubanski pseudo-vector
can be decomposed as
\begin{equation}\label{Wdec}
 \hat{W}^{\mu}\:=\:-\mathfrak{p}^{\mu}\hat{L}_{12}+e^{\mu}_{x}E\left(\hat{L}_{02}+\hat{L}_{32}\right)+e^{\mu}_{y}\left(-E\left(\hat{L}_{01}+\hat{L}_{31}\right)\right),
\end{equation}
with the above combination of the Lorentz generator components providing the generators of the transformations which leave $\mathfrak{p}^{\mu}$ invariant. More 
generally eq (\ref{Wdec}) can be written as
\begin{equation}\label{Wdec2}
 \hat{W}^{\mu}\:=\:-p^{\mu}\hat{\mathcal{H}}+\varepsilon^{\mu}_{1}\hat{\mathcal{T}}_{1}+\varepsilon^{\mu}_{2}\hat{\mathcal{T}}_{2}\:\equiv\:-p^{\mu}\hat{\mathcal{H}}-
  \frac{\varepsilon^{\mu}_{+}\hat{\mathcal{T}}_{-}+\varepsilon^{\mu}_{-}\hat{\mathcal{T}}_{+}}{\sqrt{2}},
\end{equation}
where $\hat{\mathcal{H}}$ and $\hat{\mathcal{T}}_{\pm}$ are respectively the little group rotation and translation, which satisfy the commutations relations: 
$[\hat{\mathcal{T}}_{+},\,\hat{\mathcal{T}}_{-}]\,=\,0$ and $[\hat{\mathcal{H}},\,\hat{\mathcal{T}}_{\mp}]\,=\,\mp \hat{\mathcal{T}}_{\mp}$. The second Poincar{\'e} 
Casimir turns out to be expressed in terms of $\hat{\mathcal{T}}_{\mp}$, while $\hat{\mathcal{H}}$ does not appear
\begin{equation}\label{W2b}
 \hat{W}^2\:=\:-\hat{\mathcal{T}}_{+}\hat{\mathcal{T}}_{-}.
\end{equation}
Let us now act with a general little group element on a one-particle state
\begin{equation}\label{LGel}
 \hat{\mathcal{W}}(\theta,\,\beta)|p\rangle\:\equiv\:e^{i\frac{\beta^{-}\hat{T}_{-}+\beta^{+}\hat{T}_{+}}{\sqrt{2}}}e^{-i\theta\hat{R}}|p\rangle,
 \qquad \theta\in[0,\,2\pi[,\;\beta\in\mathbb{C}.
\end{equation}
From eq (\ref{LGel}) and eq (\ref{W2b}), it is straightforward to see that one unitary representation is given by a state which is eigenfunction of 
$\hat{\mathcal{H}}$ with eigenvalue $-2h$ and on which the little group translation operators $\hat{\mathcal{T}}_{\pm}$ act trivially. In this case, the second 
Casimir acting on a state vanishes and $h$, in order to satisfy the periodicity condition, need to be integer or half-integer: this representation of the Lorentz 
little group provides the helicity states $|p,\,h\rangle$. On the other hand, one can also consider an unitary representation such that the action of 
$\hat{\mathcal{T}}_{\mp}$ is no longer trivial and, thus, the action of the second Casimir provides a scale $\rho^2$. In this case one can either choose a state 
to be eigenfunction of $\hat{\mathcal{H}}$, in which case it gets labelled by an integer $n$, or of $\hat{\mathcal{T}}_{\mp}$, in which case it gets labelled by a 
continuous parameter $\phi$. This representation, named {\it continuous spin representation} because of the possibility to provide the states with a continuous quantum
number, was  originally discussed by Wigner \cite{Wigner:1939cj}, then in the '70s \cite{Iverson:1971hq, Abbott:1976bb, Hirata:1977ss} until recently when it has been 
pointed out that, on one side, particles in this representation can mediate long-range forces \cite{Schuster:2013pxj, Schuster:2013vpr, Schuster:2013pta} and, on the 
other one, they are not present in perturbative string theory \cite{Font:2013hia}. In the rest of the review, we will not discuss these constructions and we will deal 
with helicity states only.

\subsection{Helicity amplitudes}\label{subsec:HelAmpl}

The physical information for the massless representation of the Poincar{\'e} group are encoded in the light-like momenta $p^{\mbox{\tiny $(i)$}}_{\mu}$ and in the 
polarization tensors $\varepsilon^{\mbox{\tiny $(i)$}}_{\mu_1\ldots\mu_s}$. One can also consider the fact that the Lorentz group $SO(3,1)$ is isomorphic to
$SL(2,\,\mathbb{C})$ and, thus, it is possible to map a Lorentz four-vector to a bi-spinor
\begin{equation}\label{pbispin}
 p_{\mu}\:\longrightarrow\:p_{a\dot{a}}\:=\:\sigma^{\mu}_{a\dot{a}}p_{\mu}\:=\:\lambda_{a}\tilde{\lambda}_{\dot{a}},
\end{equation}
where $\sigma^{\mu}_{a\dot{a}}\,=\,(\mathbb{I}_{a\dot{a}},\,\overrightarrow{\sigma}_{a\dot{a}})$ are the Pauli matrices, and the last equality, {\it i.e.} 
the possibility of representing a bi-spinor as a direct product of two spinors, holds because the momentum is light-like. The spinors $\lambda_{a}$ and 
$\tilde{\lambda}_{\dot{a}}$ respectively transform in the $(1/2,\,0)$ and $(0,\,1/2)$ representations of $SL(2,\,\mathbb{C})$. The spinor indices can be raised and
lowered by the two-dimensional Levi-Civita symbols $\epsilon_{ab}$ and $\epsilon_{\dot{a}\dot{b}}$, for which we take the convention 
$\epsilon_{\mbox{\tiny $12$}}\:=\:1\:=\:\epsilon_{\mbox{\tiny $\dot{1}\dot{2}$}}$ and 
$\epsilon^{\mbox{\tiny $12$}}\:=\:-1\:=\:\epsilon^{\mbox{\tiny $\dot{1}\dot{2}$}}$. They can be used to define a Lorentz invariant inner product for each of the two
representations of $SL(2,\,\mathbb{C})$:
\begin{equation}\label{SLinprod}
 \langle\lambda,\,\lambda'\rangle\:\equiv\:\epsilon^{\mbox{\tiny $ab$}}\lambda_{a}\lambda'_{b},\qquad
 [\tilde{\lambda},\,\tilde{\lambda}']\:\equiv\:\epsilon^{\mbox{\tiny $\dot{a}\dot{b}$}}\tilde{\lambda}_{\dot{a}}\tilde{\lambda}'_{\dot{b}}.
\end{equation}
The spinors $\lambda_a$ and $\tilde{\lambda}_{\dot{a}}$ carry helicity $-1/2$ and $+1/2$, respectively. This can be easily seen by considering the chiral Dirac
equations for negative/positive chiral massless fermion $\Psi^a$/$\tilde{\Psi}^{\dot{a}}$
\begin{equation}\label{ChirlEqs}
 i\sigma^{\mu}_{a\dot{a}}\partial_{\mu}\Psi^a\:=\:0\:=\:i\sigma^{\mu}_{a\dot{a}}\partial_{\mu}\tilde{\Psi}^{\dot{a}}
\end{equation}
whose solutions are the plane waves 
\begin{equation}\label{ChirlEqsSoln}
 \Psi^a\:=\:\psi^a\,e^{ix_{b\dot{b}}\lambda^b\tilde{\lambda}^{\dot{b}}},\qquad
 \tilde{\Psi}^{\dot{a}}\:=\:\tilde{\psi}^{\dot{a}}\,e^{ix_{b\dot{b}}\lambda^b\tilde{\lambda}^{\dot{b}}}
\end{equation}
if and only if the following conditions are, respectively, satisfied
\begin{equation}\label{ChirlEqsCond}
 \langle\psi,\,\lambda\rangle\:=\:0\:=\:[\tilde{\psi},\,\tilde{\lambda}].
\end{equation}
They imply that $\psi^a$ and $\tilde{\psi}^{\dot{a}}$ are respectively proportional to $\lambda^a$ and $\tilde{\lambda}^{\dot{a}}$.

Thus, the physical data about the external states of an amplitude can be encoded in the pairs of spinors 
$(\lambda^{\mbox{\tiny $(i)$}},\,\tilde{\lambda}^{\mbox{\tiny $(i)$}})$ and the helicities $h_i\:=\:\pm s_{i}$:
\begin{equation}\label{HelAmp}
 M_{n}\:=\:M_{n}\left(\{\lambda^{\mbox{\tiny $(i)$}},\,\tilde{\lambda}^{\mbox{\tiny $(i)$}};\,h_i\}\right).
\end{equation}
A further hypothesis is the possibility of defining operators which act on the amplitude as they act on the one-particle state. With such an assumption, the action of 
the helicity operator $\hat{\mathcal{H}}^{\mbox{\tiny $(i)$}}$, related to the particle labelled by $i$, is 
\begin{equation}\label{HelOpAct}
 \hat{\mathcal{H}}^{\mbox{\tiny $(i)$}}M_{n}
 \:\equiv\:
 \left(
  \lambda_a^{\mbox{\tiny $(i)$}}\frac{\partial}{\partial\lambda_a^{\mbox{\tiny $(i)$}}}-
  \tilde{\lambda}_{\dot{a}}^{\mbox{\tiny $(i)$}}\frac{\partial}{\partial\tilde{\lambda}_{\dot{a}}^{\mbox{\tiny $(i)$}}}
 \right)M_{n}
 \:=\:
 -2h_i\,M_{n}.
\end{equation}
Equivalently, the little group transformation maps $(\lambda,\,\tilde{\lambda})$ into $(t^{-1}\lambda,\,t\tilde{\lambda})$ and its action on the amplitude 
(\ref{HelAmp}) can be seen as the following scaling
\begin{equation}\label{HelScal}
 M_{n}\left(\{\lambda^{\mbox{\tiny $(i)$}},\,\tilde{\lambda}^{\mbox{\tiny $(i)$}};\,h_i\}\right)\quad\longrightarrow\quad
 t^{-2h_i}M_{n}\left(\{\lambda^{\mbox{\tiny $(i)$}},\,\tilde{\lambda}^{\mbox{\tiny $(i)$}};\,h_i\}\right).
\end{equation}

\subsection{Analyticity, locality and unitarity}\label{subsec:AnLocUn}
Up to the momentum conserving $\delta$-function, we consider the scattering amplitudes as an analytic function. This implies that their singularity structure can show
at most poles and branch points. If we also assume locality for the interactions, the analytic structure of the amplitude is further restricted: the poles can come
just from propagators $1/(\sum_k p^{\mbox{\tiny $(k)$}})^2$ and branch points are identified by those points in momentum space in which two propagators are
simultaneously singular. 

Finally, unitarity, which is given by the condition (\ref{Tunit}), implies that, when the singularities are approached, one or more particles go on-shell
and the amplitude factorizes into amplitudes with lower number of external states and/or at lower order in perturbation theory.

\subsubsection{Poles and trees}\label{subsubsec:Tree}
In a Feynman diagram language, considering just poles is equivalent to restricting to the tree-level approximation. In other words, tree level amplitudes are
rational functions of the Lorentz invariants. These poles are approached when
\begin{romanlist}[(ii)]
 \item $P^2_{ij}\:\equiv\:(p^{\mbox{\tiny $(i)$}}+p^{\mbox{\tiny $(j)$}})^2\:=\:\langle i,j\rangle[i,j]\:\longrightarrow\:0$: The amplitude factorizes as
       \begin{equation}\label{ColLimit}
        \lim_{P^2_{ij}\,\rightarrow\,0}P^2_{ij}M_n^{\mbox{\tiny tree}}\:=\:
         \sum_{h_{P}}\mbox{Split}^{\mbox{\tiny tree}}_{\mbox{\tiny $-h_{P}$}}(i,j,\zeta)M_{n-1}^{\mbox{\tiny tree}}(P_{ij}^{h_{P}},\,\mathcal{K}),
       \end{equation}
       where $\mathcal{K}$ is the set of all particles but $i$ and $j$, $\mbox{Split}^{\mbox{\tiny tree}}$ is a so called {\it splitting amplitude}, and the parameter
       $\zeta$ is such that $p^{\mbox{\tiny $(i)$}}\,\sim\,\zeta P_{ij}$ and $p^{\mbox{\tiny $(j)$}}\,\sim\,(1-\zeta)P_{ij}$. This type of singularities is referred to
       as a {\it collinear singularity}.
 \item $P^2_{\bar{\mathcal{K}}}\:\equiv\:(\sum_{k\in\bar{\mathcal{K}}}p^{\mbox{\tiny $(k)$}})^2\:\longrightarrow\:0$: Similarly to the previous case
       \begin{equation}\label{MultlLimit}
        \lim_{P^2_{\bar{\mathcal{K}}}\,\rightarrow\,0}P^2_{\bar{\mathcal{K}}}M_n^{\mbox{\tiny tree}}\:=\:
         \sum_{h_{P}}M_{\mbox{\tiny dim$\{\bar{\mathcal{K}}\}$}+1}^{\mbox{\tiny tree}}(\bar{\mathcal{K}},-P_{\mbox{\tiny $\bar{\mathcal{K}}$}}^{-h_{P}})
          M_{n-\mbox{\tiny dim$\{\bar{\mathcal{K}}\}$}+1}^{\mbox{\tiny tree}}(P_{\mbox{\tiny $\bar{\mathcal{K}}$}}^{h_{P}},\,\mathcal{Q}),
       \end{equation}
       with $\bar{\mathcal{K}}$ being the set containing more than two particles, and $\mathcal{Q}$ is the complement set of $\bar{\mathcal{K}}$: 
       $\bar{\mathcal{K}}\,\cup\,\mathcal{Q}\:=\:\left\{1,\ldots,n\right\}$. These are called {\it multi-particle singularities}.
\end{romanlist}
Which precise factorization channels are allowed depends on the details of the theory, as, for example, the helicity configurations or further symmetries.

Other interesting limits under which an amplitude may factorize are the {\it soft limits}, when the momentum of a given particle $i$ is taken to zero:
\begin{equation}\label{soft}
 M_n^{\mbox{\tiny tree}}\:\overset{p^{\mbox{\tiny $(i)$}}\,\rightarrow\,0}{\longrightarrow}\:\mbox{Soft}^{\mbox{\tiny tree}}(i)M_{n-1}^{\mbox{\tiny tree}}(\not{i}).
\end{equation}
The soft factor $\mbox{Soft}^{\mbox{\tiny tree}}(i)$ carries the helicity information related to the soft particle ({\it i.e.} it scales just under the little
group transformations related to the momentum $p^{\mbox{(i)}}$, while it is helicity blind with respect to the other particles), and the amplitude
$M_{n-1}^{\mbox{\tiny tree}}(\not{i})$ is the $(n-1)$-particle amplitude obtained from the scattering of the particle set $\{1,\,\ldots,n\}\backslash\{i\}$.

\subsubsection{Branch cuts and loops}\label{subsubsec:Loop}

Let us now consider the possibility of having both poles and branch points. The discontinuity along the branch cuts departing from each point is related to the
imaginary (dispersive) part of the amplitudes. For a generic amplitude, the branch cut structure can be very complicated since the singularities turn out to be 
nested. This can be understood thinking about the general structure of an $L$-loop amplitude
\begin{equation}\label{MnL}
 M_n^{\mbox{\tiny $(L)$}}\:=\:\int\left(\prod_{r=1}^L\frac{d^{D}l_{r}}{(2\pi)^D}\right)\sum_{v}\frac{\mathfrak{P}_{v}(\{l_r\},\,p)}{\prod_{k}P_{k}^2(\{l_r\},p)},
\end{equation}
where $\mathfrak{P}_{v}$ are polynomials in the loop momenta $l_r$ and the external ones $p$, while $P_{k}^2$ are the loop propagators which are again functions of
$l_r$ and $p$.

It is generally convenient to compute the discontinuity along the cuts rather than attacking directly the loop integration, which can be done using the unitarity
condition (\ref{Tunit}). The latter can be expanded perturbatively in the coupling constant, finding that its right-hand-side shows integration over momenta of the
intermediate states: the imaginary part of the loop amplitudes can be determined from the phase-space integrals of product of lower-order amplitudes, which are 
identified by restricting two internal propagators on-shell. The imaginary part of the loop amplitudes provides the discontinuity along the cuts, and this computation 
allows to reconstruct the amplitudes up to rational functions of the Lorentz invariants. 

The general idea is to find first a scalar integral representation of the amplitude, {\it i.e.} a representation of the amplitude as sum of integrals whose numerators 
do not have a tensor structure. For the $1$-loop case, the Passarino-Veltman reduction \cite{Passarino:1978jh, vanNeerven:1983vr, Bern:1993kr} provides a general 
method to find a basis for the amplitude and such a basis turns out to be the minimal one
\begin{equation}\label{1Lminbas}
 M_n^{\mbox{\tiny $(1)$}}\:=\:
  \sum_{i\in\mathcal{S}_4}\mathcal{C}_4^{\mbox{\tiny $(i)$}}I_4^{\mbox{\tiny $(i)$}}+
  \sum_{i\in\mathcal{S}_3}\mathcal{C}_3^{\mbox{\tiny $(i)$}}I_3^{\mbox{\tiny $(i)$}}+
  \sum_{i\in\mathcal{S}_2}\mathcal{C}_2^{\mbox{\tiny $(i)$}}I_2^{\mbox{\tiny $(i)$}}+
  \mathcal{R}^{\mbox{\tiny $(1)$}},
\end{equation}
where $I_{m}$ are scalar integral with $m$ internal propagators and the coefficients $\mathcal{C}_m$ as well as $\mathcal{R}^{\mbox{\tiny $(1)$}}$ are all rational 
functions of the Lorentz invariants. For the $L$-loop case ($L\,>\,1$), there is no general procedure to identify a basis, even if Passarino-Veltman reduction-like 
approaches can be used on case by case basis.

\begin{figure}[htbp]
 \centering 
  \[
    M_{n}^{\mbox{\tiny $(1)$}}\:=\:
    \sum_{i\in\mathcal{S}_4}\mathcal{C}_4^{\mbox{\tiny $(i)$}}{\raisebox{-.8cm}{\scalebox{.30}{\includegraphics{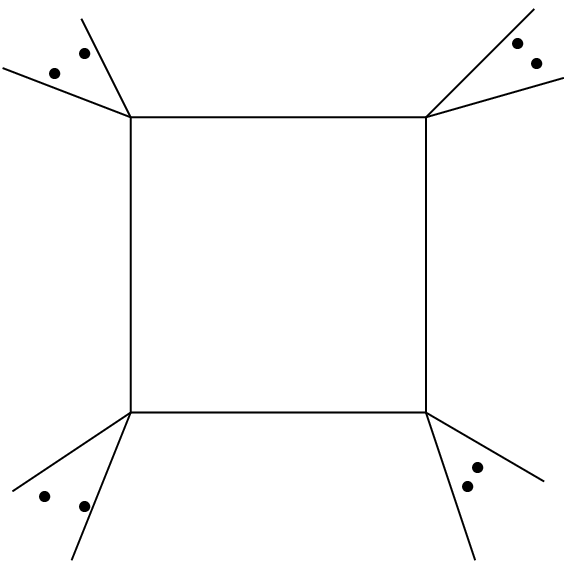}}}}+
    \sum_{i\in\mathcal{S}_3}\mathcal{C}_3^{\mbox{\tiny $(i)$}}{\raisebox{-.8cm}{\scalebox{.30}{\includegraphics{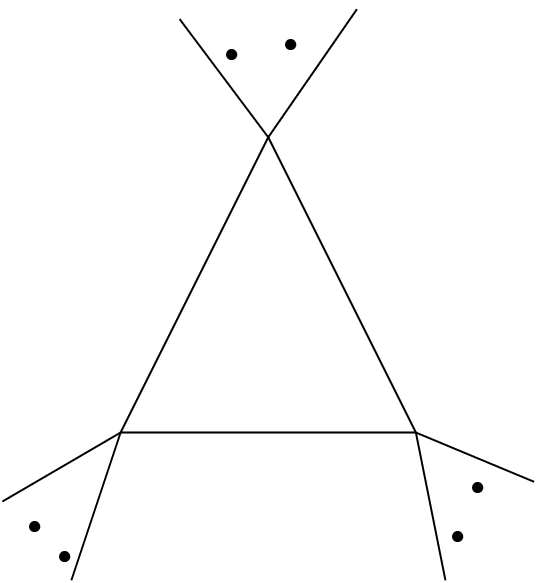}}}}+
    \sum_{i\in\mathcal{S}_2}\mathcal{C}_2^{\mbox{\tiny $(i)$}}{\raisebox{-.4cm}{\scalebox{.30}{\includegraphics{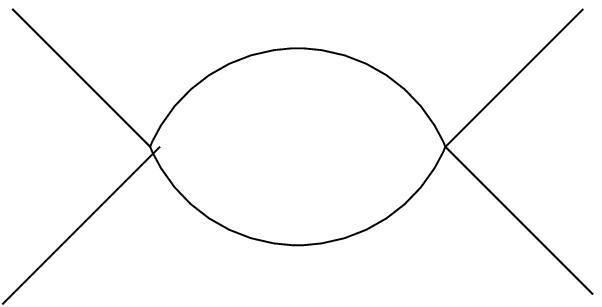}}}}+
    \mathcal{R}^{\mbox{\tiny $(1)$}}
  \]
  \caption{One-loop integral expansion. The minimal integral basis on which a one-loop amplitude in four dimensions can be expanded shows at most box-integrals, and 
           then the lower point ones. This is due to the fact that in four dimensions and in the complexified momentum space it is possible to send on-shell 
           simultaneously at most four internal propagators}\label{fig:1lIntExp}
\end{figure}

Once such a representation is known, its coefficients can be determined by unitarity through the computation of the cuts along all the possible momentum channels.
Another possibility is to consider, together with the physical singularities given by having two internal propagators on-shell, singularities with higher codimensions,
which are allowed only if the momenta are extended to be complex. This approach goes under the name {\it generalized unitarity methods} 
\cite{Bern:1994zx, Bern:1994cg, Bern:1995db, Bern:1996je, Bern:1996ja, Bern:1997sc, Bern:2004cz}.

Let us focus on the $L\,=\,1$ level. The crucial observation is that being a scattering amplitude an 
analytic function, any of its representations must share the same singularity structure with each other. In a $D$-dimensional space-time and for fixed external 
momenta, the highest codimension singularity it can show is given by the maximum number of internal propagators which can be sent on-shell. In the complexified 
momentum space, this is number is equal to the dimensions $D$ of the space-time. Therefore, a putative scalar integral decomposition of the one-loop amplitude 
in $D$-dimensions is given by a sum of scalar $k$-gon, with $k\,\le\,D$: in four dimensions one obtain exactly the expansion (\ref{1Lminbas})\cite{ArkaniHamed:2008gz}.
Now, one can perform the $k$-cuts in order to find the coefficients of the expansions: a given $k$-cut selects all those Feynman diagrams on one side and
all the scalar integral on the other, showing those $k$ internal propagators. Actually, one can proceed gradually\cite{ArkaniHamed:2008gz}: 
In four-dimensions, once one found all the box integrals which reproduce the quadruple-cuts, one can take their sum as a basis ansatz and compute the triple cuts. If 
the result is consistent, then the basis does not need lower point integrals. Otherwise, we need to complete this basis by adding scalar triangle integrals, which are 
such that do not modify the quadruple cuts but provide a non-trivial contribution to the triple cuts. Similarly, one can take the sum of boxes and triangles as a basis
ansatz and compute the double cuts: If the answer of such a computation is consistent, then this basis does not need to be completed, otherwise it is necessary to add 
the bubble integrals which do not contribute to neither the quadruple nor the triple cut, but which has a non-vanishing double cut. The cuts analysis leaves out the 
possibility of reproduce the rational terms.

\subsubsection{Quadruple cuts}\label{subsubsec:4cuts}

For $D\,=\,4$, the highest codimension singularity is identified by sending on-shell four internal propagators. This can be done just for complexified momenta.
Looking at the expansion (\ref{1Lminbas}), just the box integrals $I_4^{\mbox{\tiny $(i)$}}$ show this number of internal propagators. The set of all the external 
particles $\{1,\,\ldots,\,n\}$ can be partitioned as 
\begin{equation}\label{4cutparts}
 \left\{1,\,\ldots,\,n\right\}\:=\:\bigcup_{m=1}^{4}A_{m}^{\mbox{\tiny $(i)$}},
\end{equation}
with the different partitions on a box are identified by $(i)$. A given cut is identified by fixing $i$ in (\ref{4cutparts}) and, thus, just one box-integral 
contributes
\begin{equation}\label{4cut}
 \begin{split}
  \Delta_{4}^{\mbox{\tiny $(i)$}}M_n^{\mbox{\tiny $(1)$}}\:&\equiv\:\int_{T^4}\frac{d^4l}{(2\pi)^4}\,\prod_{m=1}^4 M_m^{\mbox{\tiny tree}}(l)
   \:=\\
  &=\:\mathcal{C}_4^{\mbox{\tiny $(i)$}}
   \int_{T^4_i}\frac{d^4l}{(2\pi)^4}\,\frac{1}{l^2(l-P^{\mbox{\tiny $(1)$}})^2(l+P^{\mbox{\tiny $(2)$}})^2(l-P^{\mbox{\tiny $(1)$}}-P^{\mbox{\tiny $(4)$}})^2}.
 \end{split}
\end{equation}
where $P^{\mbox{\tiny $(m)$}}$ are the sum of the external momenta in $A_m^{\mbox{\tiny $(i)$}}$, and the integration is performed over the $T_i^4$ which defines
the specific quadruple cut
\begin{equation}\label{T4}
 T^4_i\:=\:
  \left\{
   l\,\in\,\mathbb{C}^4\,|\,l^2\,=\,0,\:(l-P^{\mbox{\tiny $(1)$}})^2\,=\,0,\:(l+P^{\mbox{\tiny $(2)$}})^2\,=\,0,\:
   (l-P^{\mbox{\tiny $(1)$}}-P^{\mbox{\tiny $(4)$}})^2\,=\,0
  \right\}.
\end{equation}

\begin{figure}[htbp]
 \centering 
  \[
    {\raisebox{-1.4cm}{\scalebox{.50}{\includegraphics{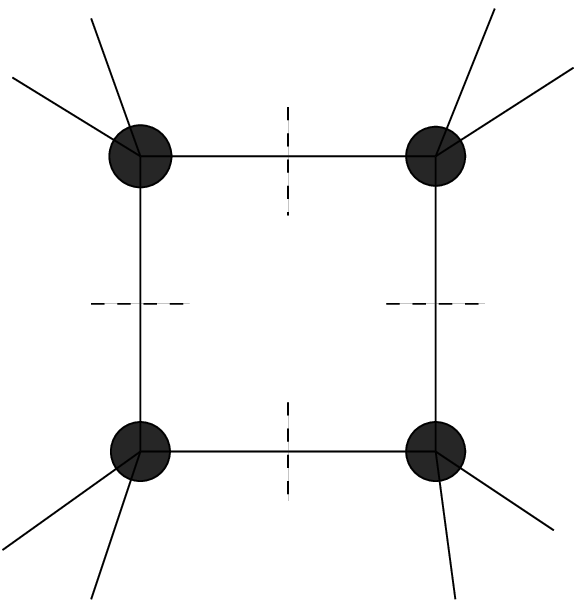}}}}\:=\:
    \mathcal{C}_4{\raisebox{-1.4cm}{\scalebox{.50}{\includegraphics{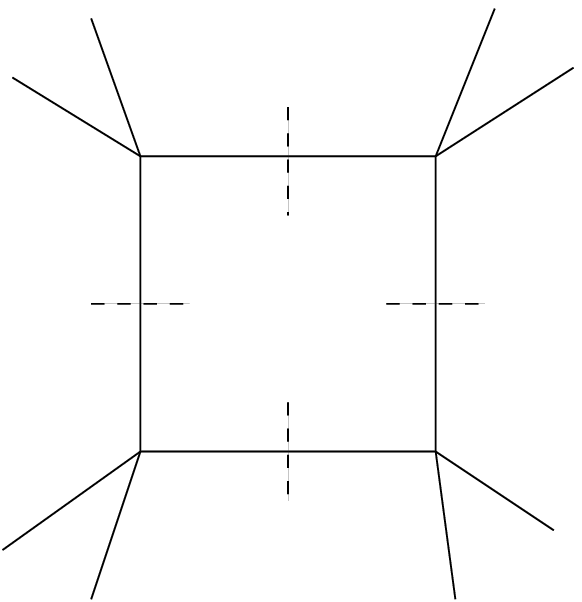}}}}
  \]
  \caption{Four-particle cut. It is obtained by sending on-shell four internal propagators. This is the highest codimension singularity in four-dimensions. Performing
           this class of cuts, one obtains the coefficients of the box-integrals as product of four tree-level amplitudes.}\label{fig:4cut}
\end{figure}

Each $T^4$ has two solutions for $l$, and thus the left-hand-side of eq (\ref{4cut}) reduces to a product of four tree-level amplitudes, summed over all the 
helicity states which can contribute and over the two solutions of the $T^4$. The coefficient $\mathcal{C}_4^{\mbox{\tiny $(i)$}}$ is therefore
\begin{equation}\label{4cutcoeff}
 \begin{split}
  \mathcal{C}_4^{\mbox{\tiny $(i)$}}\:=\:\frac{1}{2}\sum_{l_{\mbox{\tiny $\star$}}}\sum_{h}
   &M^{\mbox{\tiny tree}}_{\mbox{\tiny $1$}}\left(l_{\mbox{\tiny $\star$}}-P^{\mbox{\tiny $(1)$}}, P^{\mbox{\tiny $(1)$}}, -l_{\mbox{\tiny $\star$}}\right)
    M^{\mbox{\tiny tree}}_{\mbox{\tiny $2$}}\left(l_{\mbox{\tiny $\star$}}, P^{\mbox{\tiny $(2)$}}, -(l_{\mbox{\tiny $\star$}}-P^{\mbox{\tiny $(2)$}})\right)\times\\
   &\times
    M^{\mbox{\tiny tree}}_{\mbox{\tiny $3$}}\left(l_{\mbox{\tiny $\star$}}+P^{\mbox{\tiny $(2)$}}, P^{\mbox{\tiny $(3)$}}, 
     -(l_{\mbox{\tiny $\star$}}-P^{\mbox{\tiny $(1)$}}-P^{\mbox{\tiny $(2)$}})\right)\times\\
   &\times
    M^{\mbox{\tiny tree}}_{\mbox{\tiny $4$}}\left(l_{\mbox{\tiny $\star$}}-P^{\mbox{\tiny $(1)$}}-P^{\mbox{\tiny $(2)$}}, P^{\mbox{\tiny $(4)$}}, 
     -(l_{\mbox{\tiny $\star$}}-P^{\mbox{\tiny $(1)$}})\right).
 \end{split}
\end{equation}

\subsubsection{Triple cuts}\label{subsubsec:3cuts}

In order to be able to capture the other integrals of the basis and, therefore, compute the correspondent coefficients, we need to look at lower codimension 
singularities. Let us analyze the triple cuts:
\begin{equation}\label{3cuts}
 \Delta_3^{\mbox{\tiny $(i)$}}M_n^{\mbox{\tiny $(1)$}}\:\equiv\:\int_{T^3_i}\frac{d^4l}{(2\pi)^4}\,\prod_{m=1}^3 M_m^{\mbox{\tiny tree}}(l),i
\end{equation}
with the integration carried out over the $T_i^3$
\begin{equation}\label{T3}
 T^3_i\:=\:
  \left\{
   l\,\in\,\mathbb{C}^4\,|\,l^2\,=\,0,\:(l-P^{\mbox{\tiny $(1)$}})^2\,=\,0,\:(l+P^{\mbox{\tiny $(2)$}})^2\,=\,0.
  \right\}
\end{equation}

\begin{figure}[htbp]
 \centering 
  \[
    {\raisebox{-1.4cm}{\scalebox{.50}{\includegraphics{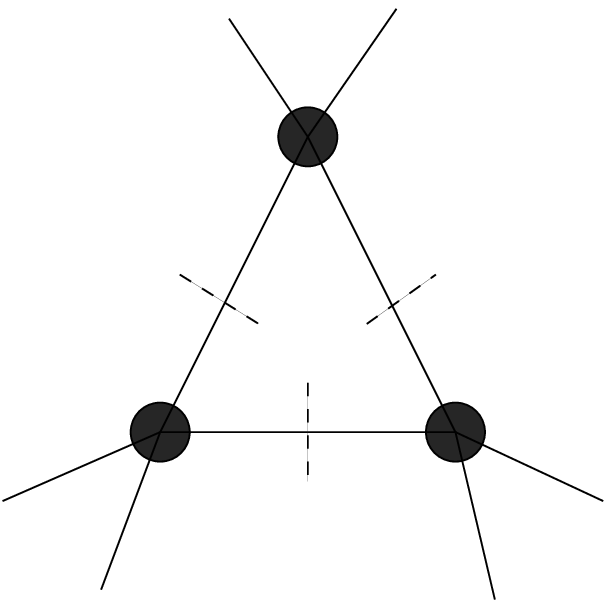}}}}\:=\:
    \sum_{i\in\mathcal{S}_4^{\mbox{\tiny $(3)$}}}\mathcal{C}_4^{\mbox{\tiny $(i)$}}{\raisebox{-1.4cm}{\scalebox{.50}{\includegraphics{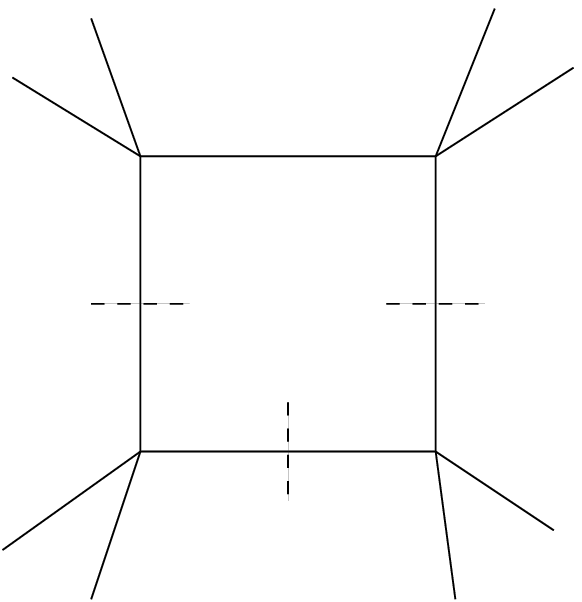}}}}+
    \mathcal{C}_3{\raisebox{-1.4cm}{\scalebox{.50}{\includegraphics{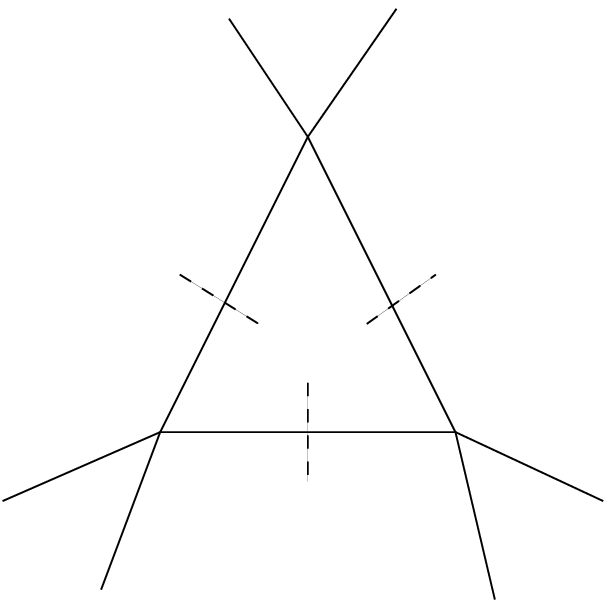}}}}
  \]
  \caption{Three-particle cut. It is obtained by sending on-shell three internal propagators.}\label{fig:3cut}
\end{figure}

The conditions in (\ref{T3}) do not fix the loop momenta $l$ but rather generate two one-parameter families of solutions\footnote{It possible to look at
this computation in a different fashion. If one replaces $\delta(l^2)$ by \cite{Mastrolia:2006ki}
\begin{equation}
 \delta(l^2)\,\longrightarrow\,\frac{1}{l^2+i\varepsilon}-\frac{1}{l^2-i\varepsilon},
\end{equation}
the problem is reduced to a double cut in the other two propagators.
}, which can be written as \cite{Forde:2007mi}
\begin{equation}\label{T3soln}
  \begin{split}
   &l\:=\:z
    \left(
     \frac{X_2^2}{z}\Lambda^{\mbox{\tiny $(1)$}}+\frac{1-X_2^2}{1-X_1^2 X_2^2}\Lambda^{\mbox{\tiny $(2)$}}
    \right)
    \left(
     \frac{1-X_1^2}{1-X_1^2X_2^2}\tilde{\Lambda}^{\mbox{\tiny $(1)$}}+\frac{X_1^2}{z}\tilde{\Lambda}^{\mbox{\tiny $(2)$}}
    \right),\\
   &l\:=\:z
    \left(
     \frac{1-X_1^2}{1-X_1^2X_2^2}\Lambda^{\mbox{\tiny $(1)$}}+\frac{X_1^2}{z}\Lambda^{\mbox{\tiny $(2)$}}
    \right)
    \left(
     \frac{X_2^2}{z}\tilde{\Lambda}^{\mbox{\tiny $(1)$}}+\frac{1-X_2^2}{1-X_1^2 X_2^2}\tilde{\Lambda}^{\mbox{\tiny $(2)$}}
    \right),
  \end{split}
\end{equation}
where $X_i^2\,=\,(P^{\mbox{\tiny $(i)$}})^2/(\langle1,2\rangle[1,2]$ ($i\,=\,1,\,2$), and $\Lambda^{\mbox{\tiny $(i)$}}\tilde{\Lambda}^{\mbox{\tiny $(i)$}}$
are the light-like projections of the momenta $P^{\mbox{\tiny $(i)$}}$.

Considering the three tree-level amplitudes in the cut, they show two poles for each $z$-dependent propagator -- this is easy to understand by looking at the
two solutions above. Those poles are connected to the box-integrals. The contributions to the triangle coefficients are the ones without any pole. Thus
\begin{equation}\label{3cutCoeff}
 \mathcal{C}_3\:=\:\frac{1}{2}\int_{\gamma_{\infty}}\frac{dz}{z}\,\sum_{l_{\star}}\prod_{m=1}^3 M_m^{\mbox{\tiny tree}}(z),
\end{equation}
where $\gamma_{\infty}$ is a contour encircling just the pole at infinity.
 
\subsubsection{Double cuts}\label{subsubsec:2cuts}

The double cuts take into account all the higher point scalar integrals. Typically, the double cuts can allow to extract the coefficients of all the integral basis.
However, the quadruple- and triple-cuts already provide a more straightforward way of computing the coefficients of the box and of the triangle. Therefore, if a given 
theory admits bubbles, we can use the double cuts just to compute the coefficient of the bubbles. It is defined as
\begin{equation}\label{DoubleCut}
 \Delta_2^{\mbox{\tiny $(i)$}}M_n^{\mbox{\tiny $(1)$}}\:=\:\int\frac{d^4l}{(2\pi)^4}\,\delta^{\mbox{\tiny $(+)$}}(l^2)\delta^{\mbox{\tiny $(+)$}}((l-\mathcal{K})^2)
  \prod_{m=1}^2M_m^{\mbox{\tiny tree}}.
\end{equation}
We postpone a deeper discussion about the double cuts in Section \ref{sec:TreeLoop}, for a description of how to extract all the coefficients of integral basis as 
well as how to combine this class of cuts with the other generalized ones, see \cite{Britto:2010xq}.

\begin{figure}[htbp]
 \centering 
  \[
    {\raisebox{-.5cm}{\scalebox{.35}{\includegraphics{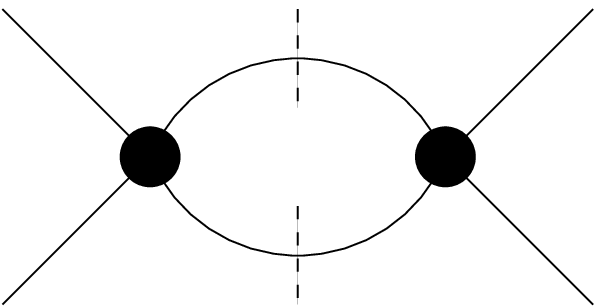}}}}\:=\:
    \sum_{i\in\mathcal{S}_4^{\mbox{\tiny $(2)$}}}\mathcal{C}_4^{\mbox{\tiny $(i)$}}{\raisebox{-.8cm}{\scalebox{.30}{\includegraphics{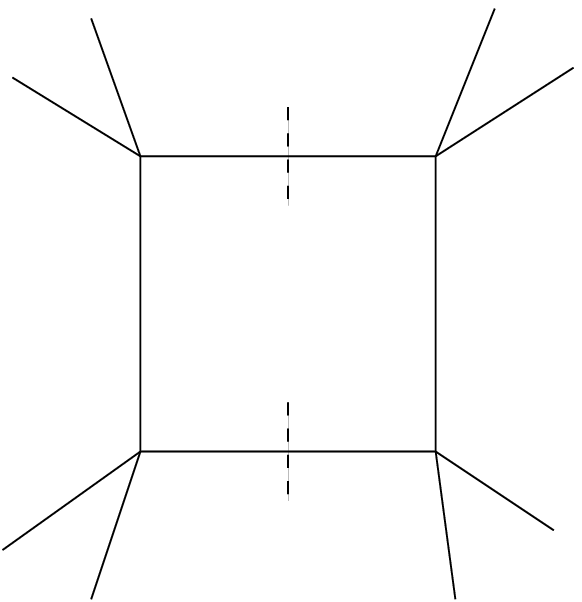}}}}+
    \sum_{i\in\mathcal{S}_3^{\mbox{\tiny $(2)$}}}\mathcal{C}_3^{\mbox{\tiny $(i)$}}{\raisebox{-.9cm}{\scalebox{.30}{\includegraphics{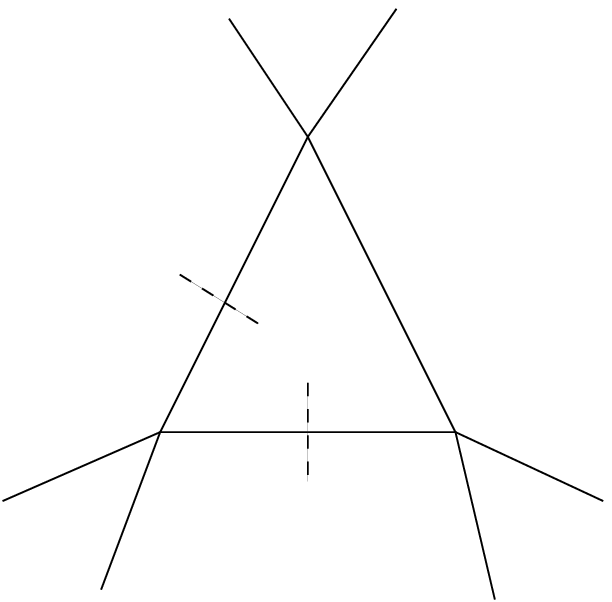}}}}+
    \mathcal{C}_2{\raisebox{-.5cm}{\scalebox{.35}{\includegraphics{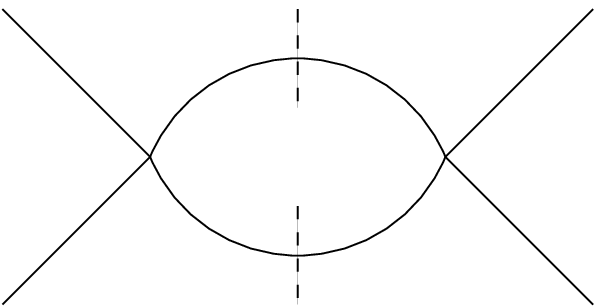}}}}
  \]
  \caption{Double cut. It is obtained by sending on-shell two internal propagators.}\label{fig:2cut}
\end{figure}

\subsection{Turning the table around}\label{subsec:FundHyp}

Let us now rewind the tape and try to extract a possible minimal set of hypothesis which may at the basis of a general S-matrix theory:
\begin{romanlist}[(ii)]
 \item \underline{Space-time isometry group invariance and existence of one-particle states}: The first step is the definition of the states which are going 
       to scatter. They can be taken as the irreducible representation of the space-time isometry group. So far, we have been discussing asymptotically Minkowski 
       space-time, for which it is natural to work in the momentum space and to classify the representations according to the Lorentz little group. Furthermore,
       the existence of one-particle states allows to define operators which acts on the scattering amplitude as they act on the single particles;
 \item \underline{Analyticity}: It restricts the singularity structure of the S-matrix to have at most poles and branch points, which we have a physical interpretation
       of;
 \item \underline{Unitarity}: It provides the factorization theorems and, therefore, the physical interpretations of the singularities;
 \item \underline{Locality of the interactions}: The singularity structure is further restricted to have at most single poles coming from standard propagators, and the
       branch cuts as well as higher codimension singularities are obtained in those points in the complexified momentum space in which two or more poles are reached,
       {\it i.e.} two or more propagators go on-shell.
\end{romanlist}
A feature which deserves attention is the absence of any type of gauge symmetry as fundamental hypothesis: If it characterizes the theory, it should arise as a sort of 
emergent property. Moreover, our perspective is to always work with on-shell amplitudes so that the result must be gauge invariant.

One can wonder how really fundamental can be the assumption of locality. Strictly speaking, we do not have any a priori reason to have a such a belief and
some recent development seems to point towards the possibility of locality to be an emergent feature \cite{ArkaniHamed:2012nw}. Furthermore, dropping such a 
requirement might allow to define consistent higher-spin coupling in Minkowski space-time 
\cite{Francia:2002aa, Francia:2002pt, Francia:2010qp, Taronna:2011kt, Benincasa:2011pg}, given that all the known no-go theorems take the locality of the
interactions for granted\cite{Johnson:1960vt, Weinberg:1964ew, Velo:1970ur, Weinberg:1980kq, Porrati:2008rm}.

The last brick we need to be able to start to build our S-matrix theory is a class of fundamental objects which can be determined from the above fundamental 
hypothesis. This will be the subject of the next section.

As a final comment, the above set of hypothesis is not meant to be the definitive minimal set. Rather, it is a good starting point which might allow to deepen
our knowledge of the perturbation theory in particle physics and, hopefully, create a framework in which it can be either redefined or reduced, or both.
\footnote{This is the direction for $\mathcal{N}\,=\,4$ supersymmetric Yang-Mills theory taken by two recent papers \cite{ArkaniHamed:2012nw, Arkani-Hamed:2013jha}.}

\section{The three particle amplitudes}\label{sec:3ptAmpl}

The simplest object we can define is a three-particle amplitude. For massless particles and considering the Lorentz group $SO(3,1;\mathbb{R})$, momentum conservation
forces it to vanish. Non-trivial three-particle amplitudes can instead be defined if one considers either the Lorentz group with signature $SO(2,2)$ or the 
complexified Lorentz group $SO(3,1;\mathbb{C})$, the last one being isomorphic to $SL(2,\mathbb{C})\times SL(2,\mathbb{C})$ \cite{Witten:2003aa}. In the last case,
the two spinors $\lambda_{a}^{\mbox{\tiny $(i)$}}$ and $\tilde{\lambda}_{\dot{a}}^{\mbox{\tiny $(i)$}}$ can be considered as transforming under different 
$SL(2,\mathbb{C})$. 

The three-particle amplitudes are fixed, up to a constant, by Poincar{\'e} invariance \cite{Benincasa:2007xk}. More precisely, momentum conservation implies that a 
three-particle amplitude can be written as a direct sum of an holomorphic and an anti-holomorphic term
\begin{equation}\label{MomCons}
 \sum_{i=1}^{3}\lambda^{\mbox{\tiny $(i)$}}_a\tilde{\lambda}^{\mbox{\tiny $(i)$}}_{\dot{a}}\:=\:0 
 \qquad\Longrightarrow\quad
 \langle i,j\rangle[i,j]\:=\:0,
\end{equation}
with the last set of equations which are solved either when all the holomorphic or the anti holomorphic inner products vanish, {\it i.e.} either all the holomorphic spinors or all the anti-holomorphic ones are proportional to each other. Thus, a three-particle amplitude has the following structure
\begin{equation}\label{3ptStr}
 M_3(\lambda,\,\tilde{\lambda})\:=\:\delta\left(\sum_{i=1}^3\lambda^{\mbox{\tiny $(i)$}}\tilde{\lambda}^{\mbox{\tiny $(i)$}}\right)
  \left[M_3^{\mbox{\tiny H}}(\lambda)+M_3^{\mbox{\tiny A}}(\tilde{\lambda})\right],
\end{equation}
with $M_3^{\mbox{\tiny H}}(\lambda)$ and $M_3^{\mbox{\tiny A}}(\tilde{\lambda})$ being the holomorphic and anti-holomorphic contribution, respectively. It is 
important to point out that such a structure characterizes the full (non-perturbative) amplitude: it is just a consequence of one of the symmetries of the theory and 
does not have anything to do with perturbation theory.

Furthermore, the covariance under the Lorentz little group $SO(2)$, together with the assumption that the Poincar{\'e} group acts on the scattering amplitudes as
it acts on individual $1$-particle states, is equivalent to state that the amplitude is an eigenfunction of the helicity operator $\mathcal{H}^{\mbox{\tiny $(i)$}}$
for particle $i$, with eigenvalue proportional to its helicity $h_i$
\begin{equation}\label{LitGp}
 \mathcal{H}^{\mbox{\tiny $(i)$}}M_3(\lambda,\,\tilde{\lambda})\:\equiv\:
  \left(
   \lambda^{\mbox{\tiny $(i)$}}_a\frac{\partial}{\partial\lambda^{\mbox{\tiny $(i)$}}_a}-
   \tilde{\lambda}^{\mbox{\tiny $(i)$}}_{\dot{a}}\frac{\partial}{\partial\tilde{\lambda}^{\mbox{\tiny $(i)$}}_{\dot{a}}}
  \right)
  M_3(\lambda,\,\tilde{\lambda})\:=\:-2h_i  M_3(\lambda,\,\tilde{\lambda}),
\end{equation}
or, equivalently considering equation (\ref{3ptStr}), 
\begin{equation}\label{LitGpHA}
  \left(\lambda^{\mbox{\tiny $(i)$}}_a\frac{\partial}{\partial\lambda^{\mbox{\tiny $(i)$}}_a}+2h_i\right)M_3^{\mbox{\tiny H}}\:=\:0
  \quad\mbox{and}\quad
  \left(\tilde{\lambda}^{\mbox{\tiny $(i)$}}_{\dot{a}}\frac{\partial}{\partial\tilde{\lambda}^{\mbox{\tiny $(i)$}}_{\dot{a}}}-2h_i\right)M_3^{\mbox{\tiny A}}\:=\:0.
\end{equation}
This equation can be exactly solved, providing a general expression for the full three-particle amplitude with states with arbitrary helicities
\begin{equation}\label{3ptFull}
 \begin{split}
  &M_3(\{\lambda^{\mbox{\tiny $(i)$}},\tilde\lambda^{\mbox{\tiny $(i)$}},h_i\})\:=\:
   \delta\left(\sum_{i=1}^3\lambda^{\mbox{\tiny $(i)$}}\tilde{\lambda}^{\mbox{\tiny $(i)$}}\right)\times\\
  &\hspace{1.5cm}\times
   \left[
    \kappa^{\mbox{\tiny H}}_{\mbox{\tiny $1+h$}} \langle1,2\rangle^{d_{3}}\langle 2,3\rangle^{d_{1}}\langle3,1\rangle^{d_{2}}+ 
    \kappa^{\mbox{\tiny H}}_{\mbox{\tiny $1-h$}} [1,2]^{-d_{3}}[ 2,3]^{-d_{1}}[3,1]^{-d_{2}}
   \right],
 \end{split}
\end{equation}
where $d_1\,=\,h_1-h_2-h_3$, $d_2\,=\,h_2-h_3-h_1$, $d_3\,=\,h_3-h_1-h_2$ and $h\,=\,h_1+h_2+h_3$. As for the amplitudes in eq (\ref{3ptStr}), the superscript $H$/$A$
in the coupling constants indicates that they are associated to the holomorphic/anti-holomorphic amplitude, while the subscript indicate their dimension
\footnote{The dimension of the coupling constant comes from a simple counting: As a consequence of the fact that a cross-section has the dimension of an area, 
an $n$-particle amplitude has dimensions $4-n$. Thus, the dimension of the three-particle couplings in eq (\ref{3ptFull}) is obtained by subtracting the dimension
of the three-particle amplitude, which is $1$, and the dimension of the kinematic dependent factor.}.

As a last requirement, the three-particle amplitudes need to vanish on the real sheet, which occurs when both the holomorphic and anti-holomorphic inner products 
simultaneously vanish. As the explicit expression (\ref{3ptFull}) shows, for $d_1+d_2+d_3\,=\,-h_1-h_2-h_3\,<\,0$ the anti-holomorphic contribution vanishes on the 
real sheet while the holomorphic term is singular, conversely for  $d_1+d_2+d_3\,=\,-h_1-h_2-h_3\,>\,0$. This means that, in order to have a non-singular behavior in 
this limit, in the first case the holomorphic coupling constant $\kappa^{\mbox{\tiny H}}_{\mbox{\tiny $1+h$}}$ needs to be zero, while in the second one this same has 
to happen for the anti-holomorphic coupling constant $\kappa^{\mbox{\tiny A}}_{\mbox{\tiny $1-h$}}$. In order words, if the helicities of the states satisfy the 
inequality $d_1+d_2+d_3\,=\,-h_1-h_2-h_3\,<\,0$, the three-particle amplitude just depends on the anti-holomorphic inner products, while if 
$d_1+d_2+d_3\,=\,-h_1-h_2-h_3\,>\,0$ is satisfied, it shows just the holomorphic term. Finally, as far as the case $d_1+d_2+d_3\,=\,-h_1-h_2-h_3\,=\,0$ is concerned,
both the holomorphic and anti-holomorphic terms contribute and the three-particle amplitude is given just by the coupling constant which, at the end of the day, turns
out to be a linear combination of $\kappa^{\mbox{\tiny H}}_{\mbox{\tiny $1$}}$ and $\kappa^{\mbox{\tiny A}}_{\mbox{\tiny $1+h$}}$.

\begin{figure}[htbp]
 \centering 
  \scalebox{.50}{\includegraphics{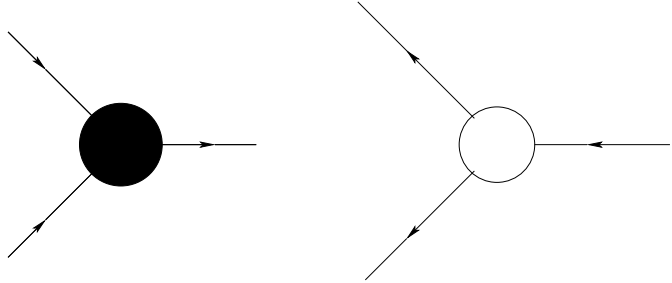}}
  \caption{Three-particle amplitudes. The black three-point vertex represents the holomorphic three-particle amplitudes, while the white vertex the anti-holomorphic
           one. The directions of the arrows represent the sign of the helicity of a given state: if the arrow is incoming, the helicity is negative, if it is outgoing
           the helicity of the given state is positive.}\label{fig:3particle}
\end{figure}

Summarizing, the three-particle amplitudes defined by states whose helicities are such that $h_1+h_2+h_3\,\neq\,0$ have defined holomorphicity, {\it i.e.} they
can be either holomorphic or anti-holomorphic, and for each holomorphic amplitude of states with helicities $(h_1,\,h_2,\,h_3)$ there exists an anti-holomorphic
counterpart whose states have helicities $(-h_1,\,-h_2,\,-h_3)$. Diagrammatically, we indicate the (anti-)holomorphic three-particle amplitude with (white)black 
three-point vertices (see Figure \ref{fig:3particle}).

The expression in (\ref{3ptFull}) can be made permutation invariant by endowing it with a structure constant $\varepsilon_{\mbox{\tiny $a_1a_2a_3$}}$
\begin{equation}\label{3ptFullSC}
 \begin{split}
  &M_3(\{\lambda^{\mbox{\tiny $(i)$}},\tilde\lambda^{\mbox{\tiny $(i)$}},h_i\})\:=\:
   \delta\left(\sum_{i=1}^3\lambda^{\mbox{\tiny $(i)$}}\tilde{\lambda}^{\mbox{\tiny $(i)$}}\right)\times\\
  &\hspace{.1cm}\times
    \left[
     \kappa^{\mbox{\tiny H}}_{\mbox{\tiny $1+h$}}\varepsilon_{\mbox{\tiny $a_1a_2a_3$}} \langle1,2\rangle^{d_{3}}\langle 2,3\rangle^{d_{1}}\langle3,1\rangle^{d_{2}}+
     \kappa^{\mbox{\tiny H}}_{\mbox{\tiny $1-h$}}\varepsilon_{\mbox{\tiny $a_1a_2a_3$}} [1,2]^{-d_{3}}[ 2,3]^{-d_{1}}[3,1]^{-d_{2}}
    \right],
 \end{split}
\end{equation}
which is equivalent to introduce internal quantum numbers for the states. For amplitudes involving just particles with the same spin $s$, these structure constants
can be either totally symmetric (for particles with even spin) or totally antisymmetric (for particles with odd spin).

If one now defines a phase-space factor $\Omega$ for a given particle as follows
\begin{equation}\label{OnShDiff}
 \Omega\:=\:\sum_{h}\frac{d^2\lambda\,d^2\tilde{\lambda}}{\mbox{Vol}\left\{GL(1)\right\}},
\end{equation}
it is then possible to think of the three-part particle amplitude as a form\cite{ArkaniHamed:2012nw}
\begin{equation}\label{OnShForm3}
 \mathcal{M}_{\mbox{\tiny $3$}}\:=\:M_{\mbox{\tiny $3$}}\Omega^{\mbox{\tiny $(i)$}}\Omega^{\mbox{\tiny $(j)$}}\Omega^{\mbox{\tiny $(k)$}},
\end{equation}
with $i,\,j,\,k$ labelling the particles. The sum in eq (\ref{OnShDiff}) takes into account all the possible helicity states allowed, while the measure is the
volume of the little group.

Some comments are now in order. Firstly, the expression in (\ref{3ptFullSC}) defines all the possible {\it fundamental} three-particle amplitudes compatible with 
Poincar{\'e} invariance. Strictly speaking, there are more singular objects which can be defined, and they can be obtained from equation (\ref{3ptFullSC}) through the 
replacement 
\begin{equation}\label{3ptSing}
 \kappa_{\mbox{\tiny $1-|h|$}}\quad\longrightarrow\quad\frac{\kappa_{\mbox{\tiny $1-|h|+2k$}}}{\left(P_{ij}^2\right)^k}.
\end{equation}
In other words, {\it fundamental} scattering amplitudes which typically are characterized by a coupling constant with dimension $1-|h|$ may appear as {\it effective} 
interactions with coupling constant $1-|h|+2k$ in correspondence of a higher order singularity. An example where this occurs is the $1$-loop gluon scattering in QCD, 
where it appears as a particular collinear limit characterized by a double pole (thus $k\,=\,1$ in (\ref{3ptSing})) \cite{Bern:2005hs}. This type of effective
interactions will not be treated in our discussion.

Furthermore, the dimensionality of the three-particle coupling can be used as a criterium to classify interacting theories 
\cite{Benincasa:2007xk, Benincasa:2011pg}. 
In particular, if we restrict ourselves to theories whose couplings within a given theory are characterized by the same dimensionality, we can identify four main
classes \cite{Benincasa:2011pg}:
\begin{romanlist}[(ii)]
 \item $[\kappa]\:=\:1-s$: it contains a self-interaction for spin $s$ and a spin-$s$/spin-$s'$ coupling, respectively with $(\mp s,\,\mp s,\,\pm s)$ and
       $(-s',\,-s',\,\mp s)$ as allowed helicity configurations;
 \item $[\kappa]\:=\:1-3s$: just a self-interacting coupling is allowed, with $(\mp s,\,\mp s,\, \mp s)$ as possible helicity configurations;
 \item $[\kappa]\:=\:1-(2s'+s)$: the helicity structure of this three particle amplitudes is $(\mp s',\, \mp s',\,\mp s)$;
 \item $[\kappa]\:=\:1-|2s'-s|$: in this case, the three-particle couplings allowed have $(\mp s',\,\mp s',\,\pm, s)$. This class has the particular
  feature that the three-particle amplitudes with a fixed helicity configuration can be either holomorphic or anti-holomorphic depending on the sign
  of $|2s'-s|$, while if this quantity is zero then both terms contribute.
\end{romanlist}
In the classification above, we have considered the possibility that at most two different spins can enter. In line of principle, there is nothing which prevents us to
consider also three-particle amplitudes of states all of them having different spin. Moreover, we are assuming the existence of fundamental three-particle couplings,
leaving out all those theories whose smallest couplings involve more than three states. However, if on one hand the classification above is indeed not complete,
on the other hand some theories, which at first sight are not included, can be reduced to one of the classes above. This is the case, we just mentioned, of theories
whose smallest coupling is not a three-particle one: higher point couplings can be splitted into effective (but still finite) three-particle coupling by introducing
massive particles with suitable spin (this procedure for $\kappa_{\mbox{\tiny $0$}}\phi^4$ has been explicitly worked out \cite{Benincasa:2007xk}). From a 
computational point of view, this might not be the best way to deal with the amplitudes of such theories; however, it becomes relevant if one is interested in 
understanding the basic structure of perturbation theory.

At last, it may be useful to compare the structure of the three-particle amplitudes (\ref{3ptFullSC}) with the ones coming from a Lagrangian formulation
\begin{equation}\label{3ptLag}
 M_3\:\sim\:\kappa_{1-F-\delta}\varepsilon_{a_F\{b_i\dot{b}_i\}}\varepsilon_{\dot{a}_F\{b_j\dot{b}_j\}}\varepsilon_{\{b_k\dot{b}_k\}}p^{\delta},
\end{equation}
where the $\varepsilon$'s are the polarization tensors of the three external particles, $a_F$ and $\dot{a}_F$ are ``extra'' spinor indices which are present
if two of the external particles are fermions (in this case $F=1$, while for $F=0$ these spinorial indices are not present and all the external particles are bosons),
and $\delta$ provides the number of derivatives of the three-particle interactions. As before, the subscript of the coupling constant indicates its mass dimension,
which, in this case, is computed by subtracting the dimension of the three-particle amplitudes (equal to $1$) by the dimensions of the polarization tensors $F$
(for bosons they are dimensionless, while for fermions a polarization vector has mass dimension $1/2$) and the number of derivative $\delta$. Notice that the
classification above can be translated into a Lagrangian language in terms of the number of derivative of the three-particle interactions: $\delta\,=\,|h|-F$. More 
interestingly, one can notice that these interactions are local if and only if $\delta\,=\,|h|-F\,\ge\,0$. However, the comparison with the classification above
shows that all those three-particle interactions are local: for the three-particle amplitudes, it seems that locality can emerge from the requirement that they vanish 
on the real sheet (which is nothing but momentum conservation on the real sheet). In other words, in order to have non-local Lagrangian three-particle interactions 
with operator $(\partial/\Box)^{\delta}$, one would need to drop the requirement that the three-particle amplitudes vanish on the real sheet.

\subsection{The supersymmetric extensions}\label{subsec:3ptSusyAmpl}

As mentioned from the very beginning, the perspective we want to take is to build up theories of interacting particles from a minimal set of assumptions. This
approach leads to rediscover some symmetries as a consequence of some consistency conditions \cite{Benincasa:2007xk, Benincasa:2011pg}, as we will explain later.
However, in some cases it may be convenient to assume those symmetries from the very beginning in order to be able to explore further the structure of certain class
of theories. This allows to study directly the scattering of coherent states, each of them containing all the helicity states which are connected by the additional 
symmetry taken into account. This is indeed the case of supersymmetric theories, for which the Super-Poincar{\'e} group defines the asymptotic states
\cite{ArkaniHamed:2008gz}.

If $Q_{\mbox{\tiny $Ia$}}$ and $\bar{Q}^{\mbox{\tiny $I\dot{a}$}}$ are the supercharges, with $a$ and $\dot{a}$ being, as usual, the spinor indices while 
$I\,=\,1,\ldots,\mathcal{N}$ is the R-symmetry index, the coherent states can be defined as \cite{ArkaniHamed:2008gz}
\begin{equation}\label{SUSYCohS}
 |\lambda,\,\tilde{\lambda};\,\eta\rangle\:=\:e^{Q_{aI}w^{a}\eta^{I}}|\lambda,\,\tilde{\lambda};\,-s\rangle,\qquad
 |\lambda,\,\tilde{\lambda};\,\tilde{\eta}\rangle\:=\:e^{\tilde{Q}^{\dot{a}I}\tilde{w}_{\dot{a}}\tilde{\eta}_{I}}|\lambda,\,\tilde{\lambda};\,+s\rangle,
\end{equation}
where $w_a$ and $\tilde{w}_{\dot{a}}$ are spinors satisfying the conditions $\langle w,\,\lambda\rangle\,=\,1$ and $[\tilde{w},\,\tilde{\lambda}]\,=\,1$ respectively,
and the Grassmann variables $\eta_{\mbox{\tiny $I$}}$ and $\tilde{\eta}^{\mbox{\tiny $I$}}$ have been introduced. The action of the supercharges on the helicity
states is conventionally defined as follows
\begin{equation}\label{Qact}
 \begin{split}
  &Q_{aI}|\lambda,\,\tilde{\lambda};\,-s\rangle\:=\:\lambda_{a}|\lambda,\,\tilde{\lambda};\,-s+\frac{1}{2}\rangle_{I},
   \quad
   Q_{aI}|\lambda,\,\tilde{\lambda};\,+s\rangle\:=\:0,\\
  &\tilde{Q}_{\dot{a}I}|\lambda,\,\tilde{\lambda};\,-s\rangle\:=\:0,
   \quad
   \tilde{Q}_{\dot{a}I}|\lambda,\,\tilde{\lambda};\,+s\rangle\:=\:\tilde{\lambda}_{\dot{a}}|\lambda,\,\tilde{\lambda};\,+s-\frac{1}{2}\rangle_{I}.
 \end{split}
\end{equation}
It is then straightforward to see that the coherent states (\ref{SUSYCohS}) are eigenstates of the supercharges $Q_{aI}$ and $\tilde{Q}_{\dot{a}I}$ respectively 
\begin{equation}\label{QactCS}
 Q_{aI}|\lambda,\,\tilde{\lambda};\,\tilde{\eta}\rangle\:=\:\tilde{\eta}_{I}\lambda_{a}|\lambda,\,\tilde{\lambda};\,\tilde{\eta}\rangle,
  \qquad
 \tilde{Q}_{\dot{a}I}|\lambda,\,\tilde{\lambda};\,\eta\rangle\:=\:\eta_{I}\tilde{\lambda}_{\dot{a}}|\lambda,\,\tilde{\lambda};\,\eta\rangle.
\end{equation}
For maximally supersymmetric theories, the coherent states contain {\it all} the helicity states, so that using the Super-Poincar{\'e} group to define the
asymptotic states for a scattering process allows to treat all the possible interactions at once. For less supersymmetric theories, the helicity states are
organized in more than one multiplet. Let us write here explicitly (and for future reference) the coherent states for $\mathcal{N}\,=\,1$ and $\mathcal{N}\,=\,4s$ 
supersymmetries (with $s\,=\,1,2$ being the highest spin in the supermultiplet -- this is the case of maximally supersymmetric Yang-Mills and gravity):
\begin{equation}\label{CSN14s}
 \begin{split}
  &\mbox{$\mathcal{N}\,=\,1$ supermultiplets}:\\
  &\hspace{1.2cm} |\lambda,\,\tilde{\lambda};\,\eta\rangle\:=\: |-s\rangle+\eta|-s+\frac{1}{2}\rangle,\\
  &\hspace{1.2cm} |\lambda,\,\tilde{\lambda};\,\tilde{\eta}\rangle\:=\: |+s\rangle+\tilde{\eta}|+s-\frac{1}{2}\rangle,\\
  &\mbox{$\mathcal{N}\,=\,4s$ supermultiplet}:\\
  &\hspace{1.2cm} |\lambda,\,\tilde{\lambda};\,\eta\rangle\:=\:|-s\rangle+\eta_{I}|-s+\frac{1}{2}\rangle^{I}+\ldots+
    \frac{1}{\mathcal{N}!}\epsilon^{I_{1}\ldots I_{\mathcal{N}}}\eta_{I_{1}}\ldots\eta_{I_{\mathcal{N}}}|+s\rangle,\\
  &\hspace{1.2cm} |\lambda,\,\tilde{\lambda};\,\tilde{\eta}\rangle\:=\:|+s\rangle+\tilde{\eta}^{I}|+s-\frac{1}{2}\rangle_{I}+\ldots+
    \frac{1}{\mathcal{N}!}\epsilon_{I_{1}\ldots I_{\mathcal{N}}}\tilde{\eta}^{I_{1}}\ldots\tilde{\eta}^{I_{\mathcal{N}}}|-s\rangle,
 \end{split}
\end{equation}
where in the right-hand-side of all the expressions above the dependence on the spinors $\lambda$ and $\tilde{\lambda}$ have been suppressed for notational 
convenience. From eqs (\ref{CSN14s}), it is clear the existence of two multiplets for $\mathcal{N}\,=\,1$, one containing the positive helicity states and the other
one the negative helicity states, while for $\mathcal{N}\,=\,4s$ the two multiplets contain all the helicity states, so that they are equivalent. Actually,
$\eta$ and $\tilde{\eta}$ provide two equivalent representations of the states and it is possible to go from one representation to the other with a Grassmann
Fourier transform
\begin{equation}\label{GFtrans}
 |\lambda,\,\tilde{\lambda};\,\eta\rangle\:=\:\int d^{\mathcal{N}}\tilde{\eta}\,e^{\tilde{\eta}\eta}|\lambda,\,\tilde{\lambda};\,\tilde{\eta}\rangle,
 \qquad
 |\lambda,\,\tilde{\lambda};\,\tilde{\eta}\rangle\:=\:\int d^{\mathcal{N}}\eta\,e^{\eta\tilde{\eta}}|\lambda,\,\tilde{\lambda};\,\eta\rangle.
\end{equation}
Through (\ref{GFtrans}), both the positive and the negative supermultiplets in the $\mathcal{N}\,<\,4s$ case can be written equivalently in the two representations.

Let us now turn to the amplitudes. First of all, the multiplets in the amplitude do not need to be in the same representation but it is possible to choose the one
which is more convenient in each particular case. Secondly, in order to determine the three-particle amplitudes one needs also to require that they transform
appropriately under supersymmetry. On a single state a supersymmetry transformation acts as follows
\begin{equation}\label{SUSY1st}
 \begin{split}
  &e^{Q_{aI}\zeta^{aI}}|\lambda,\,\tilde{\lambda};\,\eta\rangle\,=\,|\lambda,\,\tilde{\lambda};\,\eta+\langle\zeta,\,\lambda\rangle\rangle,\quad
   e^{Q_{aI}\zeta^{aI}}|\lambda,\,\tilde{\lambda};\,\tilde{\eta}\rangle\,=\,
    e^{\tilde{\eta}_{I}\langle\lambda,\zeta^{I}\rangle}|\lambda,\,\tilde{\lambda};\,\tilde{\eta}\rangle,\\
  &e^{\tilde{Q}^{\dot{a}I}\tilde{\zeta}_{\dot{a}I}}|\lambda,\,\tilde{\lambda};\,\eta\rangle\,=\,
    e^{\eta^{I}[\tilde{\lambda},\tilde{\zeta}_{I}]}|\lambda,\,\tilde{\lambda};\,\eta\rangle,\quad
   e^{\tilde{Q}^{\dot{a}I}\tilde{\zeta}_{\dot{a}I}}|\lambda,\,\tilde{\lambda};\,\eta\rangle\,=\,
     |\lambda,\,\tilde{\lambda};\,\tilde{\eta}+[\tilde{\zeta},\,\tilde{\lambda}]\rangle,
 \end{split}
\end{equation}
where $\zeta$ and $\tilde{\zeta}$ are the supersymmetry parameters. With our hypothesis of the existence of $1$-particle states, the supersymmetry transformations
act on the amplitudes as in eq (\ref{SUSY1st}). Considering all the states in the same representation as well as the fact that under the little group 
transformations $\eta$ and $\tilde{\eta}$ behave exactly as $\lambda$ and $\tilde{\lambda}$, respectively, one can easily show that the three-particle amplitudes 
acquires the form
\begin{equation}\label{3ptSusy}
 \begin{split}
  &M_{\mbox{\tiny $3$}}(\lambda,\tilde{\lambda},\eta)\:=\:
   \kappa^{\mbox{\tiny H}}\frac{\delta\left(\langle2,3\rangle\eta^{\mbox{\tiny $(1)$}}+\langle3,1\rangle\eta^{\mbox{\tiny $(2)$}}+
    \langle1,2\rangle\eta^{\mbox{\tiny $(3)$}}\right)}{\left(\langle1,2\rangle\langle2,3\rangle\langle3,1\rangle\right)^s}+
   \kappa^{\mbox{\tiny A}}\frac{\delta\left(\sum_{i\,=\,1}^3\eta^{\mbox{\tiny $(i)$}}\tilde{\lambda}^{\mbox{\tiny $(i)$}}\right)}{
    \left([1,\,2][2,\,3][3,\,1]\right)^s},\\
  &M_{\mbox{\tiny $3$}}(\lambda,\tilde{\lambda},\tilde{\eta})\:=\:
   \kappa^{\mbox{\tiny H}}\frac{\delta\left(\sum_{i\,=\,1}^3\lambda^{\mbox{\tiny $(i)$}}\tilde{\eta}^{\mbox{\tiny $(i)$}}\right)}{
    \left(\langle1,\,2\rangle\langle2,\,3\rangle\langle3,\,1\rangle\right)^s}+
   \kappa^{\mbox{\tiny A}}\frac{\delta\left([2,3]\tilde{\eta}^{\mbox{\tiny $(1)$}}+[3,1]\tilde{\eta}^{\mbox{\tiny $(2)$}}+
    [1,2]\tilde{\eta}^{\mbox{\tiny $(3)$}}\right)}{\left([1,2][2,3][3,1]\right)^s},
 \end{split}
\end{equation}
$s$ being the highest spin in the multiplets. Again, defining the super-phase space factor for a given particle as
\begin{equation}\label{SUSYphsp}
 \Omega_{\mbox{\tiny $\eta$}}\:=\:\frac{d^2\lambda\,d^2\tilde{\lambda}}{\mbox{Vol}\left\{GL(1)\right\}}d^4\eta,
 \qquad
 \Omega_{\mbox{\tiny $\tilde{\eta}$}}\:=\:\frac{d^2\lambda\,d^2\tilde{\lambda}}{\mbox{Vol}\left\{GL(1)\right\}}d^4\tilde{\eta},
\end{equation}
the three-particle amplitudes can be written as eq (\ref{OnShForm3}), with the super phase-space factors $\Omega$ given in eq (\ref{SUSYphsp}).

\begin{figure}[htbp]
 \centering 
  \scalebox{.50}{\includegraphics{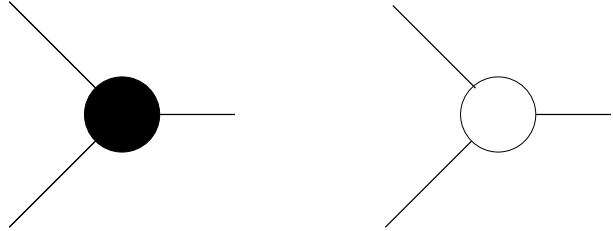}}
  \caption{Three-particle amplitudes for maximally supersymmetric theories. As in the non supersymmetric case, the black three-point vertex represents the holomorphic 
           three-particle amplitudes, while the white vertex the anti-holomorphic one. These diagrams are not decorated with arrows because each external states 
           contains all the possible helicity states of the theory. For less supersymmetric theories, where there is a notion of negative and positive coherent states,
           the three-particle amplitudes are represented as in Fig \ref{fig:3particle}.}\label{fig:3particleS}
\end{figure}

\section{Constructing the higher point amplitudes at tree level}\label{sec:HPampl}

As we just discussed, the three-particle amplitudes are determined by the Poincar{\'e} group and, when it is convenient, they can reflect some other
symmetry of the theory. In order to describe processes involving a higher number of states, one can suitably glue the three-particle amplitudes together. 
Before providing the rules to do this, it is worth to get some more insights about the structure of the amplitudes. 

\subsection{Momentum space deformations}\label{subsec:MSdef}

Scattering amplitudes are generally analytic functions of the Lorentz invariants, once the $\delta$-function implementing momentum conservation has been stripped out. 
In the complexified momentum space $\left(p^{\mbox{\tiny $(i)$}}\,\in\,\mathbb{C}^{\mbox{\tiny $D$}}\right)$, it is possible to introduce a non-trivial 
one-complex-parameter deformation such that both momentum conservation and the on-shell condition is still preserved \cite{Britto:2005aa}. 
In a general fashion, such class of transformation can be written as
\begin{equation}\label{MomDef}
 p^{\mbox{\tiny $(v)$}}\;\longrightarrow\;
 p^{\mbox{\tiny $(v)$}}(z)\:=\:p^{\mbox{\tiny $(v)$}}-z\sum_{r=1}^{\mbox{\tiny dim$\{\mathcal{D}\}$}}\alpha^{vr}q^{\mbox{\tiny $(r)$}},
 \qquad
 v\,\in\,\mathcal{D},
\end{equation}
where $z$ is the parameter of the deformation, $\mathcal{D}$ is the subset of particles whose momenta have been deformed, $\alpha^{vr}$ are coefficients fixed by 
momentum conservation and the momenta $q^{\mbox{\tiny $(r)$}}\,\in\,\mathbb{C}^{\mbox{\tiny $D$}}$, because of the on-shell condition, need to be light-like vectors 
such that
\begin{equation}\label{qconds}
 q^{\mbox{\tiny $(r_1)$}}\cdot q^{\mbox{\tiny $(r_2)$}} \:=\:0\:=\:p^{\mbox{\tiny $(v)$}}\cdot q^{\mbox{\tiny $(r)$}},
 \qquad\forall\:v\,\in\,\mathcal{D},\;\forall\:r,\,r_1,\,r_2\:=\:1,\,\ldots,\,\mbox{dim}\{\mathcal{D}\}.
\end{equation}
Notice that the coefficients $\alpha^{kr}$ do not need to be all non-vanishing, but rather their number can be at most equal to dim$\{\mathcal{D}\}$.

The deformations (\ref{MomDef}) generate a one-parameter family of amplitudes
\begin{equation}\label{FamAmpl}
  M_n\,\quad\longrightarrow\quad M_n^{\mbox{\tiny $(\mathcal{D})$}}(z),
\end{equation}
where in the notation above the superscript $\mathcal{D}$ just indicates the deformation which generates the family of amplitudes. One can now consider the family
of amplitudes as a function of $z$, study its singularity structure in $z$ and try to reconstruct the physical amplitudes -- which can be obtained by setting $z$ to 
zero -- from its singularities. The hypothesis of analyticity restricts the amplitudes to show just poles and branch cuts. In the Feynman diagram language, if we
restrict ourselves to consider just poles as characterizing the singularity structure of our amplitudes, we are restricting ourselves to consider the tree-level
approximation. At loop level, we need to consider both poles and branch points.

This analysis is completely general and it does not depend neither on the dimension of the space-time nor on the particles being massless or massive.
For the time being, we will focus on the four-dimensional massless case. 
For $D\,=\,4$, we can rephrase the deformations (\ref{MomDef}) in terms of the spinors $\lambda$ and $\tilde{\lambda}$
\begin{equation}\label{MomDef2}
 \begin{split}
  &\tilde{\lambda}^{\mbox{\tiny $(v_1)$}}\quad\longrightarrow\quad\tilde{\lambda}^{\mbox{\tiny $(v_1)$}}(z)\:=\:
    \tilde{\lambda}^{\mbox{\tiny $(v_1)$}}-z\sum_{r\le\mbox{\tiny dim}\{\mathcal{D}_{\mbox{\tiny $\lambda$}}\}}\tilde{\beta}^{v_1r}
     \tilde{\xi}^{\mbox{\tiny $(r)$}},\quad
   v_{1}\,\in\,\mathcal{D}_{\mbox{\tiny $\tilde{\lambda}$}}\\
  &\lambda^{\mbox{\tiny $(v_2)$}}\quad\longrightarrow\quad\lambda^{\mbox{\tiny $(v_2)$}}(z)\:=\:
     \lambda^{\mbox{\tiny $(v_2)$}}+z\sum_{r\le\mbox{\tiny dim}\{\mathcal{D}_{\mbox{\tiny $\tilde{\lambda}$}}\}}\beta^{v_2r}\xi^{\mbox{\tiny $(r)$}},\quad
   v_{2}\,\in\,\mathcal{D}_{\mbox{\tiny $\lambda$}}
 \end{split}
\end{equation}
with $\mathcal{D}_{\mbox{\tiny $\tilde{\lambda}$}}$ and $\mathcal{D}_{\mbox{\tiny $\lambda$}}$ being the sets of the particles whose anti-holomorphic and
holomorphic spinors have been deformed. The infinitesimal generator of the above transformation can be written as
\begin{equation}\label{InfGen}
 \mathcal{G}_{\mbox{\tiny $\mathcal{D}$}}\:=\:
  \sum_{v_2\in\mathcal{D}_{\mbox{\tiny $\lambda$}}}\sum_{r\le\mbox{\tiny dim}\{\mathcal{D}_{\mbox{\tiny $\tilde{\lambda}$}}\}}
   \beta^{v_2r}\xi^{\mbox{\tiny $(r)$}}\frac{\partial}{\partial\lambda^{\mbox{\tiny $(v_2)$}}}-
  \sum_{v_1\in\mathcal{D}_{\mbox{\tiny $\tilde{\lambda}$}}}\sum_{r\le\mbox{\tiny dim}\{\mathcal{D}_{\mbox{\tiny $\lambda$}}\}}
   \tilde{\beta}^{v_1r}\tilde{\xi}^{\mbox{\tiny $(r)$}}\frac{\partial}{\partial\tilde{\lambda}^{\mbox{\tiny $(v_1)$}}}.
\end{equation}
Under these deformations, the singularities of $M_{n}^{\mbox{\tiny $(\mathcal{G}_{\mathcal{D}})$}}(z)$ appear in correspondence of the $z$-dependent propagators. 
Varying the number of particles whose momenta get deformed varies the number of propagators which acquire a $z$-dependence and, typically, this number increases as the
number of deformed external momenta increases. If on one side this may complicate the $z$-singularity structure, on the other side it generates different 
representations which can help to unveil the properties of our theories.

In any case, if we consider just the pole structure which, as we discusses in Section \ref{subsec:AnLocUn}, corresponds to the tree level approximation, just the 
propagators involving a subset of the deformed particles (not coinciding with the full set) are $z$-dependent and, thus, provide a pole in $z$. The one-parameter
family of amplitude which gets generated is a meromorphic function of the parameter $z$. Let 
$\bar{\mathcal{D}}_{\tilde{\lambda}}$ and $\bar{\mathcal{D}}_{\lambda}$ be respectively the subgroups of the anti-holomorphic and holomorphic spinors which
gets deformed and enter in a given propagators $1/P^2_{\mbox{\tiny $\bar{\mathcal{D}}$}}$
\begin{equation}\label{zProp}
 \frac{1}{P^2_{\mbox{\tiny $\bar{\mathcal{D}}$}}}\:\longrightarrow\:\frac{1}{P^2_{\mbox{\tiny $\bar{\mathcal{D}}$}}(z)}\:=\:
 \frac{1}{P^2_{\mbox{\tiny $\bar{\mathcal{D}}$}}+z\sum_{v_1,r_1}\tilde{\beta}^{\mbox{\tiny $v_1 r_1$}}\langle v_1|P_{\mbox{\tiny $\bar{\mathcal{D}}$}}|r_1]-
   z\sum_{v_2,r_2}\beta^{\mbox{\tiny $v_2 r_2$}}\langle r_2|P_{\mbox{\tiny $\bar{\mathcal{D}}$}}|v_2]},
\end{equation}
with $v_1$ and $v_2$ taking values in $\bar{\mathcal{D}}_{\tilde{\lambda}}$ and $\bar{\mathcal{D}}_{\lambda}$, respectively. The location of the pole is therefore
given by
\begin{equation}\label{zPole}
 z_{\mbox{\tiny $\bar{\mathcal{D}}$}}\:=\:
  -\frac{P^2_{\mbox{\tiny $\bar{\mathcal{D}}$}}}{\sum_{v_1,r_1}\tilde{\beta}^{\mbox{\tiny $v_1 r_1$}}\langle v_1|P_{\mbox{\tiny $\bar{\mathcal{D}}$}}|r_1]-
   \sum_{v_2,r_2}\beta^{\mbox{\tiny $v_2 r_2$}}\langle r_2|P_{\mbox{\tiny $\bar{\mathcal{D}}$}}|v_2]}.
\end{equation}
If one consider the Riemann-sphere 
$\hat{\mathbb{C}}\:=\:\mathbb{C}\cup\{\infty\}$, then
\begin{equation}\label{RRfund}
 0\:=\:\frac{1}{2\pi i}\oint_{\hat{\mathbb{C}}}\frac{dz}{z}\,M_n^{\mbox{\tiny $(\mathcal{D})$}}(z)\:=\:M_n^{\mbox{\tiny $(\mathcal{D})$}}(0)+
   \sum_{k\in\mathcal{P}^{\mbox{\tiny $(\mathcal{D})$}}}\frac{c_k^{\mbox{\tiny $(\mathcal{D})$}}}{z_k}-\mathcal{C}_n^{\mbox{\tiny $(\mathcal{D})$}},
\end{equation}
where $M_n^{\mbox{\tiny $(\mathcal{D})$}}(0)\,\equiv\,M_n$, $\mathcal{P}^{\mbox{\tiny $(\mathcal{D})$}}$ is the set of poles generated by the deformation, 
$z_k$ are the locations of the poles (given by eq (\ref{zPole})), $c_k^{\mbox{\tiny $(\mathcal{D})$}}$ their residues and 
$\mathcal{C}_n^{\mbox{\tiny $(\mathcal{D})$}}$ the residue at infinity (or, in other words, the constant term of the Laurent series expansion of 
$M_n^{\mbox{\tiny $(\mathcal{D})$}}(z)$). 

As far as the residues $c_k^{\mbox{\tiny $(\mathcal{D})$}}$ in eq (\ref{RRfund}) are concerned, they are provided by the product of two tree-level amplitudes with 
fewer external states \cite{Britto:2005aa}: as the location of the pole in a specific channel is approached, the related momentum goes on-shell, the channel dominates
the others, and the amplitude factorizes
\begin{equation}\label{zFact}
 M_n^{\mbox{\tiny $(\mathcal{D})$}}(z)\:\overset{\mbox{\tiny $z\longrightarrow z_k$}}{\sim}\:
  M_{\mbox{\tiny L}}^{\mbox{\tiny $(\mathcal{D})$}}(z_k)\frac{1}{
   2\left(\sum_{v,r}\alpha^{vr}P_{k}\cdot q^{\mbox{\tiny $(r)$}}\right)(z_k-z)}
  M_{\mbox{\tiny L}}^{\mbox{\tiny $(\mathcal{D})$}}(z_k).
\end{equation}
From this equation, one can easily read off the residue we are looking for
\begin{equation}\label{zRes}
 -\frac{c_k^{\mbox{\tiny $(\mathcal{D})$}}}{z_k}\:=\:
  M_{\mbox{\tiny L}}^{\mbox{\tiny $(\mathcal{D})$}}(z_k)\frac{1}{P^2_k}M_{\mbox{\tiny R}}^{\mbox{\tiny $(\mathcal{D})$}}(z_k).
\end{equation}
If the amplitude vanishes as $z$ is taken to infinity, the residue of the pole at infinity $\mathcal{C}_n^{\mbox{\tiny $(\mathcal{D})$}}$ vanishes as well,
and the physical amplitude is determined by the residues (\ref{zRes}), generating a recursive relation
\begin{equation}\label{RReqs}
 M_n\:=\:\sum_{k\in\mathcal{P}^{\mbox{\tiny $(\mathcal{D})$}}} 
  M_{\mbox{\tiny L}}^{\mbox{\tiny $(\mathcal{D})$}}(z_k)\frac{1}{P^2_k}M_{\mbox{\tiny R}}^{\mbox{\tiny $(\mathcal{D})$}}(z_k).
\end{equation}
In the case that the amplitude does not vanish as $z$ is taken to infinity, the residue from the pole at infinity is non zero and, thus, the poles are not enough
to reconstruct the amplitude and further information is needed. This problem can be overcome by identifying special kinematic points for which the behavior of the
amplitude is known. Specifically, if $M_n^{\mbox{\tiny $(\mathcal{D})$}}(z)\,\sim\,z^{\nu}$ as $z\,\longrightarrow\,\infty$, one need $\nu+1$ conditions, which
can be provided by considering a subset of the zeros of the amplitudes\cite{Benincasa:2011kn}
\begin{equation}\label{zerosconds}
 0\:=\:\frac{1}{2\pi i}\oint_{\gamma_{\mbox{\tiny $0$}}^{s}}dz\,\frac{M_n^{\mbox{\tiny $(\mathcal{D})$}}(z)}{
  \left(z-z_{\mbox{\tiny $0$}}^{\mbox{\tiny $(s)$}}\right)^r},
\end{equation}
where $\gamma_{\mbox{\tiny $0$}}^{s}$ is a contour containing just the zero $z_{\mbox{\tiny $0$}}$, $r$ runs from $1$ to the multiplicity $m^{\mbox{\tiny $(s)$}}$
of the zero $z_{\mbox{\tiny $0$}}$. Using the Cauchy theorem, one obtains an algebraic system whose solution allows to reconstruct the full one-parameter family
of amplitude $M_n^{\mbox{\tiny $(\mathcal{D})$}}(z)$. In particular, for the physical amplitude one obtains
\begin{equation}\label{GenRR}
  M_n\:=\:\sum_{k\in\mathcal{P}^{\mbox{\tiny $(\mathcal{D})$}}} 
  M_{\mbox{\tiny L}}^{\mbox{\tiny $(\mathcal{D})$}}(z_k)\frac{f_k^{\mbox{\tiny $(\nu,\,n)$}}}{P^2_k}M_{\mbox{\tiny R}}^{\mbox{\tiny $(\mathcal{D})$}}(z_k),
  \quad
  f_k^{\mbox{\tiny $(\nu,\,n)$}}\:=\:
    \prod_{l=1}^{\nu+1}\left(1-\frac{P^2_k}{P_k^2(z_{\mbox{\tiny $(0)$}}^{\mbox{\tiny $(l)$}})}\right).
\end{equation}
The dressing factors $f_k^{\mbox{\tiny $(\nu,\,n)$}}$ are constrained by requiring that the amplitude factorizes correctly under the appropriate collinear and 
multi-particle limits\cite{Benincasa:2011kn}. The original discussion of this generalized on-shell representation was carried out for a specific class of deformations, 
that we will discuss later. However, the structure (\ref{GenRR}) holds for a general deformation (\ref{MomDef2}). More generally, if the behavior of the amplitude at 
specific kinematic points is known (and this already would be a case by case study), the representation (\ref{GenRR}) acquires the form\cite{Kampf:2013vha}
\begin{equation}\label{GenRR2}
  M_n\:=\:\sum_{k\in\mathcal{P}^{\mbox{\tiny $(\mathcal{D})$}}} 
  M_{\mbox{\tiny L}}^{\mbox{\tiny $(\mathcal{D})$}}(z_k)\frac{f_k^{\mbox{\tiny $(\nu,\,n)$}}}{P^2_k}M_{\mbox{\tiny R}}^{\mbox{\tiny $(\mathcal{D})$}}(z_k)
  +\sum_{l=1}^{\nu+1}M_{n}^{\mbox{\tiny $(\mathcal{D})$}}(z_{\star}^{\mbox{\tiny $(l)$}})\prod_{r=1,r\neq l}^{\nu+1}\frac{z_{\star}^{\mbox{\tiny $(r)$}}}{
   z_{\star}^{\mbox{\tiny $(r)$}}-z_{\star}^{\mbox{\tiny $(l)$}}},
\end{equation}
where the dressing factor $f_k^{\mbox{\tiny $(\nu,\,n)$}}$ takes the same expression as in eq (\ref{GenRR}), but it is computed in $z_{\star}$ rather than in a zero.
We will consider only the generalized representation (\ref{GenRR}) obtained making use of the knowledge of a subset of zeroes. It is important to notice that
the number of channels is exactly the same for eq (\ref{RReqs}) and (\ref{GenRR}): those channels provides the poles at finite location, while the dressing factors
$f_k^{\mbox{\tiny $(\nu,\,n)$}}$ take into account the contribution from the singularity at infinity through the momenta characterizing the channels but computed
at the location of the zeros.

As a final remark, it is important to stress that the recursion relations provide a notion of constructibility of a theory at tree level: if one iterates the
recursion relations, the $n$-particle amplitude can be expressed in terms of products of three-particle amplitudes, which are fixed by Poincar{\'e} invariance as
we saw in Section \ref{sec:3ptAmpl}. A theory admits an on-shell representation if and only if it is possible to identify at least one deformation which can pick
some of the poles of the amplitudes.

\subsubsection{The BCFW deformation}\label{subsubsec:BCFW}

\begin{figure}[htbp]
 \centering 
  \scalebox{.19}{\includegraphics{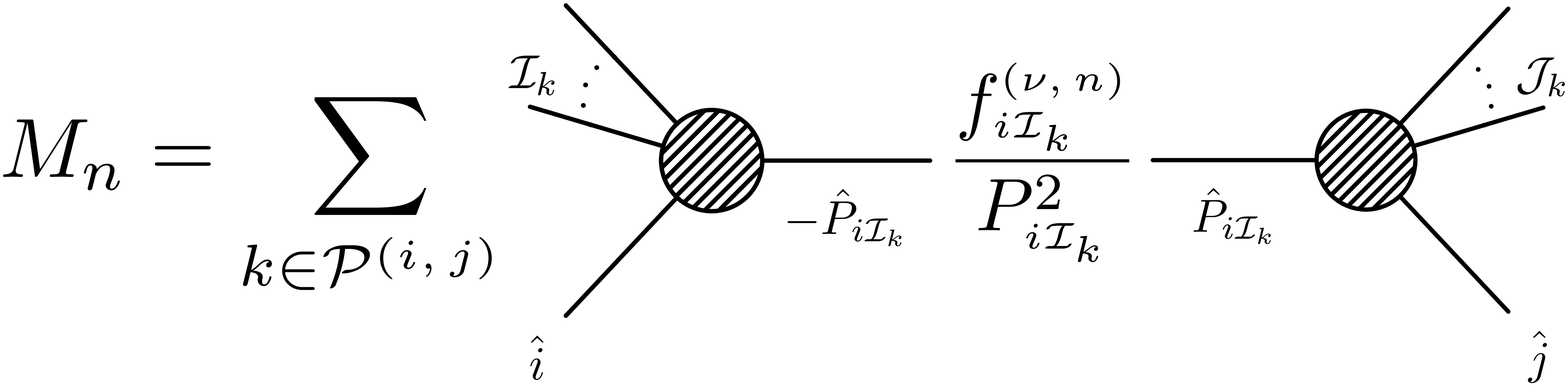}}
  \caption{Generalized on-shell recursion relation. It makes manifest that a scattering amplitude can be determined by the three-particle ones. If the
           amplitude goes to zero as two-particles become hard, then the dressing factor is equal to one, and it is determined by its pole structure. 
           Otherwise, the structure remains unchanged but the information about a subset of the zeros of the amplitudes is needed.
}\label{fig:BCFWdress}
\end{figure}

So far we have been considering a general momentum space deformation, finding a general recursive structure for an arbitrary $n$-particle amplitude. A given
deformation allows to express the amplitude as a sum over a certain number of channels, which change if we change the deformation: different deformations 
provides with different, but yet equivalent, representations. The simplest deformation is defined by shifting the momenta of two particle -- we label them by 
$(i)$ and $(j)$ -- leaving the others unchanged\cite{Britto:2005aa}
\begin{equation}\label{BCFWdef}
 \begin{split}
  &p^{\mbox{\tiny $(i)$}}(z)\,=\,p^{\mbox{\tiny $(i)$}}-zq,\quad
   p^{\mbox{\tiny $(j)$}}(z)\,=\,p^{\mbox{\tiny $(j)$}}+zq,\quad
   p^{\mbox{\tiny $(k)$}}(z)\,=\,p^{\mbox{\tiny $(k)$}},\,\forall k\,\neq\,i,\,j.
 \end{split}
\end{equation}
It manifestly respects momentum conservation, while the on-shell condition implies that $q$ is light-like and 
$(p^{\mbox{\tiny $(i)$}}\cdot q)\,=\,0\,=\,(p^{\mbox{\tiny $(j)$}}\cdot q)$. In terms of the spinors, it can be written as
\begin{equation}\label{BCFWdefspin}
 \tilde{\lambda}^{\mbox{\tiny $(i)$}}(z)\:=\:\tilde{\lambda}^{\mbox{\tiny $(i)$}}-z\tilde{\lambda}^{\mbox{\tiny $(j)$}},
 \qquad
 \lambda^{\mbox{\tiny $(j)$}}(z)\:=\: \lambda^{\mbox{\tiny $(j)$}}+z \lambda^{\mbox{\tiny $(i)$}},
\end{equation}
whose infinitesimal generator is 
\begin{equation}\label{BCFWgen}
 \mathcal{G}^{\mbox{\tiny $(i,j)$}}\:=\:
  \lambda^{\mbox{\tiny $(i)$}}_a\frac{\partial}{\partial\lambda^{\mbox{\tiny $(j)$}}_a}-
  \tilde{\lambda}^{\mbox{\tiny $(j)$}}_{\dot{a}}\frac{\partial}{\partial\tilde{\lambda}^{\mbox{\tiny $(i)$}}_{\dot{a}}}.
\end{equation}
This is the deformation which produces recursive relations with the least number of channels, which are identified by those sums of momenta involving either
$p^{\mbox{\tiny $(i)$}}$ or $p^{\mbox{\tiny $(j)$}}$. These are the only channels which acquire a $z$-dependence and, thus, show a pole in $z$.
Thus, in such an on-shell representation an amplitude can be written as
\begin{equation}\label{BCFWrr}
 M_n\:=\:\sum_{k\in\mathcal{P}^{\mbox{\tiny $(i,j)$}}}M_{\mbox{\tiny L}}^{\mbox{\tiny $(i,j)$}}(\hat{i}, \mathcal{I}_k,-\hat{P}_{i\mathcal{I}_k})
         \frac{f_{i\mathcal{I}_k}^{\mbox{\tiny $(\nu,n)$}}}{P_{i\mathcal{I}_k}^2}
         M_{\mbox{\tiny R}}^{\mbox{\tiny $(i,j)$}}(\hat{P}_{i\mathcal{I}_k},\mathcal{J}_k,\hat{j}),
\end{equation}
where the $\hat{\phantom{i}}$ indicates that the particular momentum it refers to is computed at the location of the pole of a given channel, $\mathcal{I}_k$ and 
$\mathcal{J}_k$ are subset of particles which do not contain neither particle $(i)$ nor particle $(j)$, and the dressing factors equal to $1$ for $\nu\,<\,0$ and to 
the expression in (\ref{GenRR}) for $\nu\,\ge\,0$. Diagrammatically can be written as in Figure \ref{fig:BCFWdress}. 

For this class of deformations, the limit $z\,\longrightarrow\,\infty$ has a nice physical interpretation: the momenta of the deformed particles become hard and, thus, 
it is equivalent to a hard particle going through a soft-background \cite{ArkaniHamed:2008yf}. It is important to stress that this does not provide {\it in general} a
physical interpretation for the residue at infinity, rather it is the leading term in the complex-UV which has this meaning. Only if $\nu\,=\,0$, the leading term in 
the complex-UV and the term $\mathcal{C}_n^{\mbox{\tiny $(i,j)$}}$ which contributes to the amplitude coincide. In general, the system of algebraic equations coming 
from eq (\ref{zerosconds}) provides with a general analytic expression for hard particle in a soft background
\begin{equation}\label{HardP}
 a_{\nu}^{\mbox{\tiny $(i,j)$}}\:=\:\sum_{k\in\mathcal{P}^{\mbox{\tiny $(i,j)$}}}
  \frac{M_{\mbox{\tiny L}}^{\mbox{\tiny $(i,j)$}}(\hat{i},\mathcal{I}_k,-\hat{P}_{i\mathcal{I}_k})
   M_{\mbox{\tiny R}}^{\mbox{\tiny $(i,j)$}}(\hat{P}_{i\mathcal{I}_k},\mathcal{I}_k,\hat{j})}{2 P_{iI_k}\cdot q}\prod_{l=1}^{\nu+1}
  \frac{2 P_{i\mathcal{I}_k}\cdot q}{P_{i\mathcal{I}_k}^2(z_0^{\mbox{\tiny $(l)$}})}.
\end{equation}
It is easy to verify that this expression coincides with $\mathcal{C}_n^{\mbox{\tiny $(i,j)$}}$ for the $\nu\,=\,0$.

It is now important to discuss the complex-UV limit and the behavior of the amplitudes as well as the zeros as special kinematic points. First of all, the structure
of the BCFW recursion relations show explicitly a subset of poles. It is always possible to choose the BCFW deformation whose two particles $(i)$ and $(j)$ singled
out have a non trivial collinear limit, {\it i.e.} the amplitude factorizes as $P^2_{ij}\,\longrightarrow\,0$. The representation provided by such a deformation
does not show explicitly the pole in $P^2_{ij}$, but yet the amplitude factorizes in the collinear limit in this channel. How can this puzzle be explained?

The answer to this question is that this singularity is realized as soft limit of one of the deformed momenta \cite{Schuster:2008nh, Benincasa:2011kn}. In order
to understand this, let us consider the fact that the collinear limit $P^2_{ij}\,\equiv\,\langle i,j\rangle[i,j]\longrightarrow\,0$ can be realized in two inequivalent
ways, by sending either the holomorphic or the anti-holomorphic inner product to zero. With the deformation (\ref{BCFWdefspin}), when the (anti)holomorphic limit
is taken, just the on-shell diagram with a three-particle amplitude containing particle ($i$) $j$ contributes, and the deformed spinor turns out to be proportional
to the (anti)holomorphic inner product itself. Therefore, in this limit the momentum of the deformed particle becomes soft. Let us know suppose that we do not know
the complex-UV behavior of our amplitude, and we would like to check whether the standard BCFW recursion relation -- {\it i.e.} eq (\ref{BCFWrr}) with all the dressing
factors set to one -- provides a correct representation of the amplitude. Checking the (complex) collinear limit $P^2_{ij}\,\longrightarrow\,0$, one discovers
that it is the case if and only if the soft limit of the deformed particle(s) is able to produce the correct singularity \cite{Schuster:2008nh}. Thus, the
complex-UV analysis can be reduced to the analysis of the three-particle amplitudes in a soft limit \cite{Benincasa:2011kn}, and allows to fix the parameter $\nu$
to be $\nu\,=\,1-\mbox{dim}\{\kappa\}+2h_i$ and/or $\nu\,=\,1-\mbox{dim}\{\kappa\}-2h_j$ -- dim$\{\kappa\}$ is the dimension of the three-particle coupling, while 
the ``and/or'' depends on whether the amplitude factorizes under both limits $[i,j]\,\longrightarrow\,0$ ($\hat{p}^{\mbox{\tiny $(i)$}}\,\longrightarrow\,0$) and 
$\langle i,j\rangle\,\longrightarrow0$ ($\hat{p}^{\mbox{\tiny $(j)$}}\,\longrightarrow\,0$), or just under one of them. For $\nu\,\ge\,0$, the soft limit is enhanced 
by the dressing factors, producing the correct singularity.

Finally, the dressing factors can be constrained by simply asking that the generalized on-shell representation (\ref{GenRR2}), which by itself provides with a valid
mathematical representation of the tree-level amplitudes, factorizes correctly under all the collinear/multi-particle limits. They can be divided in four classes
\footnote{Further investigation of these limits in relation to the zeros of the amplitudes was carried out by Feng et al \cite{Feng:2011jxa}}
\begin{equation}\label{CollMultLims}
 \begin{split}
  &\lim_{\mbox{\tiny $P_{i\mathcal{I}_k}^2$}\:\rightarrow\:0}
    P_{\mbox{\tiny $i\mathcal{I}_k$}}^2\,M_n\:=\:
   M(i,\,\mathcal{I}_k,\,-P_{i\mathcal{I}_k})\, M(P_{i\mathcal{I}_k}, \mathcal{J}_k, j),\\
  &\lim_{\mbox{\tiny $P_{\mathcal{K}}^2$}\:\rightarrow\:0}
    P_{\mbox{\tiny $\mathcal{K}$}}^2\,M_n\:=\:
   M_{s+1}(\mathcal{K},\,-P_{\mathcal{K}})\, M_{n-s+1}(P_{\mathcal{K}}, \mathcal{Q}, i, j),\\
  &\lim_{\mbox{\tiny $P_{k_1 k_2}^2$}\:\rightarrow\:0}
    P_{\mbox{\tiny $k_1 k_2$}}^2\,M_n\:=\:
   M_3(k_1,\,k_2,\,-P_{k_1 k_2})\, M_{n-1}(P_{k_1 k_2}, \mathcal{K}, i, j),\\
  &\lim_{\mbox{\tiny $P_{ij}^2$}\:\rightarrow\:0}
    P_{\mbox{\tiny $ij$}}^2\,M_n\:=\:
   M_{3}(i,\,j,\,-P_{ij})\, M_{n-1}(P_{ij}, \mathcal{K})
 \end{split}
\end{equation}
providing the following conditions
\begin{equation}\label{dressCond}
 \begin{split}
     &P_{ik}^2(z_0^{\mbox{\tiny $(l)$}})\:=\:\langle i,k\rangle\alpha_{ik}^{\mbox{\tiny $(l)$}}[i,j],\qquad
      P_{jk}^2(z_0^{\mbox{\tiny $(l)$}})\:=\:\langle i,j\rangle\alpha_{jk}^{\mbox{\tiny $(l)$}}[j,k],\\
     &\lim_{P_{\mathcal{K}}^2\rightarrow0}f_{\mbox{\tiny $i\mathcal{I}_k$}}^{\mbox{\tiny $(\nu,\,n)$}}\:=\:
      f_{\mbox{\tiny $i\mathcal{I}_k$}}^{\mbox{\tiny $(\nu,\,n-s+1)$}},\qquad
      \lim_{P_{i\mathcal{I}_k}^2\rightarrow0}f_{i\mathcal{I}_k}^{\mbox{\tiny $(\nu,n)$}}\,=\,1,\\
     &\lim_{[k_1,k_2]\rightarrow0}f_{\mbox{\tiny $i\bar{k}$}}^{\mbox{\tiny $(\nu,n)$}}\:=\:
       f_{\mbox{\tiny $i(k_1 k_2)$}}^{\mbox{\tiny $(\nu,n-1)$}},\qquad
      \lim_{\langle k_1,k_2\rangle\rightarrow0}f_{\mbox{\tiny $j\bar{k}$}}^{\mbox{\tiny $(\nu,n)$}}\:=\:
       f_{\mbox{\tiny $j(k_1 k_2)$}}^{\mbox{\tiny $(\nu,n-1)$}},\\
     &\lim_{[i,j]\rightarrow0}\sum_{k}(-1)^{2(h_i+h_j+h_k)+1}
      \left[
       \left(\frac{\langle i,k\rangle}{\langle i,j\rangle}\right)^{-{\mbox{\footnotesize dim}\{\kappa\}}}
       \frac{\mathcal{H}_{n-1}^{(k)}}{\prod_{l=1}^{\nu+1}\alpha_{ik}^{\mbox{\tiny $(l)$}}}\right]\:=\:1,\\
     &\lim_{\langle i,j\rangle\rightarrow0}\sum_k(-1)^{2(h_i+h_k)+1}
      \left[
       \left(\frac{[j,k]}{[i,j]}\right)^{-{\mbox{\footnotesize dim}\{\kappa\}}}
       \frac{\tilde{\mathcal{H}}_{n-1}^{(k)}}{\prod_{l=1}^{\nu+1}\alpha_{jk}^{\mbox{\tiny $(l)$}}}
      \right]\:=\:1,
    \end{split}
\end{equation}
where $\mathcal{H}_{n-1}^{(k)}$ and $\tilde{\mathcal{H}}_{n-1}^{(k)}$ are dimensionless helicity factors related to the $(n-1)$-particle scattering amplitude
emerging when, respectively, the anti-holomorphic and the holomorphic collinear limits are taken.

So far, it is not known if those limits are enough to find valid and computable expression for the dressing factors. However, for the four-particle amplitudes
the constraints above reduces to a simple equation
\begin{equation}\label{4ptDressConds}
 \prod_{r=1}^{N_P^{\mbox{\tiny fin}}}P^2_{iv_r}(z_0^{\mbox{\tiny $(s)$}})\:=\:(-1)^{N_P^{\mbox{\tiny fin}}}\left(P^2_{ij}\right)^{N_P^{\mbox{\tiny fin}}},
\end{equation}
where $N_P^{\mbox{\tiny fin}}$ is the number of poles at finite location (in the $n\,=\,4$ case, one can have either one or two poles), and $v_r$ runs on the labels
of the particles and can acquire all the values which are different from $i$ and $j$ (which label the deformed particles).

\subsubsection{Three-particle deformations}\label{subsubsec:3ptDef}

Let us now consider a different type of deformation, which involves the momenta of three particles. There is no unique way to define such a deformation. Let us
single out the particles labelled by $i$, $j$ and $k$; Then, it is possible to distinguish four classes of deformations, depending on the spinors shifted. 
In particular, one can introduce the momentum deformation by shifting the (anti)holomorphic spinors for all the three particles, the holomorphic spinors of two
particles and anti-holomorphic of the other one, and, conversely, the anti-holomorphic spinors of two particles and the holomorphic of the third one. The last two
can be used to generate a recursion relations for $\kappa_0\phi^4$-theory, after an auxiliary massive scalar has been introduced \cite{Benincasa:2007xk}, while
the first two can generate (MHV)$\bar{\mbox{MHV}}$ expansions \cite{Risager:2005vk}. Let us comment on them. In particular, let us consider the following 
three-particle deformation
\begin{equation}\label{MHVdef}
 \tilde{\lambda}^{\mbox{\tiny $(i)$}}(z)\:=\:\tilde{\lambda}^{\mbox{\tiny $(i)$}}+z\langle j,k\rangle\tilde{\omega},\hspace{.2cm}
 \tilde{\lambda}^{\mbox{\tiny $(j)$}}(z)\:=\:\tilde{\lambda}^{\mbox{\tiny $(j)$}}+z\langle k,i\rangle\tilde{\omega},\hspace{.2cm}
 \tilde{\lambda}^{\mbox{\tiny $(k)$}}(z)\:=\:\tilde{\lambda}^{\mbox{\tiny $(k)$}}+z\langle i,j\rangle\tilde{\omega},
\end{equation}
where $\tilde{\omega}$ is a reference spinor, the three particles whose momenta have been shifted have negative helicity, and the coefficients given by the
holomorphic inner products are fixed by momentum conservation. Its infinitesimal generator can be written as
\begin{equation}\label{MHVgen}
 \mathcal{G}_{\mbox{\tiny $(i,j,k)$}}^{\dot{a}}\:=\:
  \langle j,k\rangle\frac{\partial}{\partial\tilde{\lambda}^{\mbox{\tiny $(i)$}}_{\dot{a}}}+
  \langle k,i\rangle\frac{\partial}{\partial\tilde{\lambda}^{\mbox{\tiny $(j)$}}_{\dot{a}}}+
  \langle i,j\rangle\frac{\partial}{\partial\tilde{\lambda}^{\mbox{\tiny $(k)$}}_{\dot{a}}}.
\end{equation}
The fact that this type of deformation induces an MHV representation of the amplitude, {\it i.e.} the amplitude can be constructed by gluing together just
holomorphic amplitudes with lower external states, has been proved just for gluon amplitudes \cite{Risager:2005vk}, providing with an on-shell version of the original
CSW expansion \cite{Cachazo:2004kj}. More precisely, under the deformation (\ref{MHVdef}) gluons scattering amplitudes vanish in the complex-UV limit, so that it
can be expressed in terms of the residues of its poles at finite location, which can be recast as products of MHV amplitudes with fewer external states. 

However, already if we consider the graviton scattering amplitudes -- which in the BCFW case are better behaved in the complex-UV -- this behavior no longer holds.
In particular, the NMHV $n$-graviton amplitude goes at infinity as $z^{n-12}$, which was firstly proved numerically \cite{Bianchi:2008pu} and then analytically
\cite{Benincasa:2007qj}. In this specific case, the object obtained by summing all the residues at finite location show all the physical poles of the full amplitude.
but it turns out that it does not factorize correctly under two classes of collinear limits \cite{Conde:2012ik}: the holomorphic collinear limit involving two 
particles with positive helicity and the holomorphic collinear limit involving two particle with different helicities. Furthermore, together with the physical poles,
there are others which depend on the reference spinor $\tilde{\omega}$ and thus are unphysical.

The boundary term can be computed by defining a new object
\begin{equation}\label{MHVanom}
 \mathcal{A}_n^{\mbox{\tiny $(i,j,k)$}}\:=\:M_n-M_n^{\mbox{\tiny $(i,j,k)$}}\:=\:
  \frac{\mathcal{P}_n}{\prod_{a,l_i,n_{A_k}<n-8}\langle a|P_{a A_k}|\tilde{\omega}]^{n-8-n_{A_k}}\langle a,l_0\rangle\langle l_1,l_2\rangle},
\end{equation}
where $M_n$ is, as usual, the physical amplitudes while $M_n^{\mbox{\tiny $(i,j,k)$}}$ is the result provided by the sum of the residues at finite location under
the deformation (\ref{MHVdef}). The last expression in (\ref{MHVanom}) shows the schematic structure of $\mathcal{A}_n$, with $\mathcal{P}_n$ being a polynomial in the 
Lorentz invariants, $a$ runs over the negative helicity particles while $l_i$ over the positive helicity ones. This newly defined object can be explicitly computed 
using the BCFW-deformation discussed in Section \ref{subsubsec:BCFW}, with the main difference that the unphysical pole is generally a multiple pole. 
In the complex-UV limit $\mathcal{A}_n^{\mbox{\tiny $(i,j,k)$}}$ vanishes, so it can be reconstructed from its poles. As far as the residue of the unphysical multiple
pole, it is identified with
\begin{equation}\label{UnphRes}
 M^{(\mbox{\tiny unph})}_n \:=\:M_{\mbox{\tiny L}}^{\mbox{\tiny $(i,j,k)$}}(\hat{a},A_l,-\hat{P}_{a A_l})\frac{1}{P^2_{aA_l}}
  M_{\mbox{\tiny R}}^{\mbox{\tiny $(i,j,k)$}}(\hat{P}_{a A_l}, \hat{b}, \hat{c}, B_l).
\end{equation}

\subsubsection{Multi-particle deformations}\label{subsubsec:MultDef}

Finally, we discuss the possibility of deforming the momenta of a high number of particles. The need to resort to this type of deformations comes from the higher 
probability to generate a one-parameter family of amplitudes which is good behaved in the complex-UV, {\it i.e.} it vanishes as the deformed momenta are taken to
infinity and, thus, there is no contribution from the boundary term. Indeed, there are many ways in which one can choose to implement a multi-particle deformation. 
A first example was used to generate an on-shell representation for graviton amplitudes which later served to prove that, under a BCFW-deformation, these amplitudes 
behaves as $z^{-2}$ when $z$ is taken to infinity \cite{Benincasa:2007qj}. In that case, if the number of external gravitons with positive helicity is higher than
the number of negative helicity gravitons, then one singles out one particle with negative helicity, shifting the anti-holomorphic spinor, and all the positive
helicity graviton, shifting their holomorphic spinors. Conversely, if the number of positive helicity gravitons is smaller than the one of the negative helicity 
gravitons, one singles out one positive helicity graviton, shifting its holomorphic spinor, and all the negative helicity gravitons, shifting their anti-holomorphic
spinors.

Here we briefly discuss a deformation which is called in the literature {\it all line shift} \cite{Cohen:2010mi}. It is defined as
\begin{equation}\label{AllS}
 \begin{split}
  &\mbox{anti-holomorphic: }
   \tilde{\lambda}^{\mbox{\tiny $(i)$}}(z)\:=\:\tilde{\lambda}^{\mbox{\tiny $(i)$}}-z\tilde{\beta}^{\mbox{\tiny $(i)$}}\tilde{\omega},\qquad i\:=\:1,\ldots,n,\\
  &\mbox{holomorphic: }\qquad
   \lambda^{\mbox{\tiny $(i)$}}(z)\:=\:\lambda^{\mbox{\tiny $(i)$}}+z\beta^{\mbox{\tiny $(i)$}}\omega
 \end{split}
\end{equation}
with the Lorentz invariant coefficients $\beta^{\mbox{\tiny $(i)$}}$ and $\tilde{\beta}^{\mbox{\tiny $(i)$}}$ fixed by momentum conservation, and
$\omega/\tilde{\omega}$ are reference spinors. Under such a deformation, the dependence on $z$ comes
just from the inner products of a given holomorphicity (depending on the deformation chosen), each of them with a linear dependence. Thus, as $z$ is taken to infinity,
the one-parameter family of amplitudes behaves at worst as $z^{\mathfrak{s}}$ for the anti-holomorphic shift or $z^{\mathfrak{a}}$ for the holomorphic one, 
$\mathfrak{s}$ and $\mathfrak{a}$ being the difference between the number of inner products with a fixed holomorphicity in the numerator and in the denominator 
($\mathfrak{s}$ in the anti-holomorphic case, $\mathfrak{a}$ in the holomorphic one). The powers $\mathfrak{s}$ and $\mathfrak{a}$ can be expressed in term of known 
parameters via a simple dimensional analysis. First of all, the dimension of an $n$-particle amplitude is $4-n$; However, the general structure of the amplitude allows
also to write it in terms of the dimension of the $n$-point coupling $\kappa_n$, and of the difference between the number of inner products in the numerator and in the
denominator 
\begin{equation}\label{DimAmpl}
 4-n\:=\:\mathfrak{a}+\mathfrak{s}+\mbox{dim}\{\kappa_n\}.
\end{equation}
Secondly, the helicity scaling implies that $\mathfrak{a}-\mathfrak{s}\:=\:-\sum_i h_i$. This equation, together with eq (\ref{DimAmpl}), can be solved for 
$\mathfrak{s}$ and $\mathfrak{a}$, providing the worst large-$z$ behavior under the all-line shifts
\begin{equation}\label{AllLlargez}
 2\mathfrak{s}\:=\:4-n-\mbox{dim}\{\kappa_n\}+\sum_i h_i,\qquad
 2\mathfrak{a}\:=\:4-n-\mbox{dim}\{\kappa_n\}-\sum_i h_i
\end{equation}
which need to be negative in order to admit an on-shell representation with no boundary terms. 
In the case that the different terms in the recursion relation are characterized by coupling constant with different dimensions, then the (worst) large-$z$ behavior
is given by the same expression (\ref{AllLlargez}) but with $\mbox{dim}\{\kappa_n\}\,\longrightarrow\,\mbox{min}\{\mbox{dim}\{\kappa_n\}\}$. 

Let us discuss some example, starting with those theories characterized by fundamental three-particle couplings with the same dimensionality. In this case, the 
dimension of the $n$-particle coupling constant can be expressed in terms of the dimension of the three-particle one: 
$\mbox{dim}\{\kappa_n\}\:=\:(n-2)\mbox{dim}\{\kappa_3\}$. If we further restrict to the case of self-interacting spin-$s$ theories, then we have:
\begin{equation}\label{AllLlargezSelf}
 \begin{split}
  &2\mathfrak{s}\:=\:4-n-(n-2)(1-s)-n_{-}s+n_{+}s\:<\:0\quad\Longrightarrow\quad (s-1)n_{+}-n_{-}\:<\:s-3,\\
  &2\mathfrak{a}\:=\:4-n-(n-2)(1-s)+n_{-}s-n_{-}s\:<\:0\quad\Longrightarrow\quad (s-1)n_{-}-n_{+}\:<\:s-3,
 \end{split}
\end{equation}
where $n_{+}$ and $n_{-}$ being respectively the number of particles with positive and negative helicity ($n\,=\,n_{+}+n_{-}$). Therefore, for $s\,=\,0$
($\kappa_{1}\phi^3$-theory), the amplitudes are constructible under both the all line shifts (\ref{AllS}) for any $n$; for $s\,=\,1$ (pure Yang-Mills)
the anti-holomorphic/holomorphic deformation works if the amplitudes involve at least three negative/positive helicity gluons; finally, for $s\,=\,2$ (pure gravity)
the two conditions (\ref{AllLlargezSelf}) can be respectively rewritten as $n_{-}-n_{+}\,>\,1$ and $n_{+}-n_{-}\,>\,1$, so that the smallest amplitudes which
admit an on-shell representation generated by an all-line deformation is the six-positive amplitude with two positive/negative helicity gravitons.

\subsubsection{Supersymmetric BCFW-deformations}\label{subsubsec:SuperBCFW}

The classes of deformations discussed so far allow to show a recursive structure for tree-level amplitudes, provided that either the amplitude vanishes as 
the momenta are taken to infinity along some complex direction, or we can handle the possible boundary term. In the case of supersymmetric theories, those
deformations do not respect supersymmetry and, furthermore, not all the amplitudes of the theory vanish in the complex-UV. It is possible to define
a supersymmetric extension of the BCFW-deformation \cite{ArkaniHamed:2008gz}
\begin{equation}\label{BCFWsuper}
 \tilde{\lambda}^{\mbox{\tiny $(i)$}}(z)\:=\:\tilde{\lambda}^{\mbox{\tiny $(i)$}}-z\tilde{\lambda}^{\mbox{\tiny $(j)$}},
 \quad
 \lambda^{\mbox{\tiny $(j)$}}(z)\:=\: \lambda^{\mbox{\tiny $(j)$}}+z \lambda^{\mbox{\tiny $(i)$}},
 \quad
 \left\{
  \begin{array}{l}
   \eta^{\mbox{\tiny $(j)$}}(z)\:=\:\eta^{\mbox{\tiny $(j)$}}+z\eta^{\mbox{\tiny $(i)$}}\\
   \phantom{\ldots}\\
   \tilde{\eta}^{\mbox{\tiny $(i)$}}(z)\:=\:\tilde{\eta}^{\mbox{\tiny $(i)$}}-z\tilde{\eta}^{\mbox{\tiny $(j)$}}
  \end{array}
 \right. ,
\end{equation}
where either $\eta^{\mbox{\tiny $(j)$}}$ or $\tilde{\eta}^{\mbox{\tiny $(i)$}}$ are deformed, depending on whether the amplitude is chosen to be in the $\eta$-
or $\tilde{\eta}$-representation. Under such deformations, the supersymmetric $\delta$-functions do not change, supersymmetry is preserved, and the complex-UV
behavior is $\mathcal{O}(z^{-s})$ ($s$ being the highest spin in the multiplet). Therefore, the amplitudes admit the following recursive structure
\begin{equation}\label{SuperRR}
 \begin{split}
  &M_n(\{\lambda,\tilde{\lambda},\eta\})\:=\sum_{k\in\mathcal{P}^{\mbox{\tiny $(i,j)$}}}\int d^{\mathcal{N}}\eta
   \left\{\phantom{\frac{1}{P_{i\mathcal{I}_k}^2}}\right.\\
  &\left.
   \hspace{.2cm}M_{\mbox{\tiny L}}^{\mbox{\tiny $(i,j)$}}
    \left(\hat{i},\eta^{\mbox{\tiny $(i)$}};\mathcal{I}_k,\{\eta\}_{\mathcal{I}_k};-\hat{P}_{i\mathcal{I}_k},\eta\right)
    \frac{1}{P_{i\mathcal{I}_k}^2}
    M_{\mbox{\tiny R}}^{\mbox{\tiny $(i,j)$}}\left(\hat{P}_{i\mathcal{I}_k},\eta; \mathcal{J}_k,\{\eta\}_{\mathcal{J}_k};\hat{j},\hat{\eta}^{\mbox{\tiny $(j)$}}\right)
   \right\},
 \end{split}
\end{equation}
where, as usual, the hat $\hat{\phantom{i}}$ indicates that the related quantity is evaluated at the location of the pole -- similarly, for the amplitudes in the
$\tilde{\eta}$-representation. Notice that the lower-point amplitudes are computed at $\hat{j},\hat{\eta}^{\mbox{\tiny $(j)$}}$, which means that these recursion 
relations involves amplitudes with different external states. Note also that, in maximally supersymmetric theories, both the two possible two-particle deformations 
({\it i.e.} the one in eq (\ref{BCFWsuper}) and the one obtained from it by the exchange $i\,\longleftrightarrow\,j$) have the same complex-UV behavior and, therefore,
induce a recursive structure such as eq (\ref{SuperRR}), contrarily to what happens in the non-supersymmetric BCFW-deformation for which it depends on the helicity
of the particles whose momenta get deformed.

\subsection{Tree level constructibility}\label{subsec:TreeConstr}

The existence of on-shell recursion relations provide us with the notion of constructibility: a theory is constructible if its scattering amplitudes can be built
recursively from some fundamental object. These recursion relations can be generated in different ways and can provide different on-shell representations for the
same physical quantity. In general, this type of structure is easily determined if the deformations generating it is well-behaved in the complex-UV: in this case
a scattering amplitude can be expressed in terms of products of two scattering amplitudes with fewer external states. However, the starting point for building the
higher point amplitudes may change in different on-shell representations. As an example, the all-line deformations for gluon and graviton amplitudes respectively need
the four- and five-particle amplitudes. It is true that those amplitudes are simple enough to be determined in other ways: for example, the four-gluon amplitudes can 
be expressed in terms of the Parke-Taylor formula \cite{Parke:1986gb}, while the five-graviton amplitude can be represented by the Berends-Giele-Kuijf one 
\cite{Berends:1988zp}. However, if we want to be able to reduce the construction of a general S-matrix theory to a minimum amount of assumptions, all the non-trivial 
scattering amplitudes need to be connected {\it through the same rules} to a building-block which should be determined from first principles. Said in another way, the 
isometry group of the space-time defines a class of fundamental objects on which we want to build consistent rules to construct and describe scattering processes.

Indeed, this issue would be solved in many cases if we had a clear understanding of the boundary term connected to each deformation. For the multi-particle 
deformations, there is no real knowledge about it: Neither a definite physical interpretation is available, nor (even a partial) prescription about how to compute it. 
More or less, a similar situation concerns the three-particle deformations: in the case we discussed in Section \ref{subsubsec:3ptDef}, first of all, the smallest 
amplitude the deformation can be applied is a five-particle amplitude, and, secondly, the computation of the missed term is possible making use of the good-behavior in
the complex-UV under a BCFW-deformation.

As far as the BCFW-deformations are concerned, well-behaved amplitudes in the complex-UV can be naturally built by gluing together our building blocks. If on one side
it is true that there is no complete prescription which allows to compute the boundary terms, on the other side it is also true that the physical interpretation
of this limit is cleaner as well as it is clear why the standard BCFW-representation fails to hold: the collinear singularity which is not manifest in this 
representation appears as a soft singularity connected to one of the deformed external momenta, and the failure of this soft limit to provide the correct singularity 
(and, consequently, the failure to produce the correct factorization) implies the failure of the standard BCFW-representation and the need of some extra term. 
More specifically, all this depends on the behavior of the three-particle amplitudes of the theory when one of its particles becomes soft: this type of deformation 
allows to recast even the complex-UV behavior of the amplitudes in terms of the three-particle amplitudes.

Furthermore, the on-shell representation generated with a BCFW-deformation has been completed to include the boundary terms, even if several issues still need to be 
addressed. In particular, this completion is nothing but a dressed version of the standard BCFW-representation. The prescription to compute the dressing factors
is not easily solvable in general: it can be solved exactly in the four-particle case, and just in very specific cases for higher point amplitudes. Notwithstanding all
the open issues, it provides a useful extension of the notion of constructibility.

\subsection{Tree level consistency}\label{subsec:TreeCons}

So far we have been discussing the possibility of reconstructing the scattering amplitudes of an arbitrary theory recursively. However, none of the on-shell 
representations discussed guarantees by itself the physical consistency of the object one can construct. In a constructible theory, the physical amplitudes
should be independent of the deformation which generates it. Thus, we can consider the simplest non-trivial amplitude, compute it via two different deformations
and impose that these objects are equal \cite{Benincasa:2007xk}
\begin{equation}\label{4ptTest}
 M_4^{\mbox{\tiny $(i,j)$}}(0)\:=\:M_4^{\mbox{\tiny $(i,k)$}}(0).
\end{equation}
Such a requirement returns non-trivial constraints on the scattering amplitudes. In particular, it is possible to use the dressed on-shell
representation (\ref{BCFWrr}) and the prescription (\ref{4ptDressConds}) to implement the consistency condition (\ref{4ptTest}) 
\cite{Benincasa:2011pg}.

Among the interesting results of this consistency requirement, it is important to stress that
\begin{romanlist}[(ii)]
 \item for self-interacting spin-$1$ particles, a non-trivial interaction is admitted if and only if the following relation holds
       \begin{equation}\label{Jacobi}
        \sum_{a_P}\varepsilon_{a_i a_k a_P}\varepsilon_{a_P a_l a_j} + \sum_{a_P}\varepsilon_{a_i a_l a_P}\varepsilon_{a_P a_k a_j} + 
        \sum_{a_P}\varepsilon_{a_i a_j a_P}\varepsilon_{a_P a_l a_k} \:=\:0,
       \end{equation}
       which is nothing but the Jacobi identity, and $\varepsilon$'s are the structure constant of a Lie algebra;
 \item the same happens when scattering amplitudes of gluons and matter ($s'\,\le\,1/2$);
 \item considering the external states with arbitrary spin (but each three-particle amplitudes having at most two particles with different spin), the consistency
       condition returns all the known theories: the parity preserving interactions are characterized by three particle amplitudes with helicity configurations
       $(-s,+s,\mp s)$ and $(-s',+s',s)$, while parity violating ones (essentially the Yukawa coupling) are characterized by the helicity configurations 
       $(-s',-s',+s)$ and $(+s',+s',-s)$
 \item in particular, it returns the coupling between graviton and gravitino with the correct relation between the self-interacting spin-$2$ coupling and the
       graviton-gravitino one, as dictated in $\mathcal{N}\,=\,1$ supergravity;
 \item providing the spin-$2$ with some internal structure, the consistency requirement implies a reducible algebra so that we have just a collection of 
       self-interacting gravitons which do not interact with each other.
\end{romanlist}

Some comments are now in order. First of all, in order for the consistency condition (\ref{4ptTest}) to be applied, it is necessary that the singularity structure
of the scattering amplitude is characterized by at least two channels in momentum space, so that the two different BCFW-deformations can pick (at least) one of them.
This is the case for all the theories just mentioned. However, as it is possible to argue from the classification of the three-particle coupling of Section 
\ref{sec:3ptAmpl}, there are theories that, in principle, are characterized by a single factorization channel. Let us fix for concreteness this factorization
channel to correspond to the momenta $P^2_{il}$ going on-shell. First, considering the $(i,l)$-deformation does not pick any pole and the full amplitude coincides with
the boundary term: such a representations would be useless since there would be no way to accede to this residue. Instead, both the deformations considered in eq 
(\ref{4ptTest}) pick this pole; however, there is no other 
factorization channel to be considered, so that all the analysis which connects the complex-UV behavior of an amplitude to the soft-one of the three-particle
building block breaks down. One can think about the limits in which the S-matrix becomes trivial as a generic property and, if this were the case, it may be 
reasonable to assume that the condition on the zeros (\ref{4ptDressConds}) still holds. With this extra assumption, it seems that the consistency conditions
(\ref{4ptTest}) may allow for interactions of particles with spin higher than two. Indeed, the assumption of a certain ``universality'' of the zeros of the amplitude
is a very strong assumption (which, for the time being, we do not intend to add to our set of hypothesis, and that we mention just for the sake of completeness) with
a very interesting implication: the boundary terms do not show any pole in momentum space (which is consistent with the fact that the helicity configurations
for these classes of theories allow for just one factorization channel) and have a structure of a contact interaction which a simple dimensional analysis show
to have higher derivatives with respect to the three-particle couplings. Not only, this number of derivatives seem to increase with the growth of the number of
external states. This can be interpreted as a signature of non-locality for these theories. However, it is important to stress that given the lack of control on the
complex-UV behavior for this class of theories, it is not guaranteed at all that consistency at four-particle level implies consistency at higher-point level.

Another way to look at the four-particle consistency has been recently proposed \cite{McGady:2013sga}. A first selection of potentially consistent theories is 
based on pole counting. More specifically, in a schematic way, a four particle amplitude has the form
\begin{equation}\label{4ptForm}
 M_4(1^{h_1}, 2^{h_2}, 3^{h_3}, 4^{h_4})\:=\:\kappa^2\frac{\mathcal{H}(1,2,3,4)}{\mathcal{F}(s,t,u)},
\end{equation}
where $\mathcal{H}$ carries the helicity information, and $\mathcal{F}(s,t,u)$ contains the pole structure, considering that the four particle amplitude can show at 
most three poles and the number of poles can be written as $N_p\,=\,2\hat{h}+1-h\,\le\,3$ ($\hat{h}\,\equiv\,\mbox{max}\{|h_1|,\,|h_2|,\,|h_3|\}$). Amplitudes with
minimal numerators are considered.

Then, a further selection imposing unitarity and locality requiring that the four-particle amplitudes factorize under complex factorization 
$\langle i,j\rangle\,\longrightarrow\,0$ and $[i,j]\,\longrightarrow\,0$. 

This approach indeed provides an handle on the scattering amplitudes with just one factorization channel and on high-spin interactions. In particular, it has been
found that there might exist theories which are consistent (with unitarity and locality) and they satisfy the constraint $\hat{h}\,\in\,[h/3,\,h/2]$, with $h\,>\,3$.
These theories do not couple neither to YM nor to GR, however they may consistently interact with spin-$1$ and spin-$2$ particles whose self-interaction is 
characterized by the helicity configurations $(\mp s,\,\mp s,\, \mp s)$ -- {\it i.e.} $F^3$ and $R^3$ operators.

\section{Trees and loops}\label{sec:TreeLoop}

The extensive analysis of the on-shell representations at tree level taught us that the pole structure of the tree level S-matrix contains poles at finite location, 
which signal the possibility of producing a new particle on-shell in its two helicity states, and a pole at infinity connected with the soft behavior of the 
three-particle amplitudes. Now a legitimate question is {\it what can the tree-level tell us about the loop structure?} or equivalently {\it how the tree-level 
structure reflects itself in the loop one?}

In order to address this question, let us begin with considering a concrete example at one-loop. Consider the three-particle cut identified by the following $T^3$
\begin{equation}\label{TreeLoopT3}
 T^3\:=\:\left\{l\,\in\,\mathbb{C}^4\,|\,l^2\,=\,0,\:(p^{\mbox{\tiny $(i)$}}-l)^2\,=\,0,\:(p^{\mbox{\tiny $(j)$}}+l)^2\,=\,0\right\}.
\end{equation}
It has two families of one-parameter solutions
\begin{equation}\label{TreeLoopT3soln}
 \begin{split}
  &\mbox{solution 1: }\\
  &\hspace{.45cm}
   l\:=\:z\lambda^{\mbox{\tiny $(i)$}}\tilde{\lambda}^{\mbox{\tiny $(j)$}},\quad
   p^{\mbox{\tiny $(i)$}}-l\:=\:\lambda^{\mbox{\tiny $(i)$}}(\tilde{\lambda}^{\mbox{\tiny $(i)$}}-z\tilde{\lambda}^{\mbox{\tiny $(j)$}}),\quad
   p^{\mbox{\tiny $(j)$}}+l\:=\:(\lambda^{\mbox{\tiny $(j)$}}+z\tilde{\lambda}^{\mbox{\tiny $(i)$}})\tilde{\lambda}^{\mbox{\tiny $(j)$}},\\
  &\mbox{solution 2: }\\
  &\hspace{.45cm}
   l\:=\:z\lambda^{\mbox{\tiny $(j)$}}\tilde{\lambda}^{\mbox{\tiny $(i)$}},\quad
   p^{\mbox{\tiny $(i)$}}-l\:=\:(\lambda^{\mbox{\tiny $(i)$}}-z\tilde{\lambda}^{\mbox{\tiny $(j)$}})\tilde{\lambda}^{\mbox{\tiny $(i)$}},\quad
   p^{\mbox{\tiny $(j)$}}-l\:=\:\lambda^{\mbox{\tiny $(j)$}}(\tilde{\lambda}^{\mbox{\tiny $(j)$}}+z\tilde{\lambda}^{\mbox{\tiny $(i)$}}),
 \end{split}
\end{equation}
each of them contributing to the three-particle cut with a term of the following form
\begin{equation}\label{TreeLoop3cut}
 \begin{split}
  \left.\Delta_{3}M_n^{\mbox{\tiny $(1)$}}\right|_{l_{\star}}\:
   &=\:\int_{T^3_{\star}}\frac{d^4l}{(2\pi)^4}\,M_{3}^{\mbox{\tiny tree}}(-(p^{\mbox{\tiny $(i)$}}-l),\,p^{\mbox{\tiny $(i)$}},\,-l)
    M_3^{\mbox{\tiny tree}}(l,\,p^{\mbox{\tiny $(j)$}},\,-(p^{\mbox{\tiny $(j)$}}+l))\times\\
   &\hspace{1cm}\times M_n^{\mbox{\tiny tree}}(p^{\mbox{\tiny $(j)$}}+l,\,\mathcal{K},\,p^{\mbox{\tiny $(i)$}}-l)\:=\\
   &=\:\int\frac{dz}{z}\,M_{3}^{\mbox{\tiny tree}}(-(p^{\mbox{\tiny $(i)$}}-zq),\,p^{\mbox{\tiny $(i)$}},\,-zq)
    M_3^{\mbox{\tiny tree}}(zq,\,p^{\mbox{\tiny $(j)$}},\,-(p^{\mbox{\tiny $(j)$}}+zq))\times\\
   &\hspace{1cm}\times M_n^{\mbox{\tiny tree}}(p^{\mbox{\tiny $(j)$}}+zq,\,\mathcal{K},\,p^{\mbox{\tiny $(i)$}}-zq),
 \end{split}
\end{equation}
where $q\,=\,\lambda^{\mbox{\tiny $(i)$}}\tilde{\lambda}^{\mbox{\tiny $(j)$}}$ and $q\,=\,\lambda^{\mbox{\tiny $(j)$}}\tilde{\lambda}^{\mbox{\tiny $(i)$}}$
respectively for the solution 1 and 2 of the $T^3$ (\ref{TreeLoopT3}).
Probably, the bottom line of this example is already clear. First, the two families of one-parameter solutions of the $T^3$ (\ref{TreeLoopT3}) are nothing but
the two possible BCFW-deformations which can be defined on particle $i$ and $j$. Secondly, for definiteness, let us focus on solution 1 (for solution 2
the reasoning follows the exact same flow). It is straightforward to notice that the holomorphic spinors of $l$ and $p^{\mbox{\tiny $(i)$}}-l$ are proportional to
each other, while the anti-holomorphic spinor of $l$ is proportional to the one of $p^{\mbox{\tiny $(j)$}}+l$. Moreover, we already know that the 
three-particle amplitudes have definite holomorphicity if $|h|\,\neq\,0$. 
Finally, the $n$-particle amplitude appearing in the integral (\ref{TreeLoop3cut}) is nothing but the one-parameter family of tree-level amplitudes generated
by a BCFW-deformation. To make it even more explicit, let us rewrite the integral (\ref{TreeLoop3cut}) as
\begin{equation}\label{TreeLoop3cut2}
 \begin{split}
  \left.\Delta_{3}M_n^{\mbox{\tiny $(1)$}}\right|_{l_{\star}}\:
    &=\:\int\frac{dz}{z}\,M_{3}^{\mbox{\tiny tree}}(-(p^{\mbox{\tiny $(i)$}}-zq),\,p^{\mbox{\tiny $(i)$}},\,-zq)
     M_3^{\mbox{\tiny tree}}(zq,\,p^{\mbox{\tiny $(j)$}},\,-(p^{\mbox{\tiny $(j)$}}+zq))\times\\
    &\times
     \left[
      \sum_{k\in\mathcal{P}^{\mbox{\tiny $(i,j)$}}}
      M_{\mbox{\tiny L}}^{\mbox{\tiny tree}}(\hat{i},\mathcal{I}_k,-\hat{P}_{i\mathcal{I}_k})
      \frac{1}{P^2_{i\mathcal{I}_k}(z)}
      M_{\mbox{\tiny R}}^{\mbox{\tiny tree}}(\hat{P}_{i\mathcal{I}_k},\mathcal{J}_k,\hat{j})+
      \mathcal{C}_n^{\mbox{\tiny $(i,j)$}}(z)
     \right].
 \end{split}
\end{equation}
The terms in the sum are characterized by one further propagator each, showing the structure of a box-diagram: when the integration picks the poles corresponding to
these propagators, a further propagator gets cut (goes on-shell) and the resulting products of the amplitudes provides the coefficients of the box-integrals.
If the tree-level $n$-particle amplitude vanishes in the complex-UV, the term $\mathcal{C}_n^{\mbox{\tiny $(i,j)$}}(z)$ is not there. Let us consider instead that 
the $n$-particle amplitude does not vanish. 
It is straightforward to see that the structure that the tree-level boundary term induces is the structure of
a triangle diagram, and the related coefficient can be computed in terms of the boundary terms of the tree-level amplitudes.

Summarizing, the presence of triangle-integrals is related to a non-vanishing boundary term at tree-level. In other words, if a theory admits a generalized on-shell
representation at tree-level then it must show triangle-integrals at one-loop level, while if it admits a standard BCFW representation (the dressing factors
are all equal to one), then the one-loop level reveals just box-integrals and the triple cut is determined by the quadruple-cuts.

Strictly speaking, we draw this conclusion by looking at a specific type of diagrams. This is however more general and it amounts to state that the triangle
coefficient is given by the product of three tree-level amplitudes evaluated at infinite $z$ \cite{ArkaniHamed:2008gz}.

Let us now consider the double cuts
\begin{equation}\label{TreeLoopBubble}
 \begin{split}
  \Delta_2 M_n^{\mbox{\tiny $(1)$}}\:=\:\int d^4l_1\,d^4l_2\,&\delta^{\mbox{\tiny $(+)$}}\left(l_1^2\right)\delta^{\mbox{\tiny $(+)$}}\left(l_2^2\right)
    \delta^{\mbox{\tiny $(4)$}}\left(l_1+l_2-\mathcal{K}_{\mbox{\tiny R}}\right)\times\\
   &\times M_{\mbox{\tiny L}}^{\mbox{\tiny tree}}\left(-l_2,\mathcal{K}_{\mbox{\tiny L}},-l_1\right)
   M_{\mbox{\tiny R}}^{\mbox{\tiny tree}}\left(l_1,\mathcal{K}_{\mbox{\tiny R}},l_2\right),
 \end{split}
\end{equation}
and on the (fixed) on-shell momenta $l_1$ and $l_2$ one can perform a BCFW-deformation: $l_1(z)\,=\,l_1-zq$, $l_2(z)\,=\,l_2+zq$. First, we can write the two deformed 
on-shell tree-level amplitude in as a contribution from the poles at finite location and a polynomial in $z$
\begin{equation}\label{TreeLoopBubble2}
  \begin{split}
  \Delta_2 &M_n^{\mbox{\tiny $(1)$}}\:=\:\int d^4l_1\,d^4l_2\,\delta^{\mbox{\tiny $(+)$}}\left(l_1^2\right)\delta^{\mbox{\tiny $(+)$}}\left(l_2^2\right)
    \delta^{\mbox{\tiny $(4)$}}\left(l_1+l_2-\mathcal{K}_{\mbox{\tiny R}}\right)\times\\
  &\times\left[
   \sum_{k_1\in\mathcal{P}_{\mbox{\tiny L}}}M_{\mbox{\tiny L}}^{\mbox{\tiny (L)}}(\hat{l}_1,\mathcal{L}_k^{\mbox{\tiny $(1)$}},-\hat{P}_{1\mathcal{L}_k})
    \frac{1}{P_{1\mathcal{L}_k}^2(z)}M_{\mbox{\tiny L}}^{\mbox{\tiny (R)}}(\hat{P}_{1\mathcal{L}_k},\mathcal{L}_k^{\mbox{\tiny $(2)$}},\hat{l}_2)+
   \mathcal{C}_{\mbox{\tiny L}}(z)
  \right]\times\\
  &\times\left[
   \sum_{k_2\in\mathcal{P}_{\mbox{\tiny R}}}M_{\mbox{\tiny R}}^{\mbox{\tiny (L)}}(\hat{l}_1,\mathcal{R}_k^{\mbox{\tiny $(1)$}},-\hat{P}_{1\mathcal{R}_k})
    \frac{1}{P_{1\mathcal{R}_k}^2(z)}M_{\mbox{\tiny R}}^{\mbox{\tiny (R)}}(\hat{P}_{1\mathcal{R}_k},\mathcal{R}_k^{\mbox{\tiny $(2)$}},\hat{l}_2)+
   \mathcal{C}_{\mbox{\tiny R}}(z)
  \right],
 \end{split}
\end{equation}
where $\mathcal{K}_{\mbox{\tiny L}}\,=\,\mathcal{L}_k^{\mbox{\tiny $(1)$}}\,\cup\,\mathcal{L}_k^{\mbox{\tiny $(2)$}}$ and 
$\mathcal{K}_{\mbox{\tiny R}}\,=\,\mathcal{R}_k^{\mbox{\tiny $(1)$}}\,\cup\,\mathcal{R}_k^{\mbox{\tiny $(2)$}}$, and $\mathcal{C}_{\mbox{\tiny L/R}}(z)$
are two polynomial in $z$ whose zeroth-order coefficients provide the tree-level boundary term for $M_{\mbox{\tiny L/R}}^{\mbox{\tiny tree}}$. 

It is straightforward to see that the coefficient of the bubble integral can be written as\cite{ArkaniHamed:2008gz}
\begin{equation}\label{TreeLoopBubbleCoeff}
 \begin{split}
  \mathcal{C}_2\:=\:\int d^4l_1\,d^4l_2\,&\delta^{\mbox{\tiny $(+)$}}\left(l_1^2\right)\delta^{\mbox{\tiny $(+)$}}\left(l_2^2\right)
     \delta^{\mbox{\tiny $(4)$}}\left(l_1+l_2-\mathcal{K}_{\mbox{\tiny R}}\right)\times\\
    &\times\int_{\gamma_{\infty}}\frac{dz}{z}\,M_{\mbox{\tiny L}}^{\mbox{\tiny tree}}\left(-l_2(z),\mathcal{K}_{\mbox{\tiny L}},-l_1(z)\right)
   M_{\mbox{\tiny R}}^{\mbox{\tiny tree}}\left(l_1(z),\mathcal{K}_{\mbox{\tiny R}},l_2(z)\right),
 \end{split}
\end{equation}
$\gamma_{\infty}$ being a contour which contains just the pole at infinity. Actually, eq (\ref{TreeLoopBubble2}) is providing us with more information:
as expected, it reveals the structure of the box- and triangle-integrals, but it is also stating that whether or not a theory at one loop has triangles 
and/or bubbles depends on the tree-level structure, {\it i.e.} on the presence of the tree-level boundary terms under certain BCFW-deformations. 

\subsection{More on the analytic structure at loop level}\label{subsec:AnStrLoop}

Let us now try to get more insights on the singularities at loop level, starting with considering the structure of the integrands. First of all, there are
kinematic points at which propagators, independent of the loop momenta, go on-shell. They correspond to factorization channels similar to the ones observed
at tree-level: when these poles are approached, the integrands factorize into a product of two lower-loop integrands. 

Another class of singularities is identified by all those propagators containing the loop momenta going on-shell. When these points are approached the original 
$L$-loop integrand is mapped into an $(L-1)$-loop $(n+2)$-point one with the two new on-shell states evaluated in the forward limit \cite{CaronHuot:2010zt}. 
In general, these single cuts return contributions which have a singular kinematics and thus are not well-defined. So, using a BCFW-deformation to try to obtain 
recursion relations for loop amplitudes is not really straightforward: together with the problem just mentioned, it is also necessary to define unambiguously the loop 
momenta -- in order to consistently study the singularity of the integrand and its residues, one would need to be able to write all the contributions in terms of a 
single integral.

Both of these issues have been solved for maximally supersymmetric theories in the planar limit \cite{CaronHuot:2010zt, ArkaniHamed:2010kv}. Specifically,
it was observed that all the singular terms cancel once the sum over the supermultiplet is performed and the single cut corresponds to the forward limit of
a $(n+2)$-particle amplitudes at $(L-1)$-loops, and this occurs for theories which have at least $\mathcal{N}\,=\,1$ in the massless case, or $\mathcal{N}\,=\,2$ 
supersymmetries in the massive one  \cite{CaronHuot:2010zt}. Furthermore, the ambiguity of defining the loop momenta can be easily fixed in the dual coordinate space 
\cite{ArkaniHamed:2010kv}, which is defined as \cite{Drummond:2006rz}
\begin{equation}\label{DualSpace}
 x^{\mbox{\tiny $(i)$}}_{a\dot{a}}-x^{\mbox{\tiny $(i+1)$}}_{a\dot{a}}\:=\:p^{\mbox{\tiny $(i)$}}_{a\dot{a}}, \qquad
 x^{\mbox{\tiny $(1)$}}\,=\,x^{\mbox{\tiny $(n+1)$}}.
\end{equation}
The definition (\ref{DualSpace}) reflects the color ordering, while the last condition guarantees momentum conservation. The ambiguity in the definition of the
loop momenta, {\it i.e.} the possibility to redefine it through a shift, is reflected as a translation in the dual space. Therefore, fixing the coordinates 
$\{x^{\mbox{\tiny $(i)$}}\}$ resolves this issue.

The full understanding of the loop singularities and their residues, as well as the fixing of the other ambiguities allow the possibility of applying a BCFW-procedure,
which is better done in the momentum-twistor space \cite{ArkaniHamed:2010kv}. This allowed to obtain an all-loop BCFW-recursion relations for planar supersymmetric
theories. The discussion of the BCFW recursion relation in momentum-twistor space goes beyond the purpose of the present review. We redirect the interested reader to
the original paper \cite{ArkaniHamed:2010kv} as well as the review \cite{Elvang:2013cua}.

\section{On-shell diagrammatics}\label{sec:OnShDiag}

Our discussion has been so far devoted to the exploration of the singularities of the scattering amplitudes and their physical meaning. The main lesson is that,
at least for some classes of theories, the physical data determining the scattering amplitudes are encoded in the three-particle amplitudes. So, the question is now
how, turning the table around, one can start from the three-particle amplitudes themselves and build up physical processes from them. It is important to point out
that any diagram which can be created by gluing together {\it on-shell} three-particle amplitudes is always characterized by being a representation of an 
(gauge-invariant) {\it on-shell} process, and, thus, of a physical process. Furthermore, any object built in this way does not show neither IR nor UV divergences.

\begin{figure}[htbp]
 \centering 
 \[
  \scalebox{.50}{\includegraphics{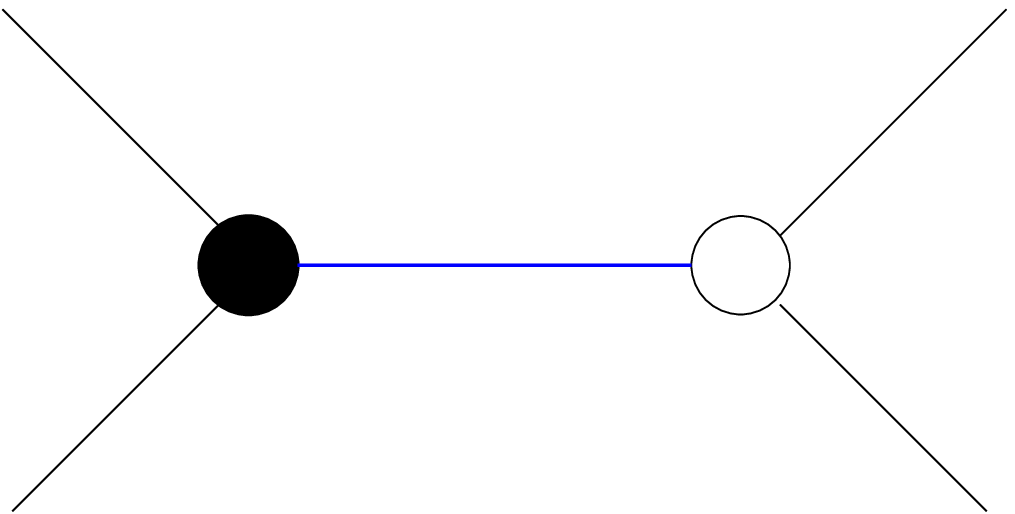}} \qquad
  \scalebox{.50}{\includegraphics{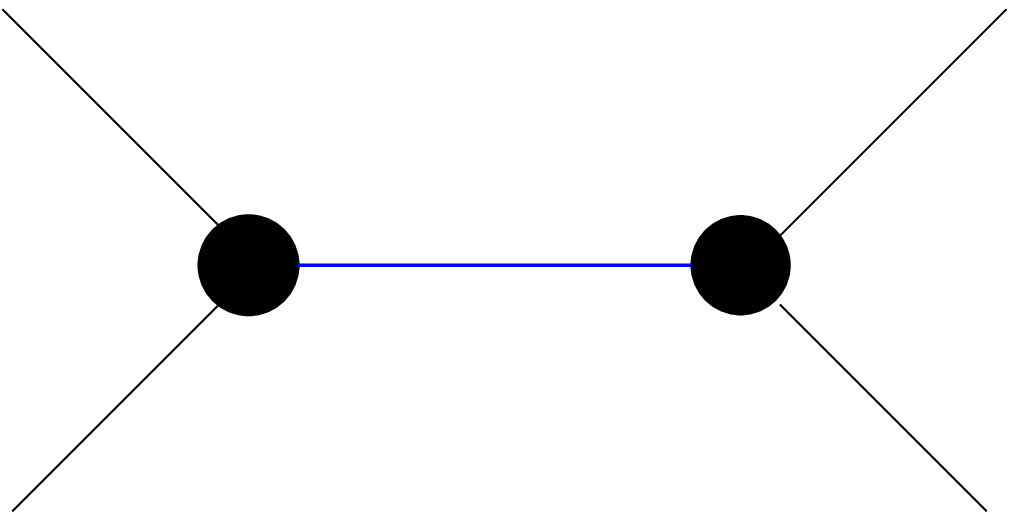}}
 \]
  \caption{On-shell diagram formed by two three-particle amplitudes. The first diagram is made by three-particle amplitudes with different holomorphicity. It 
           corresponds to a singularity given that having the internal line on-shell implies having a $\delta$-function that makes two of the external momenta 
           collinear. The second diagrams is built from two three-particle amplitudes of the same holomorphicity.}\label{fig:Gluing}
\end{figure}

When these building blocks are glued together, the ``internal'' states are actually the sum of all the possible (on-shell) states available in the theory, with
momentum conservation enforced. An intriguing way to perform the gluing operation among three-particle amplitudes is by considering the latter as the 
{\it on-shell form}\cite{ArkaniHamed:2012nw} defined in eq (\ref{OnShForm3}) -- let us recall that this definition holds for non-supersymmetric theories, while the 
supersymmetric version of the on-shell form is obtained by substituting the on-shell phase-space forms with their supersymmetric extensions (\ref{SUSYphsp}).
For convenience, we rewrite such forms here
\begin{equation}\label{OnShForms}
 \begin{split}
  &\hspace{2cm}\mathcal{M}_3\:=\:\delta^{\mbox{\tiny $(4)$}}\left(\sum_{i=1}^3\lambda^{\mbox{\tiny $(i)$}}\tilde{\lambda}^{\mbox{\tiny $(i)$}}\right)
   M_3\Omega^{\mbox{\tiny $(i)$}}\Omega^{\mbox{\tiny $(j)$}}\Omega^{\mbox{\tiny $(k)$}}\\
   &\mbox{non-supersymmetric case: }\hspace{.75cm}\Omega\:=\:\sum_{h}\frac{d^2\lambda\,d^2\tilde{\lambda}}{\mbox{Vol}\left\{GL(1)\right\}}\\
   &\mbox{supersymmetric case: }\hspace{1.5cm}\Omega\:=\:\left\{
    \begin{array}{l}
     \frac{d^2\lambda\,d^2\tilde{\lambda}}{\mbox{Vol}\left\{GL(1)\right\}}d^I\eta,\\
     \phantom{\ldots}\\
     \frac{d^2\lambda\,d^2\tilde{\lambda}}{\mbox{Vol}\left\{GL(1)\right\}}d^I\tilde{\eta}.
    \end{array}
   \right.
 \end{split}
\end{equation}
where we explicitly wrote down the momentum conserving $\delta$-function. Let us try to build up some simple on-shell diagram. 
As a very first step we can glue together just two three-particle amplitudes. In the first diagram
of Figure \ref{fig:Gluing}, a holomorphic and an anti-holomorphic three-particle amplitudes are glued together. As we already mentioned, all the lines (internal and
external) are on-shell. In the case of the internal (blue) line, this implies that a delta-function involving the two external momenta is enforced. Thus, this
diagram represents a singularity.

\begin{figure}[htbp]
 \centering 
 \[
  {\raisebox{-.8cm}{\scalebox{.30}{\includegraphics{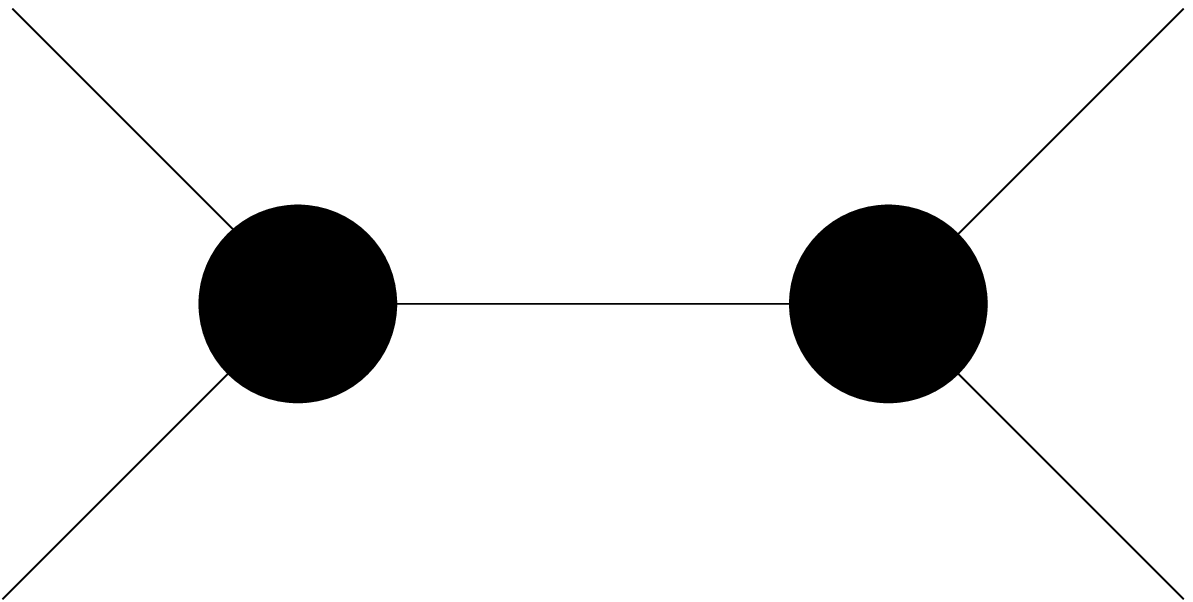}}}}\qquad\Longleftrightarrow\qquad
  {\raisebox{-.8cm}{\scalebox{.30}{\includegraphics{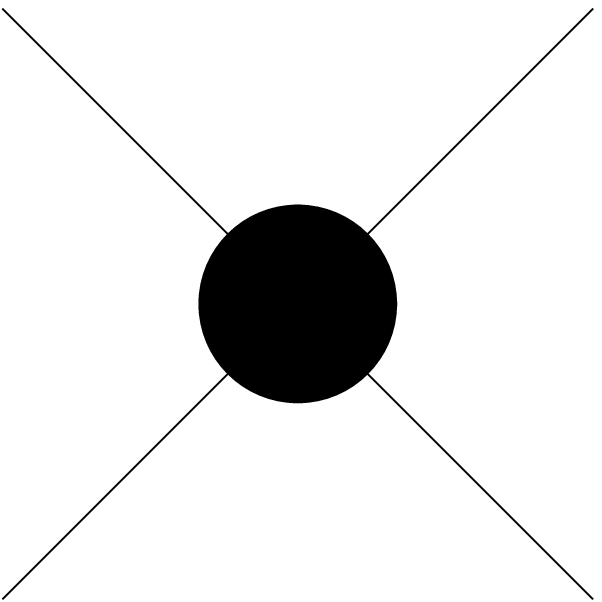}}}}\qquad\Longleftrightarrow\qquad
  {\raisebox{-1.6cm}{\scalebox{.30}{\includegraphics{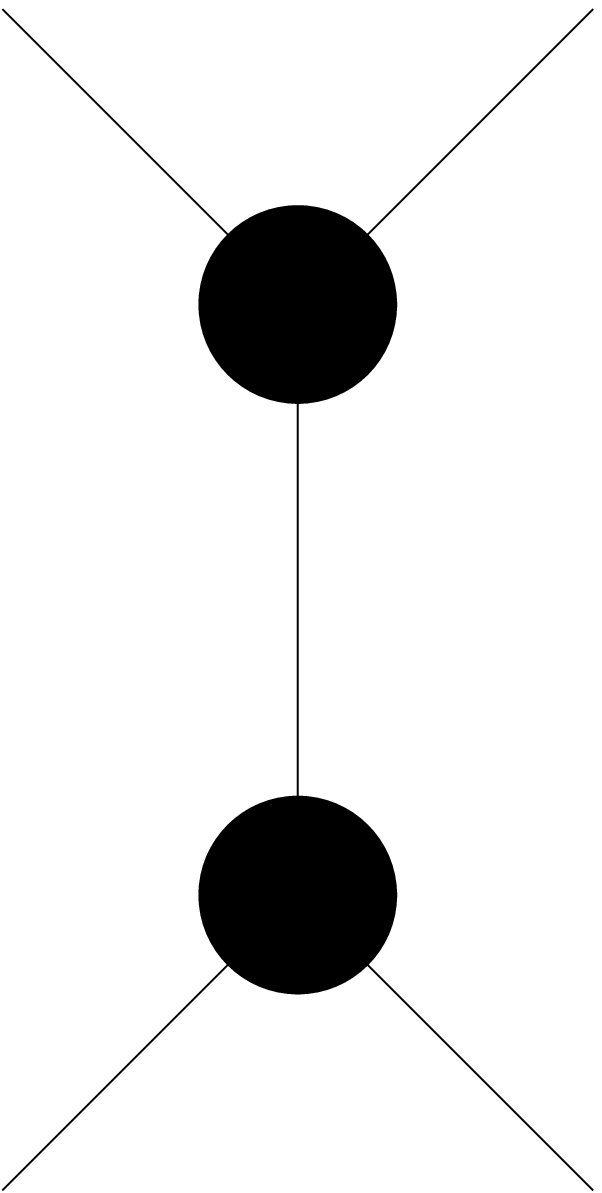}}}}
 \]
  \caption{Merger operation. The two black amplitudes have all the anti-holomorphic spinors proportional to each other. Thus, the two amplitudes can
           be glued together in several equivalent ways. This is equivalent to the statement that the initial diagram can be collapsed and then extended
           again along a different channel but still providing a representation of the same physical process.}\label{fig:Merger}
\end{figure}

As far as the second diagram is concerned, it has been built by joining two amplitudes with the same holomorphicity. As a consequence, all the spinors with the
same holomorphicity (in the case of Figure \ref{fig:Gluing}, the anti-holomorphic spinors) are proportional to each other. Therefore, the two three-particle
amplitudes can be connected to each other in several physically equivalent ways -- see Figure \ref{fig:Merger}. This equivalence operation, called {\it merger} 
can be seen as a contraction of the two three-particle amplitude in a four-particle object and then its expansion along a ``different channel''. 
Especially when this type of diagram is contained within a bigger on-shell diagram, two on-shell diagrams connected by a merger operation describe the same
physical process.

\begin{figure}[htbp]
 \centering 
  \scalebox{.50}{\includegraphics{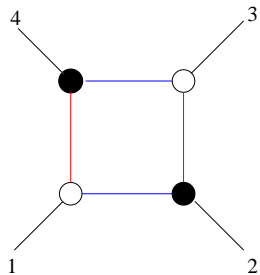}}
  \caption{Four particle on-shell diagram. }\label{fig:OnShellD4pt}
\end{figure}

Let us move forward. Let us complete the first diagram of Figure \ref{fig:Gluing} by attaching a white amplitude to the existing black one and a black amplitude
to the white one in such a way to form a square (see Figure \ref{fig:OnShellD4pt}). Notice that the four internal lines enforce four $\delta$-functions. 
First, recall that the white (anti-holomorphic) amplitudes are characterized by having all the holomorphic spinors proportional to each other, while the black
(holomorphic) amplitudes by having all the anti-holomorphic spinors proportional to each other. Now, starting the analysis of Figure \ref{fig:OnShellD4pt} from
the lower blue line, one can immediately see that the $\delta$-function along this line forces the related momentum to have the holomorphic spinor proportional
to $\lambda^{\mbox{\tiny $(1)$}}$ and the anti-holomorphic spinor proportional to $\tilde{\lambda}^{\mbox{\tiny $(2)$}}$. Thus,
\begin{equation}\label{OnShDiagMomD}
 p^{\mbox{\tiny (D)}}\:=\:z\lambda^{\mbox{\tiny $(1)$}}\tilde{\lambda}^{\mbox{\tiny $(2)$}},
\end{equation}
where the superscript (D) indicates the lower blue line of Figure \ref{fig:OnShellD4pt}, and we took its flow from the white amplitude to the black one.

Taking the flow of the momenta along the red lines upwards, following the same reasoning as before, the two momenta are given by
\begin{equation}\label{OnShDiagMoms}
 p^{\mbox{\tiny (L)}}\:=\:\lambda^{\mbox{\tiny $(1)$}}(\tilde{\lambda}^{\mbox{\tiny $(1)$}}-z\tilde{\lambda}^{\mbox{\tiny $(2)$}}),\qquad
 p^{\mbox{\tiny (R)}}\:=\:(\lambda^{\mbox{\tiny $(2)$}}+z\lambda^{\mbox{\tiny $(1)$}})\tilde{\lambda}^{\mbox{\tiny $(2)$}},
\end{equation}
where $p^{\mbox{\tiny (L)}}$ and $p^{\mbox{\tiny (R)}}$ are the momenta of the red line on the left and on the right respectively. From the form of these momenta,
it is already clear that attaching the two amplitudes induces a BCFW-deformation. Attaching the connected white-black amplitudes to an on-shell diagram has been
referred to as attaching a {\it BCFW-bridge} \cite{ArkaniHamed:2012nw}. If one is dealing with supersymmetric three-particle amplitudes, also the Grassmann
variables get deformed: if one is choosing to represent all the amplitudes in the $\eta$ representation, then 
$\eta^{\mbox{\tiny $(2)$}}\,\longrightarrow\,\eta^{\mbox{\tiny $(2)$}}+z\eta^{\mbox{\tiny $(1)$}}$; If instead the $\tilde{\eta}$ representation has been chosen,
then $\tilde{\eta}^{\mbox{\tiny $(1)$}}\,\longrightarrow\,\tilde{\eta}^{\mbox{\tiny $(1)$}}-z\tilde{\eta}^{\mbox{\tiny $(2)$}}$.

This is indeed not the end of the story. There is still one more momentum conserving $\delta$-function, which is the one related to the upper blue line. It fixes the 
parameter $z$ to
\begin{equation}\label{OnShDiagZ}
 z\:\longrightarrow\:z_{14}\:=\:\frac{[1,4]}{[2,4]},
\end{equation}
and thus the anti-holomorphic spinor of $p^{\mbox{\tiny (L)}}$ at this point is proportional to $\tilde{\lambda}^{\mbox{\tiny $(4)$}}$, while the holomorphic spinor
of  $p^{\mbox{\tiny (R)}}$ is proportional to $\lambda^{\mbox{\tiny $(3)$}}$.

Let us now attach a BCFW bridge to a more general on-shell diagram that we indicate with $\mathcal{M}_n$. Following the same reasoning as before, one gets exactly
the same value for the momenta in the internal lines as in eqs (\ref{OnShDiagMomD}) and (\ref{OnShDiagMoms}): the on-shell form related to the on-shell diagram 
$\mathcal{M}_n$ is mapped into a new on-shell form via the BCFW-bridge
\begin{equation}\label{BCFWbrForm}
 \begin{split}
  &\mbox{non-supersymmetric case: }\\
  &\hspace{.5cm}
   \hat{\mathcal{M}}_n(\lambda^{\mbox{\tiny $(1)$}},\tilde{\lambda}^{\mbox{\tiny $(1)$}},\lambda^{\mbox{\tiny $(2)$}},\tilde{\lambda}^{\mbox{\tiny $(2)$}})\:=\:
   \frac{dz}{z}\mathcal{M}_n(\lambda^{\mbox{\tiny $(1)$}},\tilde{\lambda}^{\mbox{\tiny $(1)$}}-z\tilde{\lambda}^{\mbox{\tiny $(2)$}},
    \lambda^{\mbox{\tiny $(2)$}}+z\lambda^{\mbox{\tiny $(1)$}},\tilde{\lambda}^{\mbox{\tiny $(2)$}}),\\
  &\mbox{supersymmetric case: }\\
  &\hspace{.5cm}
   \hat{\mathcal{M}}_n(\lambda^{\mbox{\tiny $(1)$}},\tilde{\lambda}^{\mbox{\tiny $(1)$}},\lambda^{\mbox{\tiny $(2)$}},\tilde{\lambda}^{\mbox{\tiny $(2)$}};
   	\tilde{\eta}^{\mbox{\tiny $(1)$}})\:=\\
  &\hspace{2.5cm}=\:
   \frac{dz}{z}\mathcal{M}_n(\lambda^{\mbox{\tiny $(1)$}},\tilde{\lambda}^{\mbox{\tiny $(1)$}}-z\tilde{\lambda}^{\mbox{\tiny $(2)$}},
    \lambda^{\mbox{\tiny $(2)$}}+z\lambda^{\mbox{\tiny $(1)$}},\tilde{\lambda}^{\mbox{\tiny $(2)$}}; 
    \tilde{\eta}^{\mbox{\tiny $(1)$}}-z\tilde{\eta}^{\mbox{\tiny $(2)$}})
 \end{split}
\end{equation}

\begin{figure}[htbp]
 \centering 
  \[
   {\raisebox{-2cm}{\scalebox{.50}{\includegraphics{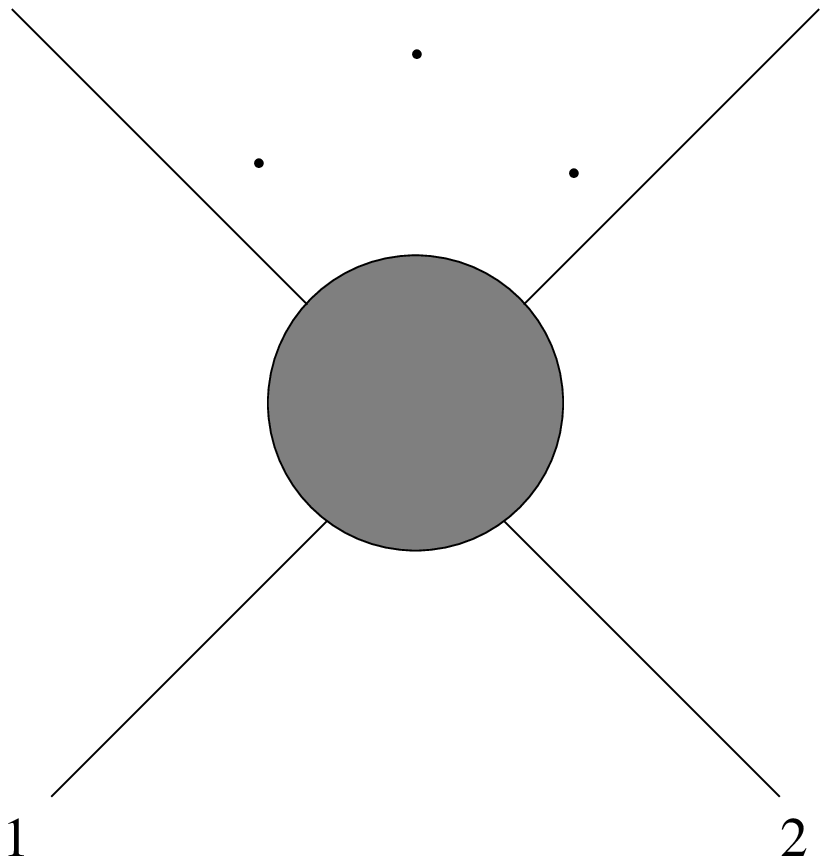}}}}\qquad\Longrightarrow\qquad
   {\raisebox{-2cm}{\scalebox{.50}{\includegraphics{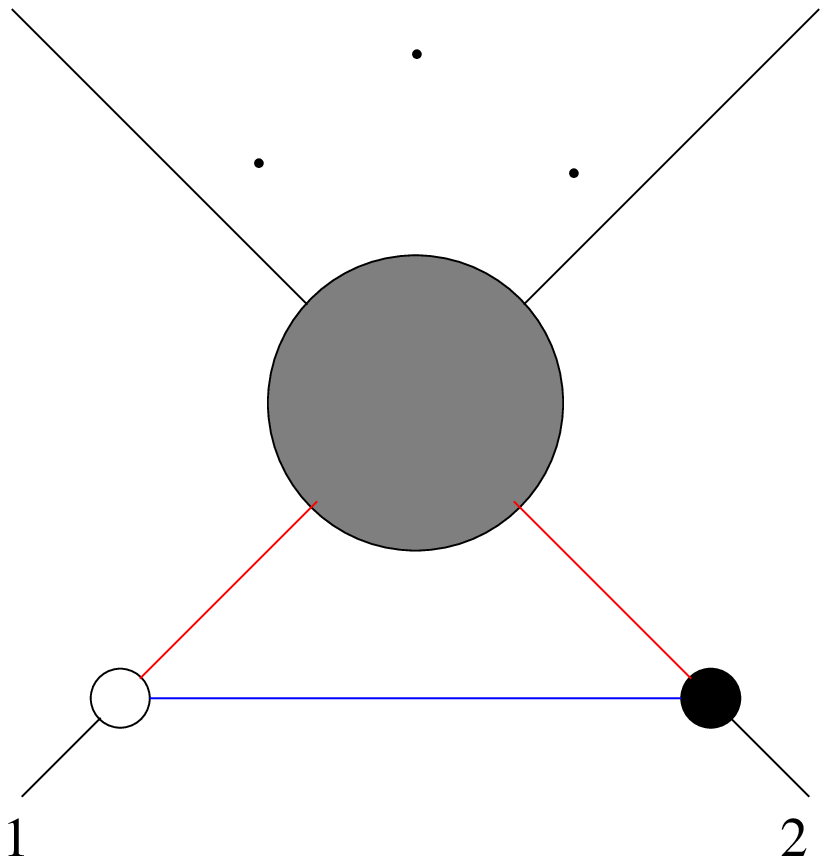}}}}
  \]
  \caption{BCFW bridge}\label{fig:BCFWbrdg}
\end{figure}

\begin{figure}[htbp]
 \centering 
  \[
   {\raisebox{-1.4cm}{\scalebox{.50}{\includegraphics{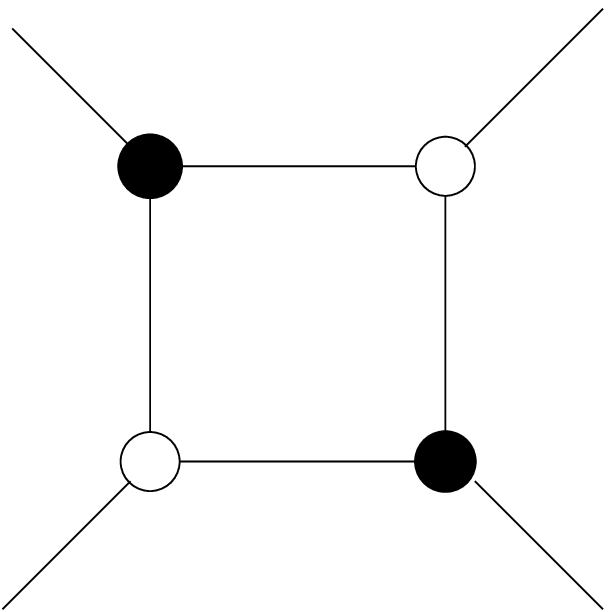}}}}\qquad\Longleftrightarrow\qquad
   {\raisebox{-1.4cm}{\scalebox{.50}{\includegraphics{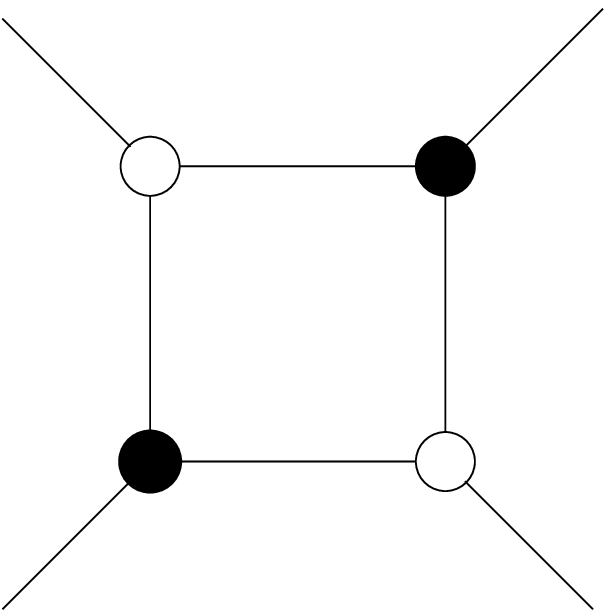}}}}
  \]
  \caption{Square moves}\label{fig:SqMoves}
\end{figure}

Let us comment further on the four-particle on-shell diagram of Figure \ref{fig:OnShellD4pt}. In the case of (color ordered)  $\mathcal{N}\,=\,4$  supersymmetric 
Yang-Mills theory, for which each on-shell state is the full supermultiplet, one can deduce that it does not really matter how the three-particle amplitudes are glued 
to form the square, as long as the white and black amplitude are alternate. This leads to another equivalence operation, named squared move (see Figure 
\ref{fig:SqMoves}): an on-shell diagram containing a square diagram with alternate white and black three-particle amplitudes is equivalent to the on shell diagram 
obtained by exchanging white and black three-particle amplitudes. Furthermore, the square diagram in Figure \ref{fig:OnShellD4pt} is the only inequivalent on-shell 
diagram one can write, {\it i.e.} any other non-trivial square diagram is equivalent to this one. In $\mathcal{N}\,=\,4$ supersymmetric Yang-Mills theory, this 
on-shell diagram represents the four-particle amplitude at tree-level\cite{Britto:2004nj}.

Another interesting feature to notice is the following. Let us stick to $\mathcal{N}\,=\,4$ supersymmetric Yang-Mills for simplicity. In the previous discussion, we 
look at this on-shell diagram as generated by attaching the $(1,2)$-BCFW-bridge to the on-shell diagram made up by gluing together just the upper black and white 
amplitudes. However, we can look at it also as attaching the $(1,4)$-BCFW-bridge to the on-shell diagram made up by gluing together the up-right white amplitude 
with the down-right black amplitude. Thus, it is easy to understand that this is nothing but the statement that it satisfies the consistency condition 
eq (\ref{4ptTest}).

Let us generalize the discussion to an arbitrary theory. The first consideration is that, contrarily to what happens in maximally supersymmetric theories for which
the states are supersymmetric coherent states, for a generic theory the helicity configurations of the amplitudes play an important role. Thus, the on-shell
diagrams can be decorated with an arrow as in Figure \ref{fig:3particle}: as a convention, we indicate the negative helicity states with incoming arrows and
the positive helicity states as outgoing ones \footnote{It is important not to confuse this ``helicity direction'' with the momentum flow in each state.}. 
Secondly, one should consider all the possible ways to glue the three-particle amplitudes, consistent with
the helicity configurations. For simplicity, let us work out the example of pure Yang-Mills theory (for which we consider color-ordered amplitudes), and let
us fix the helicities of the external states to be $(-,+,-,+)$ starting from the bottom-left and going counterclockwise.

\begin{figure}[htbp]
 \centering 
  \[
   {\raisebox{-1.1cm}{\scalebox{.50}{\includegraphics{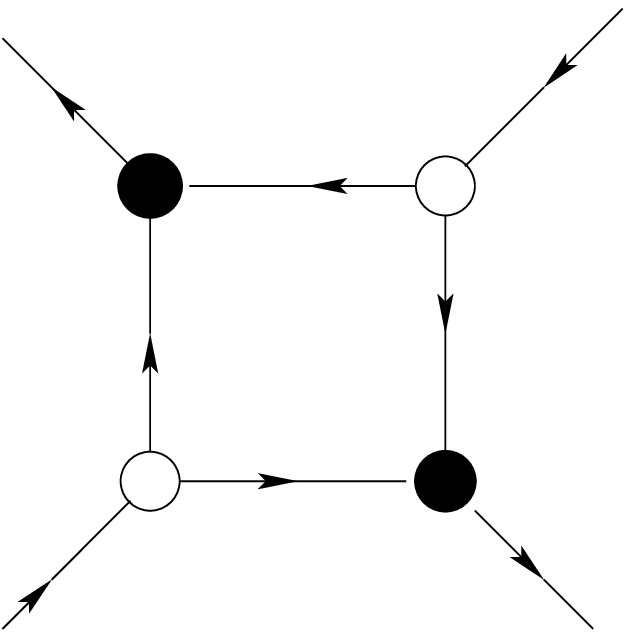}}}}\qquad
   {\raisebox{-1.4cm}{\scalebox{.50}{\includegraphics{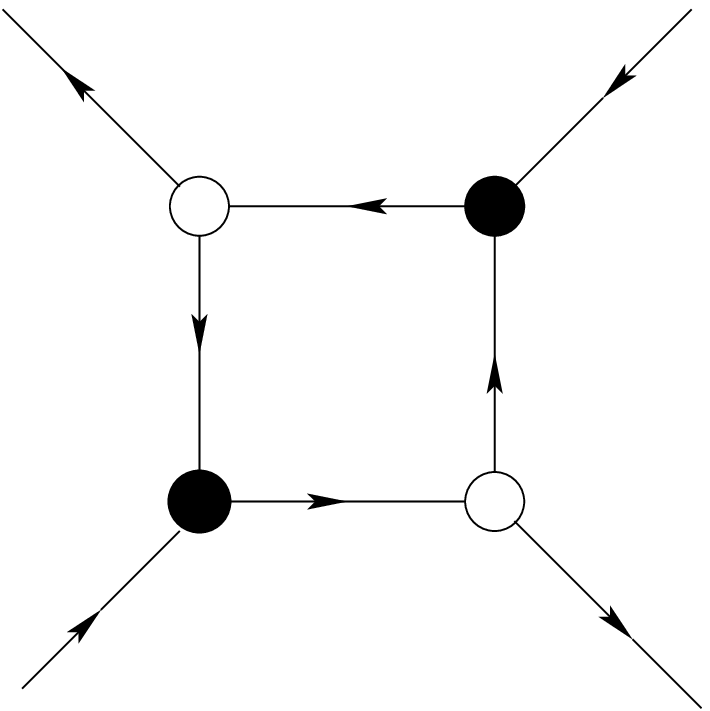}}}}\qquad
   {\raisebox{-1.4cm}{\scalebox{.50}{\includegraphics{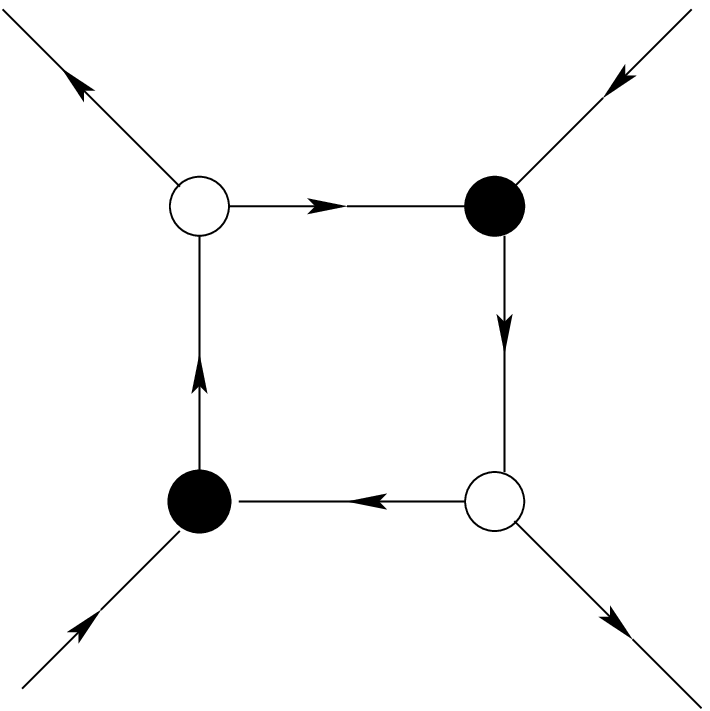}}}}
  \]
  \caption{Decorated diagrams. The ingoing/outgoing arrows represent negative/positive helicity states.}\label{fig:DecDiag}
\end{figure}

We can apply a BCFW-bridge in two different ways: either the negative helicity state is on the white amplitude and the positive helicity state on the black one 
(first diagram in Figure \ref{fig:DecDiag}), or vice-versa (second and third diagrams  in Figure \ref{fig:DecDiag}). However, the two turns out not to be equivalent
with each other. For the first BCFW-bridge, there is a unique way to glue the four on-shell forms. Furthermore, the helicity flow does not 
change independently of the fact that we look at it from the $s$-channel or from the $t$-channel. This does not occur for each of the other two diagrams.
It is easy to see that introducing the first BCFW-bridge is equivalent to induce a BCFW-deformation under which the amplitude is well-behaved in the complex-UV,
while the second BCFW-bridge induces BCFW-deformation under which the amplitude is not. Thus, the first diagram corresponds to a tree-level four particle
amplitude, while the (sum of the) other two diagrams do not.

As a final comment, a square move can be defined also for decorated on-shell diagrams if the exchange between black and white amplitude occurs together with
a flip of the direction of the helicity arrows.

\subsection{All-loop amplitudes in planar supersymmetric theories}\label{subsec:AllLoop}

So far we have been describing how to glue three-particle amplitudes together and we stressed that this way of implementing it is equivalent to consider 
certain factorization channels. It is therefore not a surprise that, at best, we could obtain amplitudes at tree-level. 

In order to provide a potentially full
on-shell diagrammatic for loop amplitudes, we need to be able to introduce further singularities. As we previously discussed, there is not a clear understanding
of all the singularities of the loop-integrands for a general theory. The only context in which we do know the physical meaning of the residues of those
singularities are supersymmetric theories in the planar limit: as already mentioned, those residues corresponds to $(n+2)$ $(L-1)$-loop amplitudes in the forward
limit. 

\begin{figure}[htbp]
 \centering 
  \[
   {\raisebox{-.9cm}{\scalebox{.25}{\includegraphics{AllLoopSYM1.eps}}}}\:=\:
   \sum_{\mbox{\tiny $\mathcal{P}^{\mbox{\tiny $(1,n)$}}$}}{\raisebox{-.9cm}{\scalebox{.25}{\includegraphics{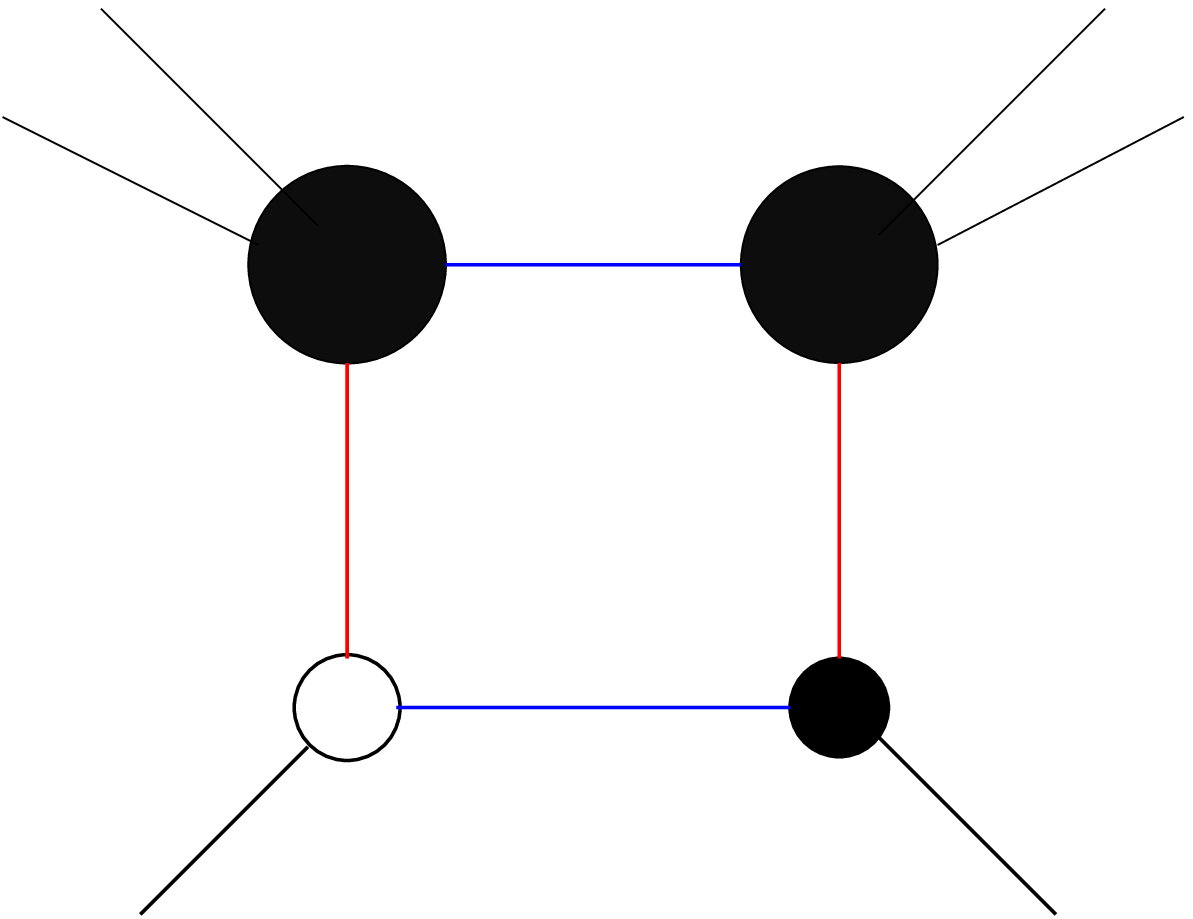}}}}+
   {\raisebox{-.9cm}{\scalebox{.25}{\includegraphics{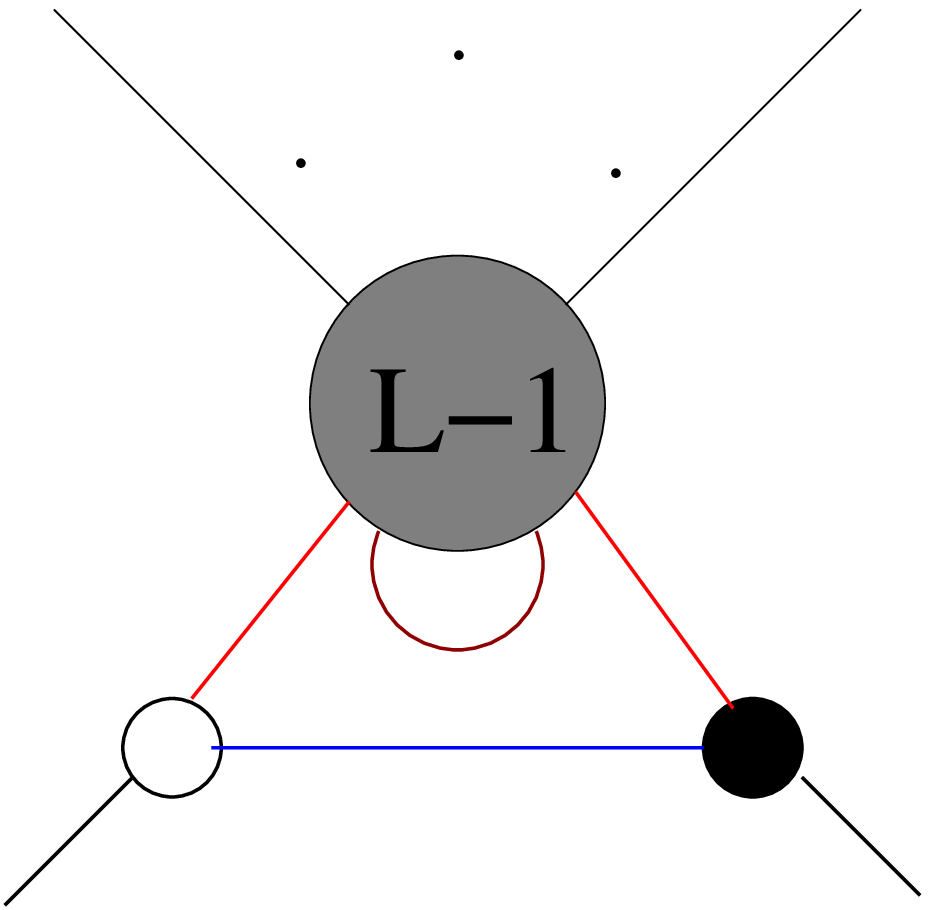}}}}
  \]
  \caption{All loop recursion relation for planar $\mathcal{N}\,=\,4$ SYM.}
  \label{fig:AllLoopSYMb}
\end{figure}

Thus, considering all the singularities, it is possible to write \cite{ArkaniHamed:2012nw}:
\begin{figure}[htbp]
 \centering 
  \[
   \partial\left[{\raisebox{-.9cm}{\scalebox{.25}{\includegraphics{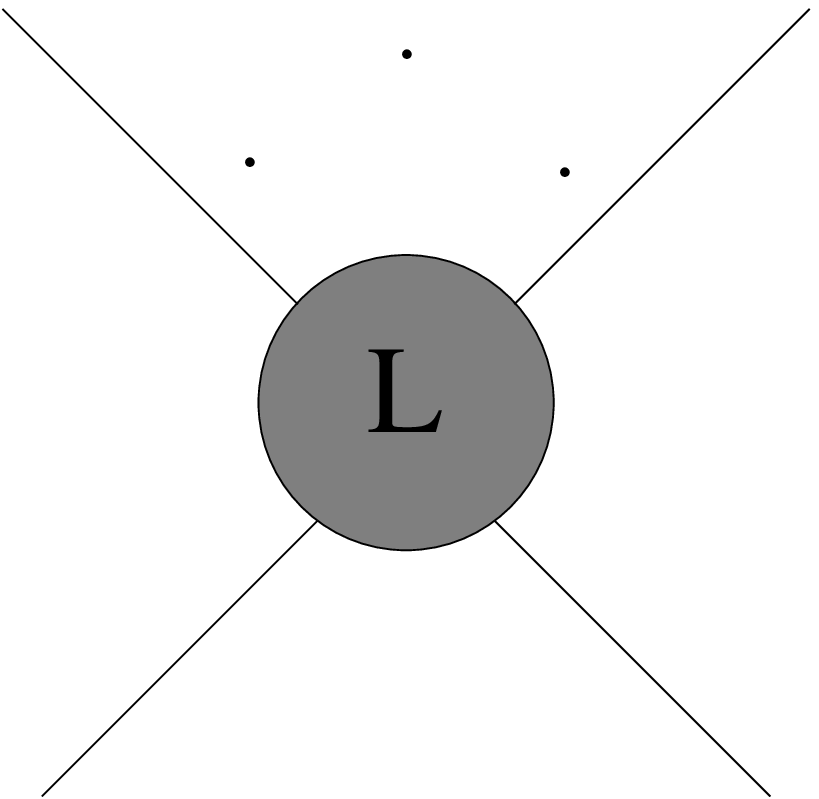}}}}\right]\:=\:
   \sum_{\mbox{\tiny poles}}{\raisebox{-1cm}{\scalebox{.40}{\includegraphics{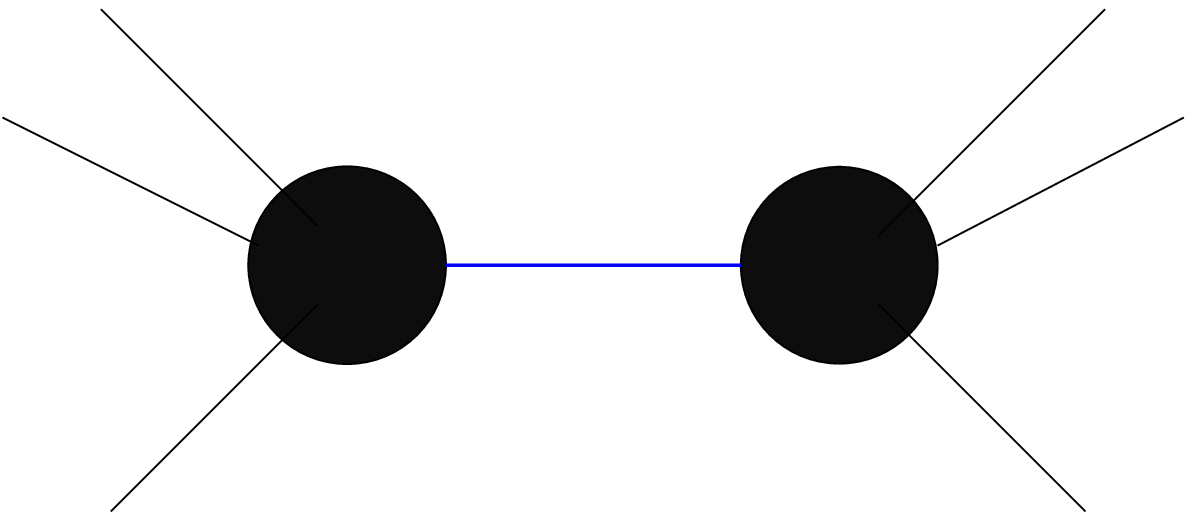}}}}+
   \sum{\raisebox{-.9cm}{\scalebox{.25}{\includegraphics{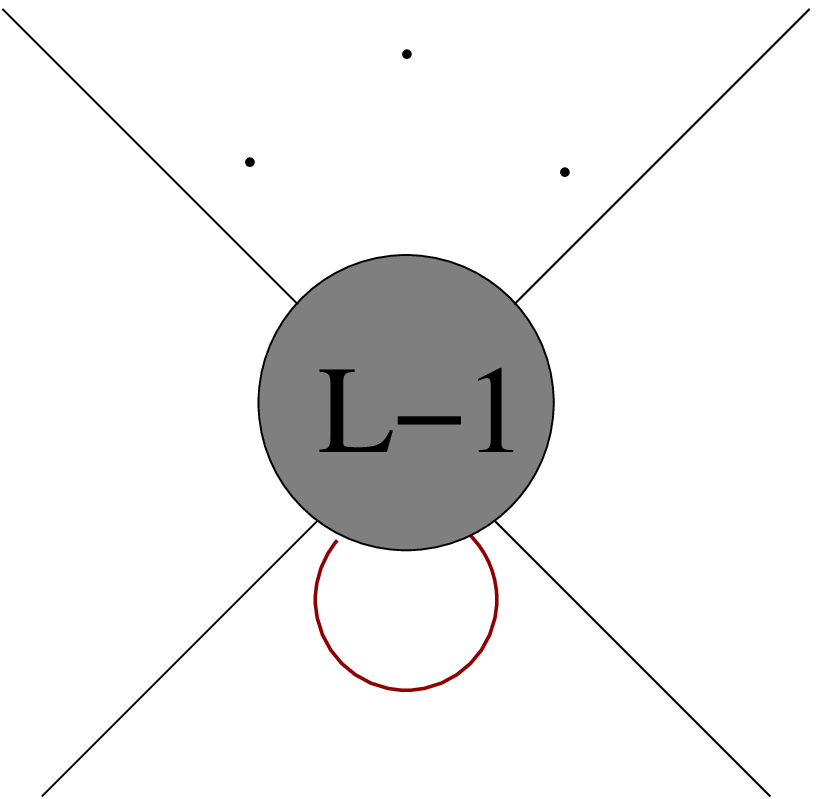}}}}
  \]
  \label{fig:AllLoopSYM}
\end{figure}

\noindent where the last diagram indicates the $(n+2)$ $(L-1)$-loop amplitudes in the forward limit. The diagrammatic equation above can be {\it integrated} through
a BCFW-bridge, which selects the channels and the forward amplitude (see Figure \ref{fig:AllLoopSYMb}).

A diagrammatic proof of this statement can be done by induction \cite{ArkaniHamed:2012nw} and it boils down to prove that the boundary have all the correct
factorization channels as well as forward limits.


\section*{Acknowledgments}

I would like to thank Freddy Cachazo and Eduardo Conde for collaboration on some of the results reported in this review as well as for many valuable discussions during
the years. I am also in debt with Fiorenzo Bastianelli who pushed me to give lectures about this subject, with Mauricio Martinez and Yacine Mehtar-Tani, and 
with Jos{\'e} Edelstein, Nicolas Grandi and the network Strings@ar, who gave me the opportunity to repeat them and, thus, keep them updated. 
I would like to thank as well for many valuable discussions about various aspects of this and related topics: Nicolas Boulanger, Xian O. Camanho, Dario Francia, 
Euihun Joung, Gustavo Lucena G{\'o}mez, Miguel Paulos, Rakibur Rahman, Augusto Sagnotti.

Most of this work has been done while affiliated to the Universidade de Santiago de Compostela and thus is funded in part by MICINN under grants FPA2008-01838 and 
FPA2011-22594, by the Spanish Consolider-Ingenio 2010 Programme CPAN (CSD2007-00042) and by Xunta de Galicia (Conseller{\'i}a de Educaci{\'o}n, grant INCITE09 206 121 
PR and grant PGIDIT10PXIB206075PR) and by FEDER. I was also supported as well by the MInisterio de Ciencia e INNovaci{\'o}n through the Juan de la Cierva program.
Its final stages have been instead carried out with the current affiliation to Instituto de F{\'i}sica Te{\'o}rica, Universidad Aut{\'o}noma de Madrid / CSIC, where it
has been supported in part by Plan Nacional de Altas Energ{\'i}as (FPA2011-25948 and FPA2012-32828), Spanish MICINN Consolider-Ingenio 2010 Program CPAN 
(CSD2007-00042), Comunidad de Madrid HEP-HACOS S2009/ESP-1473 and the Spanish MINECO's Centro de Excelencia Severo Ochoa Programme under grants SEV-2012-0234 and 
SEV-2012-0249.



\bibliographystyle{ws-ijmpa}

\bibliography{amplitudesrefs}	

\end{document}